\documentclass[a4paper,11pt]{article}
\pdfoutput=1 

\usepackage{jheppub}
\usepackage{color} 
\usepackage{float}

\usepackage{amsmath,graphicx,amssymb,subfigure,bbm}
\usepackage[all]{xy}

\newcommand{\be}{\begin{equation}}
\newcommand{\ee}{\end{equation}}
\newcommand{\bea}{\begin{eqnarray}}
\newcommand{\eea}{\end{eqnarray}}

\begin{document}

\title{\boldmath A universal order parameter for Inverse Magnetic Catalysis}

\author[a]{Alfonso Ballon-Bayona,}
\author[b,c]{Matthias Ihl,}
\author[d,e]{Jonathan P Shock}
\author[b]{and Dimitrios Zoakos}
\affiliation[a]{Instituto de F\'isica Te\'orica, Universidade Estadual Paulista,\\
01140-070 S{\~a}o Paulo, SP, Brazil.}       
\affiliation[b]{Centro de F\'isica do Porto e Departamento de F\'isica e Astronomia,\\
Faculdade de Ci{\^e}ncias da Universidade do Porto, \\�Rua do Campo Alegre 687, 4169-007 Porto, Portugal.}
\affiliation[c]{African Institute for Mathematical Sciences,\\ 6-8 Melrose Rd, Muizenberg, Cape Town 7945, South Africa.}
\affiliation[d]{Department of Mathematics and Applied Mathematics,
University of Cape Town,\\
Private Bag, Rondebosch, Cape Town 7700, South Africa.}
\affiliation[e]{National Institute for Theoretical Physics,\\
Private Bag X1, Matieland, Stellenbosch 7602, South Africa.}
\emailAdd{aballonb@ift.unesp.br}
\emailAdd{matthias.ihl@fc.up.pt}
\emailAdd{jonathan.shock@uct.ac.za}
\emailAdd{zoakos@gmail.com}

\abstract{We revisit the chiral transition in the finite density Sakai-Sugimoto model and find that, at fixed temperature $T$, the magnetisation  near the critical line $\mu_c(B)$ acts as an order parameter to distinguish Inverse Magnetic Catalysis from Magnetic Catalysis. Moreover, we propose a universal relation between $\mu_c (B)$ and the magnetisation that allows us to predict the behaviour of the former from the behaviour of the latter. We find that a similar relation holds, at fixed chemical potential $\mu$, for the critical line $T_c(B)$. Our results are obtained by investigating a fully numerical solution to the relevant equations. At low temperatures our results reduce to those obtained by Preis, Rebhan and Schmitt [JHEP 1103 (2011) 033], based on a semi-analytic approximation. }

\maketitle
\flushbottom

\section{Introduction}

One of the ultimate aims of holography is to construct a string dual of quantum chromodynamics (QCD). This would provide a powerful tool for the investigation of non-perturbative QCD at strong coupling. While this goal has remained elusive for various reasons, over the past few years steady progress has been made in improving our understanding of holographic QCD, a term that has been used collectively for a variety of gravity duals describing the large $N_c$ limit of QCD-like theories. The most successful top-down holographic QCD model has been the Sakai-Sugimoto (S-S) model \cite{Sakai:2004cn,Sakai:2005yt}. It can be considered a prototype for the many different models of holographic QCD that are available in the literature. While some such models may have specific features that are more realistic when comparing with real-world QCD, the advantage of the S-S model lies in its simplicity and applicability to a broad range of complex problems.  

The S-S model is particularly well suited for investigating the qualitative aspects of the QCD phase diagram in the large $N_c$ limit. The first important step in this direction was taken by Aharony, Sonnenschein and Yankielowicz, who developed a holographic description of the finite temperature deconfinement and chiral transition \cite{Aharony:2006da}. Subsequently baryonic and isospin density were introduced in \cite{Bergman:2007wp,Rozali:2007rx} and \cite{Aharony:2007uu} respectively. Interestingly, at non-zero density non-homogeneous phases also appear in the phase diagram of the S-S model \cite{Ooguri:2010xs,Bayona:2011ab,Kaplunovsky:2012gb}. Many aspects of the S-S model at finite temperature and density have been studied in the past few years; recent examples include its Fermi liquid behaviour and the corresponding collective excitations, most notably the zero sound mode \cite{DiNunno:2014bxa}.

\subsection{(Inverse) Magnetic Catalysis of chiral symmetry breaking}

Strong magnetic fields are expected to play an important role in two observationally accessible laboratories: non-central relativistic heavy ion collisions (large temperature and small density) and magnetars (small temperature and large density). In both situations, the magnetic field can be of the same order as $\Lambda_{\mathrm QCD}$ and will conceivably influence the physics governed by the strong interactions. In recent years this has motivated the investigation of the influence of an external magnetic field on the QCD phase diagram \cite{Andersen:2014xxa}.

Chiral symmetry breaking is an inherently non-perturbative feature of QCD. The enhancement of chiral symmetry breaking due to the presence of a magnetic field at zero temperature and zero chemical potential, known as magnetic catalysis (MC), is by now well understood \cite{Shovkovy:2012zn,Miransky:2015ava}. The physical picture is the following: A  magnetic field leads to a dimensional reduction from $3+1 \to 1+1$ and the chiral condensate becomes a measure of quark-antiquark pairing in the lowest Landau level (LLL). Since the magnetic field also generates a dynamical quark mass in the LLL, it becomes a catalyst of chiral symmetry breaking. According to MC one would expect, at finite temperature, that a non-zero magnetic field $B$ should increase the critical temperature for chiral symmetry restoration $T_c$. Similarly, at zero temperature and finite density, one would expect a critical chemical potential $\mu_c$ increasing with $B$. 

However, it turns out that there are new subtle phenomena that work against MC leading, in some cases and for a certain range of parameters, to the opposite behaviour for the critical temperature $T_c$ (or critical chemical potential $\mu_c$); this effect has been dubbed inverse magnetic catalysis (IMC). In fact, IMC has already been observed in lattice QCD simulations \cite{Bali:2011qj,D'Elia:2012tr,Ilgenfritz:2013ara} at finite temperature and zero chemical potential. The first IMC result at finite chemical potential was obtained in an effective Nambu-Jona-Lasinio (NJL) model of QCD \cite{Inagaki:2003yi}. The physical mechanism behind this effect was investigated by Preis, Rebhan and Schmitt (PRS) in \cite{Preis:2010cq,Preis:2012fh} within the framework of holographic QCD, where it was called Inverse Magnetic Catalysis (IMC) for the first time. These authors considered the deconfined phase of the S-S model and found, for small fixed temperatures, that the critical chemical potential $\mu_c$ decreases with the magnetic field $B$. 

It is important to remark that there are two very different physical mechanisms associated with IMC.  At zero density and finite temperature, the lattice results in \cite{Bali:2011qj,D'Elia:2012tr,Ilgenfritz:2013ara} can be interpreted in terms of the chiral condensate. Although those results differ from most of the predictions in effective models, an explanation of the discrepancy was provided in  \cite{Bruckmann:2013oba}. The reasoning is the following: There are two different contributions to the chiral condensate $ \langle \bar \psi \psi \rangle$, namely the so-called valence and sea effects. The valence effect is associated with dynamical mass generation and the Dirac operator favouring MC, whereas the sea effect measures the magnetic influence on the quark determinant and favours IMC. At zero temperature the valence effect dominates leading to MC and at finite temperature this effect is overtaken by the sea effect leading to IMC.  An alternative explanation for IMC at zero density can be given in terms of magnetic inhibition of confinement, which is found after considering both the gluon and quark contributions to the free energy \cite{Fraga:2012ev}. In holographic QCD, zero density IMC appears in models that take into account backreaction of the magnetic field in the background dual to QCD. At finite density, the physical mechanism behind IMC  was described in \cite{Preis:2010cq,Preis:2012fh}. Here, the picture is slightly different: although the magnetic field generates a dynamical mass that increases the chiral condensate, at finite density it also contributes (along with the chemical potential) to the energy cost to form such a condensate. At small values of the magnetic field the energy cost is higher than the gain from condensation and thus IMC occurs. 

MC in gauge/gravity models was extensively investigated in various setups \cite{Filev:2010pm}.
The first MC studies in the framework of the S-S model were carried out in \cite{Bergman:2008sg,Johnson:2008vna}. In \cite{Johnson:2008vna} it was found that, at vanishing density, the critical temperature $T_c$ for chiral symmetry restoration increases with $B$. It is in this sense that the magnetic field catalyses chiral symmetry breaking. This effect, however, dissappears in the antipodal limit (i.e. for massless quarks). Moreover, taking into account backreaction effects in the antipodal S-S model leads to IMC as a consequence of magnetic inhibition of confinement \cite{Ballon-Bayona:2013cta}. The first studies of the S-S model at non-zero magnetic field and density were carried out in \cite{Thompson:2008qw, Bergman:2008qv}, within the antipodal limit. As mentioned above, IMC at finite density was observed in \cite{Preis:2010cq} (see also  \cite{Preis:2012fh, Preis:2012dw, Preis:2011sp}) by considering the full phase diagram $(T,\mu,B)$ for the deconfined phase of the S-S model. The results in \cite{Preis:2010cq} rely on a semi-analytic approximation that is valid either at small temperatures or large constituent quark masses. IMC at zero density has also been investigated recently in bottom-up approaches \cite{Mamo:2015dea,Rougemont:2015oea,Dudal:2015wfn,Fang:2016cnt,Evans:2016jzo, Gursoy:2016ofp}.

\subsection{An order parameter for IMC}

In this paper we find the exact numerical solutions for the chirally broken and chirally symmetric profiles in the deconfined S-S model and use those solutions to investigate the  $(T,\mu,B)$ phase diagram. At small temperatures our results agree with those of  \cite{Preis:2010cq}, confirming IMC at non-zero density. We propose a novel order parameter that can distinguish the normal effect of magnetic catalysis (MC) from the inverse effect (IMC). Our proposal for the order parameter is, at fixed $T$, the magnetisation near the critical line $\mu_c(B)$. The magnetisation $M$ exhibits a jump across the phase transition from the chirally broken to the chirally symmetric phase. For a given temperature $T$, the magnetisation jump $\Delta M$ will be either positive (IMC) or negative (MC). The corresponding magnetic susceptibility diverges at the phase transition. In turn, there will be a critical value $B_c(T)$ for which $\Delta M = 0$ at the phase transition, signifying the onset of MC (or the end of IMC). We will show that  $B_c(T)$ decreases with growing $T$ until a certain critical $T_c$ is reached for which $B_c(T_c)=0$. Above that critical temperature, IMC is not possible and only the normal MC effect remain. 

We will find a useful relation between the critical chemical potential $\mu_c (B)$ and the magnetisation jump $\Delta M (B)$. That relation enables us to track the transition between the two phases at arbitrary $B$. At fixed chemical potential $\mu$, we will find a similar relation that allows us to track the critical temperature $T_c(B)$ from  $\Delta M (B)$.  It should be noted that our proposed order parameter for IMC will be universal and can be used in any phenomenological or holographic model for the chiral phase transition in the presence of a magnetic field; in particular, it will be free of model-dependent mechanisms. Last but not least, the order parameter proposed in this paper should be very useful in phenomenological or holographic models where a chiral condensate is either difficult to calculate or not well defined. 

The rest of the paper is organized as follows. In section \ref{sec:SSmodel} we review the zero density S-S model focusing on the deconfinement transition. Then, in section \ref{sec:DeconfSS}, we turn on a magnetic field and density in the deconfined phase. Our numerical results for the chiral transition are presented in section \ref{sec:Numerics}. The description of the magnetisation as an order parameter for IMC is provided in section \ref{sec:magnetisation}, and we finish with our conclusions in \ref{sec:Conclusions}.
In appendix \ref{AppA} we present identities useful for section \ref{sec:DeconfSS}, whereas appendix \ref{AppB} describes our analytic results for small magnetic field and temperature.  

\section{The S-S model  at zero density and zero magnetic field}

\label{sec:SSmodel}

We start this section by briefly reviewing confinement and chiral symmetry breaking in the S-S model\footnote{For a recent review of the S-S model, cf. \cite{Rebhan:2014rxa}.}. Then we review specifically the deconfined phase in the absence of a magnetic field and chemical potential. In section \ref{sec:DeconfSS} we will investigate the effect of a non-zero magnetic field and a non-zero chemical potential in the chiral transition for the deconfined phase.

\subsection{Confinement and chiral symmetry breaking in the  S-S model}

The S-S model is the flavoured version of Witten's model \cite{Witten:1998zw}, which arises from a stack of coincident $N_c$ $D4$ (color) branes in Type IIA String Theory. At weak coupling, the $D4$ brane model reduces to 5-d $SU(N_c)$ Super Yang-Mills theory. Compactifying one of the spatial coordinates with anti-periodic boundary conditions for the fermions, the theory reduces to 4-d (non-supersymmetric) Yang-Mills theory.  At strong coupling, the description is that of a $D4$-brane background given by the 10-d metric 
\begin{equation}
ds^2 = \frac{u^{3/2}}{R_{D4}^{3/2}} \left [ - dt^2 + d x_i^2 + f(u) d \tau^2  \right ]
+ \frac{R_{D4}^{3/2}}{u^{3/2}} \left [ \frac{du^2}{f(u)}
+ u^2 d \Omega_4^2 \right ] \, , \quad f(u) = 1 - \frac{u_{KK}^3}{u^3} \, ,  
\label{eq:D4confined}
\end{equation}
with a dilaton and RR 4-form given by
\bea
e^{\phi}  &=& g_s  \frac{u^{3/4}}{R_{D4}^{3/4}} \quad , \quad
F_4 = \frac{2 \pi N_c}{V_{S^4}} \epsilon_4 \,, \label{eq:D4confinedII}
\eea
where $t$ and $x^i$ are the  4-d coordinates and $\tau$ is the compactified direction on the $D4$-brane world-volume. $V_{S^4}$ denotes the volume of the unit four-sphere with volume form $\epsilon_4$. The parameter $g_s$ denotes the string coupling. The $D4$-brane parameter $R_{D4}$ is given by
\bea
R_{D4}^3 = \pi g_s N_c \, \ell_s^3 \, , \label{eq:RD4}
\eea
where $\ell_s$ is the fundamental string length. The submanifold spanned by $\tau$ and $u$ has the shape of a cigar with tip located at $u=u_{KK}$; for the tip to be non-singular, we need to impose a periodicity condition on $\tau$, namely
\bea
2 \pi R = \delta \tau = \frac{4 \pi}{3} \frac{R_{D4}^{3/2}}{u_{KK}^{1/2}} \, ,
\eea
where $R$ is the radius of the compactified circle.  The background \eqref{eq:D4confined} and \eqref{eq:D4confinedII}, both at zero and (with slight modifications \cite{Aharony:2006da}) at low temperatures,  describes the confined phase of the S-S model. 
As usual, the parameters of the gauge theory, i.e., the glueball mass scale $M_{KK}$, the 5-d gauge coupling $g_5$ and the low-energy 4-d gauge coupling $g_{YM}$ are obtained with the identifications
\begin{equation}
g_5^2 = 4 \pi^2 g_s \ell_s \ , \quad 
g_{YM}^2 = \frac{g_5^2}{2 \pi R} 
\quad \mathrm{and} \quad 
M_{KK} = \frac{1}{R} \  . \label{eq:couplings}
\end{equation}
From \eqref{eq:RD4} and \eqref{eq:couplings}  we can express the D4-brane parameter $R_{D4}$ in terms of the 4-d 't Hooft coupling $\lambda = g_{YM}^2 N_c$ as
\bea
R_{D4}^3 = \frac{R}{2} \ell_s^2 \, \lambda  \,. 
\eea

On the other hand, following the dictionary \cite{Kinar:1998vq}, the confining string tension $\sigma$ associated with the background \eqref{eq:D4confined}, \eqref{eq:D4confinedII} takes the form 
\begin{equation}
\sigma = \frac{1}{2 \pi \ell_s^2} \frac{u_{KK}^{3/2}}{R_{D4}^{3/2}} = 2 \frac{ \bar \lambda}{ R^2}
\quad { \rm where} \quad 
\bar \lambda = \frac{\lambda}{27 \pi}  \,. 
\end{equation}
In this paper we will use the units of \cite{Preis:2010cq} where 
\begin{equation}
R_{D4} = u_{KK} = \frac32 R  \quad \Rightarrow \quad 
\frac{1}{2 \pi \ell_s^2} = 2 \frac{ \bar \lambda}{R^2} \,.  \label{eq:PRSunits}
\end{equation}

Sakai and Sugimoto incorporated $N_f$ (flavour) $D8/\overline{D8}$-brane pairs localised at different points on the compact circle, which provide $N_f$ left-handed and $N_f$ right-handed quarks coupled to the gauge theory in the dual picture \cite{Sakai:2004cn, Sakai:2005yt}. The flavour branes span the coordinates $\left(t, x_i, \tau , \Omega_4\right)$, and follow a trajectory $\tau (u)$ in the $(u,\tau)$-submanifold. In the UV, i.e., for  $u \to \infty$, the stack of $N_f$ $D8$-branes is located at $\tau = - L/2$, and the $N_f$ $\overline{D8}$-branes are located at $\tau= L/2$ with $L \leq  \pi R $. The parameter $L$ will be related to the constituent quark mass in the theory, as discussed below.

An important observation is that, geometrically, the flavour $D8$ and $\overline{D8}$-branes do not have a locus to end on in the background described by \eqref{eq:D4confined}-\eqref{eq:D4confinedII}, which already hints at the possible shapes of the trajectories $\tau(u)$: Namely, the trajectories must be such that the flavour branes and anti-branes connect smoothly at some minimal value, $u_0$, in the IR.
Given that the dynamics of the fields on the two stacks of branes become coupled as they connect at $u_0$, the merging of the branes is a geometrical realisation of dynamical chiral symmetry breaking, from $U(N_f)_L \times U(N_f)_R$ to the diagonal subgroup $U(N_f)_V$.

To determine the specific flavour brane configuration, we need to solve the Dirac-Born-Infeld (DBI) equations on the $D8$-branes. The induced metric for the $D8$-branes reads 
\bea
ds^2_{D8} &=&  \frac{ u^{3/2}}{R_{D4}^{3/2}} \left[ -\text{d}t^2 +d x_i^2  \right]+ \frac{ u^{3/2}}{R_{D4}^{3/2}} \left[ f(u) + \frac{R_{D4}^3}{u^3} \frac{\left(\partial_{\tau} u\right)^2}{f(u)}\right] \text{d}\tau^2  + R_{D4}^{3/2} u^{1/2} \Omega_4^2 \ . 
\eea 
In the absence of a background gauge field on the branes, the Chern-Simons (CS) term is not necessary to calculate the classical equations of motion and thus it is not included in the present discussion. This will be introduced in the following section, where finite gauge fields are turned on on the brane. Here, we restrict to the DBI action
\bea 
S_{DBI, D8}= - \frac{\mu_8}{g_s} \, \mathcal{C} \int d \tau \; u^4 \sqrt{ f(u) + \frac{R_{D4}^3}{u^3}  \frac{\left(\partial_{\tau}  u \right)^2}{f(u)} } \ , 
\eea 
where $\mathcal{C}$ collects various factors from the integrations over the remaining world volume coordinates. It is straightforward to solve the resulting DBI equations for the embedding profile of the flavour branes. In general, the profile can only be obtained numerically. There is a one-to-one correspondence between the minimal value $u_0$ and $L$ which can be given as
\be
L = \int \text{d}\tau = 2 \int_{u_0}^{\infty} \frac{\text{d}u}{\left(\partial_{\tau} u \right)}= 2 R_{\text{D4}}^{3/2} \int_{u_0}^{\infty} \text{d} u \left( f(u) u^{3/2} \sqrt{\frac{f(u) u^8}{f(u_0) u_0^8}-1}\right)^{-1} \ .
\ee
In the limiting case $L = \pi R$, we find the original antipodal model of \cite{Sakai:2004cn, Sakai:2005yt}, where the two stacks of branes join smoothly at $u_0 = u_{KK}$. In the opposite limit, $u_0$ large, 
we get $L \sim \left( \frac{R_{\text{D4}}^3}{u_0}\right)^{1/2}$. The general, non-antipodal configurations lead to a richer phase structure. For example, the model at finite temperature \cite{Aharony:2006da}, with all other fields turned off, features a critical value $L_{\text{cr.}}/R$, below which the chiral phase transition occurs for temperatures above the deconfinement temperature, $T_{\chi}> T_{\text{dec.}} = 1/(2\pi R)$. Moreover, the extra scale 
$u_0 - u_{KK}$ has been associated with a mass scale for mesons or a constituent quark mass \cite{Sakai:2004cn,Sakai:2005yt,Casero:2005se,Erdmenger:2007cm}, and also plays an important role in generating an attractive potential for (holographic) baryons.\footnote{However, there are some important shortcomings with the traditional approach. For instance, the Goldstone boson of spontaneous chiral symmetry breaking remains massless even in the non-antipodal case. These issues are addressed, e.g.,  in the recent paper \cite{Sonnenschein:2016pim}.}
It should be noted that both the antipodal and non-antipodal configurations are stable. This can be confirmed by a perturbative analysis  of the backreaction of the D8-branes \cite{Burrington:2007qd,Bigazzi:2014qsa}. The fluctuations around the flavour brane embedding do not become tachyonic, at least in the perturbative regime. This can be attributed to an intricate cancellation between the DBI and CS parts of the $D8$-brane action.\\


\subsection{Finite temperature and the deconfinement transition in the S-S model} 

It is expected that at finite temperature in large $N_c$ Yang-Mills theory, the gluons should undergo a deconfinement transition. This transition maps holographically to a Hawking-Page (HP) phase transition on the gravity side in Witten's D4-brane model\footnote{Note that there is some discussion over whether the confinement/deconfinement transition is a Hawking-Page phase transition, or whether the transition is actually realised as a Gregory-Laflamme instability \cite{Mandal:2011ws}.}. The HP transition describes the transition from the cigar manifold \eqref{eq:D4confined} to the black brane manifold
\begin{equation} \label{eq:D4metric}
ds^2 = \frac{u^{3/2}}{R_{D4}^{3/2}} \left [ -h(u) dt^2+dx_i^2+d\tau^2\right ] + 
\frac{R_{D4}^{3/2}}{u^{3/2}} \left [\frac{du^2}{h(u)}+u^2 d\Omega_4^2\right ] \, , 
\quad 
h(u)=1- \frac{u_T^3}{u^3} \, . 
\end{equation}
The dilaton and 4-form are still given by \eqref{eq:D4confinedII}. The temperature is obtained by taking an 
imaginary time period, while the absence of conical singularities constrains the horizon position $u_T$ to 
be related to the temperature $T$ by 
\begin{equation}
\frac{1}{T} = \delta t = \frac{4 \pi}{3} \frac{R_{D4}^{3/2}}{u_T^{1/2}} \, . \label{eq:uT}
\end{equation}
As described in \cite{Aharony:2006da}, evaluating the renormalised on-shell actions for \eqref{eq:D4confined} and \eqref{eq:D4metric}, with the dilaton and 4-form given by \eqref{eq:D4confinedII}, one finds that the HP transition occurs at $T_{\text{dec.}} = 1/(2\pi R)$ which is the holographic realisation of the deconfinement transition. Adding $N_f$ $D8$/$\overline{D8}$ branes in the antipodal configuration to the black brane background \eqref{eq:D4metric}, one finds that in the deconfined phase, chiral restoration is automatically achieved \cite{Aharony:2006da}, leading to a very simple brane embedding. The non-antipodal scenario offers a richer phase structure where  the chiral transition occurs at a temperature $T_{\chi}$, which may be higher than $T_{\text{dec.}}$, depending on the parameter $u_0$, associated with the constituent quark mass.  

\begin{figure}[t] 
   \centering
   \includegraphics[width=5in]{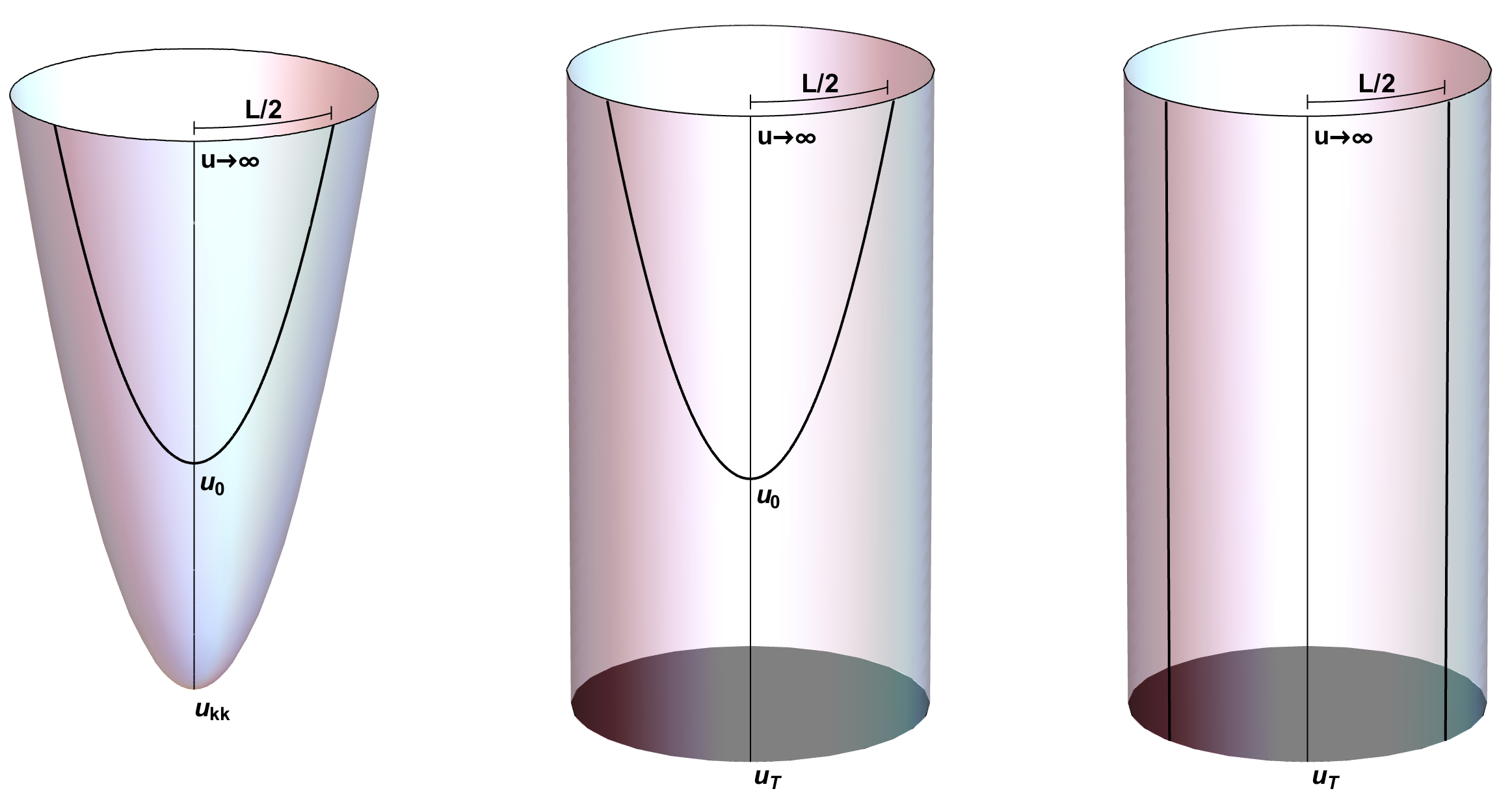} 
   \caption{\small Left figure: The confined geometry where chiral symmetry breaking is guaranteed, with the $D8$/$\overline{D8}$ branes joining at radial position $u_0$. Middle figure: The deconfined phase, also with chiral symmetry breaking, which only happens for non-antipodal brane configurations. Right figure: The deconfined phase with chiral symmetry restored. }
  \label{fig:transitions}
\end{figure}

The three different configurations (one in the confined case and two in the deconfined case) can 
be seen in Figure \ref{fig:transitions}.

In the rest of the paper, we will study the phase structure of the deconfined phase of the S-S model in the presence of a magnetic field and chemical potential. We will describe the chirally broken and chirally symmetric regions in the phase diagram. Then we will investigate the behaviour of the magnetisation and density near the chiral transition and argue that the magnetisation is the right observable to distinguish IMC from MC.


\section{The deconfined S-S model at finite density and magnetic field}

\label{sec:DeconfSS}

In this section we describe the dynamics of a $D8$/$\overline{D8}$-brane ($N_f=1$) in the deconfined phase of the S-S 
model with non-zero magnetic field and chemical potential. We start by writing the DBI-CS equations 
for a general field configuration and then specify the ansatz for the problem at hand. We finish the section describing 
the details of the chirally broken and chirally symmetric profiles that appear as solutions of the DBI-CS equations. 

\subsection{DBI-CS equations from probe branes in the deconfined S-S model}

A probe $D8$/$\overline{D8}$-brane pair is described by a DBI action $S_{\text{DBI}} = S_{\text{DBI}}^L+S_{\text{DBI}}^R$ where 
\bea
S_{\text{DBI}}^{L(R)}=-\mu _8\int d^5x \,  d^4\Omega \, e^{-\Phi }\sqrt{-\det \left[G_{MN}^{L(R)}+2\pi \alpha' F_{MN}^{L(R)}\right]} \,.
\eea
Consider profiles described by $\tau=\tau(u)$ of a probe $D8$/$\overline{D8}$-brane in the black brane background \eqref{eq:D4metric}. The induced metric on either brane reads
\bea \label{eq:D8metric}
ds^2_{D8} = \frac{u^{3/2}}{R_{D4}^{3/2}} \left [ -h(u) dt^2+dx_i^2 \right ] + \left [ \frac{R_{D4}^{3/2}}{u^{3/2} h(u)} + \frac{u^{3/2}}{R_{D4}^{3/2}} \left (\partial_u  \tau \right )^2 \right ] du^2+ R_{D4}^{3/2} u^{1/2} d\Omega_4^2  \ . 
\eea 
We will here consider gauge fields which do not have components in the $S^4$ directions. In this case, one finds the effective action  
\bea 
S_{\text{DBI}}^{L(R)} &=&-\int d^4x \int_{u_0}^{\infty }du \, \gamma (u)\sqrt{-\det \left[ g_{mn}^{L(R)}+\beta F_{mn}^{L(R)}\right ]},
\eea
where the effective 5-d metric is 
\bea
g_{mn}^{L(R)} d x^m dx^n =g_{tt}dt^2 + g_{xx}dx_i^2+g_{uu}^{L(R)}du^2 \, ,  
\eea
\bea
g_{tt} = -h(u) \frac{u^{3/2}}{R_{D4}^{3/2}} \, , \quad 
g_{xx} =\frac{u^{3/2}}{R_{D4}^{3/2}} \, , \quad
g_{uu}^{L(R)} = \frac{1}{\left|g_{tt}\right|}+g_{xx}\left(\partial_u\tau_{L(R)}\right)^2 \, ,
\eea
and 
\bea
\gamma (u)= \frac{\mu _8}{g_s}V_{S^4} R_{D4}^{15/4}u^{1/4} \, ,  \quad \beta =2\pi \alpha '.
\eea
Additionally, the one-flavour Chern-Simons action is given by 
$S_{\text{CS}}=S_{\text{CS}}^L-S_{\text{CS}}^R$ with
\begin{equation}
S_{\text{CS}}^{L(R)} = \mu _8 \frac{(2\pi \alpha ')^3}{3!}\int _{\text{D8}\left(\overline{\text{D8}}\right)}\omega
_5^{L(R)}\wedge P\left[F_4\right] =
\frac{\alpha }{4}\epsilon ^{lmnpq}\int d^4x \int_{u_0}^{\infty }du \, A_l^{L(R)}F_{mn}^{L(R)}F_{pq}^{L(R)} \, , 
\end{equation}
and $\alpha = N_c/(24 \pi^2)$. Defining the tensor
\begin{equation}
E_{mn}^{L(R)}= g_{mn}^{L(R)}+\beta F_{mn}^{L(R)} 
\quad {\rm with} \quad
E_{L(R)} = \det \left [ E_{mn}^{L(R)} \right ] \, , 
\end{equation}
the DBI-CS action takes the form 
\begin{equation}
S=\int d^4x \int _{u_0}^{\infty }\text{d}u\, \Big \{-\gamma (u)\left[\sqrt{-E_L}+\sqrt{-E_R}\right]
+\frac{\alpha }{4}\epsilon^{\ell mnpq}\left[A_\ell^LF_{mn}^LF_{pq}^L-A_\ell^RF_{mn}^RF_{pq}^R\right]\Big \} \, . 
\end{equation}
The variation of the action with respect to the gauge fields $A_m^{L(R)}$ and the scalar field $\tau _{L(R)}$ leads to the DBI-CS equations 
\bea \label{eq:DBICS}
\partial _u\left[\frac{\gamma }{2\sqrt{-E_{L(R)}}}\frac{\partial E_{L(R)}}{\partial \left(\partial _u\tau _{L(R)}\right)}\right]&=&0 \ ,  \cr
\partial _m\left[\beta
\gamma \sqrt{-E_{L(R)}}E_{L(R)}^{\langle m \ell \rangle}\right]\mp\frac{3}{4}\alpha \epsilon ^{\ell mnpq}F_{mn}^{L(R)}F_{pq}^{L(R)}&=&0 \, ,
\eea
where $E^{<m \ell>} = \frac12 (E^{m \ell} - E^{\ell m})$ and $E^{m \ell}$ is the inverse of $E_{m \ell}$. Decomposing the DBI-CS equations \eqref{eq:DBICS} into $(u,0,i)$ components we get for the left sector 
\bea \label{eq:DBICSv2}
&&\partial _u\left[\frac{\gamma }{\sqrt{-E}}g_{xx}E_0\left(\partial _u\tau \right)\right]=0 \ , \cr
&&\partial _0\left[\beta \gamma \sqrt{-E}E^{<0u>}\right]+\partial
_i\left[\beta \gamma \sqrt{-E}E^{<iu>}\right]-3\alpha \epsilon ^{ijk}F_{0i}F_{jk}=0 \ , \nonumber \\
&&\partial _u\left[\beta \gamma \sqrt{-E}E^{<u0>}\right]+\partial
_i\left[\beta \gamma \sqrt{-E}E^{<i0>}\right]+3\alpha \epsilon ^{ijk}F_{ui}F_{jk}=0 \ , \nonumber \\
&&\partial _u\left[\beta \gamma \sqrt{-E}E^{<ui>}\right]+\partial
_0\left[\beta \gamma \sqrt{-E}E^{<0i>}\right]+\partial _j\left[\beta \gamma \sqrt{-E}E^{<ji>}\right] \nonumber \\ && \qquad \qquad  \qquad \qquad \qquad \qquad -3\alpha \epsilon^{ijk}F_{u0}F_{jk}+6\alpha
\epsilon^{ijk}F_{uj}F_{0k}=0 \ , 
\eea
where $E_0 =  \det \left[E_{\mu \nu }\right]$ where $\mu=(0,i)$. The results for the right sector are obtained by taking 
$\alpha \to - \alpha$. In appendix \ref{AppA} we provide a list of useful identities for the DBI-CS equations. 
Note that at the classical level, the fields on the left and right branes do not couple. 
It is only at the level of the fluctuations that we must take into account the coupled boundary conditions 
at the point where the branes join. 


\subsection{Turning on the magnetic field and chemical potential}

In order to introduce a chemical potential and a magnetic field, we consider the ansatz
\begin{equation} \label{eq:ansatzmfcp}
\tau _{L(R)}=\pm \tau (u) \, , \quad
A_u^{L(R)}=0  \, , \quad 
 A_0^{L(R)} = f_0(u) \,  , \quad
\vec{A}_{L(R)}=\frac{1}{2}\vec{B}\times \vec{x} \pm \vec{f}(u) \, ,  
\end{equation}
where $\vec{f}(u)\times \vec{B}=0$ (parallel vectors) and the symmetry in the left and right brane 
configurations allows for the simplifying profile ansatz. 
This ansatz is motivated by the fact that $A_0$ is dual to a baryonic/quark chemical potential and $\vec{f}$ is the dual of an axial current. Taking $\vec{B}$ and $\vec{f}$ along the $x_3$ direction, i.e. $\vec{B}=B\hat{x}_3 \, , \, \vec{f}=f_3\hat{x}_3$, the DBI-CS equations (\ref{eq:DBICSv2}) reduce to 
\bea \label{eq:DBICSv3}
\partial _u\left[\gamma \sqrt{\frac{Q_0}{Q_2}}\left(g_{xx}\right)^{3/2}g_{tt}g_{xx}\left(\partial _u\tau \right)\right]&=&0 \ , \nonumber \\
\partial _u\left[\beta^2\gamma \sqrt{\frac{Q_0}{Q_2}}\left(g_{xx}\right)^{3/2}\left(\partial_u f_0\right)\right]+6\alpha B\left(\partial_u f_3\right)&=&0 \ , \nonumber \\
\partial _u\left[\beta^2\gamma \sqrt{\frac{Q_0}{Q_2}}\left(g_{xx}\right){}^{3/2}\left(\frac{g_{tt}}{g_{xx}}\right)\left(\partial_u f_3\right)\right]-6\alpha B\left(\partial _uf_0\right)&=&0 \ ,
\eea
where 
\bea 
 Q_0 = 1+\beta ^2\left(g^{xx}\right)^2 B^2 
 \quad \mathrm{and} \quad 
Q_2 = -g_{uu}g_{tt}-\beta^2\left(\partial_u f_0\right)^2-\beta^2\left(\frac{g_{tt}}{g_{xx}}\right)\left(\partial_u f_3 \right)^2  \,. 
\eea

Note that the last two equations in \eqref{eq:DBICSv3} show a dependence on the sign of $B$. This trivial dependence tells us that $f_3$ becomes negative when $B$ is negative. This is in accordance with the expectation that an axial current generated by a nonzero magnetic field and chemical potential should align in the same direction as the magnetic field.

We adopt the units \eqref{eq:PRSunits} and as in \cite{Preis:2010cq}   redefine the coordinate and fields as follows
\bea
v = \frac{u}{u_{KK}} \ , \quad
\hat{f}_{0,3}=f_{0,3} \, \frac{2\pi \ell_s^2}{R_{D4}} \ , \quad 
\hat{\tau} =\frac{\tau}{R_{D4}} \ , \quad 
b=2\pi \ell_s^2 \, B\, . \label{eq:FieldsTransf}
\eea
Note from \eqref{eq:PRSunits} and \eqref{eq:FieldsTransf} that the original gauge fields $f_{0,3}$ as well as the magnetic field $B$ are actually of order $\bar \lambda$ with $\bar \lambda = \lambda/(27 \pi)$. This means that although we are considering magnetic fields of order of $\Lambda_{QCD}$, and thus many times stronger than any magnetic field occuring on Earth, they are actually small when compared to $\lambda$. Moreover, since we always work in the 't Hooft limit, where $N_c$ is much larger than $\lambda$, the gauge fields will not backreact on the black brane background  \eqref{eq:D4metric}. This justifies the use of the probe approximation.

The DBI-CS equations (\ref{eq:DBICSv3}) take the form 
\bea \label{eq:DBICSv4}
\partial _v\left[\sqrt{\frac{Q_0}{Q_2}}v^{11/2}h(v)\left(\partial _v\hat{\tau }\right)\right]&=&0\  , \cr
\partial _v\left[\sqrt{\frac{Q_0}{Q_2}}v^{5/2}\left(\partial_v\hat{f}_0\right)\right]+ 3 b \left(\partial _v\hat{f}_3\right)&=&0  \, , \cr
\partial _v\left[\sqrt{\frac{Q_0}{Q_2}}v^{5/2}h(v)\left(\partial_v\hat{f}_3\right)\right]+ 3 b \left(\partial _v\hat{f}_0\right)&=& 0 \, , 
\eea
with 
\begin{equation}
Q_0=1+\frac{b^2}{v^3} \quad \mathrm{and} \quad
Q_2 = 1+v^3h(v)\left(\partial_v \hat{\tau }\right)^2 -\left(\partial_v \hat{f}_0\right)^2+
h(v)\left(\partial _v\hat{f}_3\right)^2 \, . \label{eq:Q0Q2}
\end{equation} 
and
\begin{equation}
h(v) = 1 - \frac{v_T^3}{v^3} \, , \quad v_T = \left ( \frac43 \pi T \right )^2 \,. 
\end{equation}
Integrating the differential equations in \eqref{eq:DBICSv4}, we get the first order differential equations 
\bea 
\sqrt{\frac{Q_0}{Q_2}}v^{11/2}h(v)\left(\partial _v\hat{\tau }\right)&=&\hat{k} \, , \label{eq:Tau} \\
-\sqrt{\frac{Q_0}{Q_2}}v^{5/2}\left(\partial _v\hat{f}_0\right)=3 b\hat{f}_3+\hat{c} &=& \widetilde{f_3} \, , \label{eq:f0} \\
\sqrt{\frac{Q_0}{Q_2}}v^{5/2}h(v)\left(\partial _v\hat{f}_3\right)=-3 b \hat{f}_0+\hat{d}&=& -\widetilde{f_0}  \, . \label{eq:f3} 
\eea
where $\hat{k}, \hat{c}$ and $\hat{d}$ are integration constants. Using (\ref{eq:Tau}) and (\ref{eq:f0}) as well as the definition of $Q_2$, we arrive at 
\begin{equation} \label{eq:Q0overQ2}
\frac{Q_0}{Q_2}=\left[Q_0-\frac{\hat{k}^2}{v^8h(v)}+\frac{\tilde{f}_3^2}{v^5}\right]\left[1+h(v)\left(\partial _v\hat{f}_3\right)^2\right]^{-1}.
\end{equation}
Combining \eqref{eq:f3} and \eqref{eq:Q0overQ2} we find another useful expression,
\begin{equation} \label{eq:Q0overQ2v2}
\frac{Q_0}{Q_2}=Q_0-\frac{\hat{k}^2}{v^8h(v)}+\frac{1}{v^5h(v)}\left[h(v)\widetilde{f}_3^2-\widetilde{f}_0^2\right] \ .
\end{equation}
On the other hand, plugging \eqref{eq:f0} into the last equation of \eqref{eq:DBICSv4}, results in a decoupled second order differential equation for $\tilde{f}_3$, namely
\begin{equation} \label{eq:Master}
\sqrt{\frac{Q_0}{Q_2}}v^{5/2}\partial _v\left[\sqrt{\frac{Q_0}{Q_2}}v^{5/2}h(v)\left(\partial _v\widetilde{f}_3\right)\right]=\left(3 b\right)^2\widetilde{f}_3 \, .
\end{equation}
In order to investigate the thermodynamics, we need to evaluate the Hamiltonian for the different profiles that satisfy the DBI-CS equations\footnote{We work in the Lorentzian prescription where the Hamiltonian dictates the thermodynamics instead of the Euclidean prescription where one defines the free energy from a Euclidean action.}. First we evaluate the on-shell action 
\bea
S_{DBI}+S_{CS} = \mathcal{N}\int d^4x\int \text{d}v\left\{-v^{5/2}\sqrt{Q_0}\sqrt{Q_2}+ b \left[\left(\partial _v\hat{f}_0\right)\hat{f}_3-\hat{f}_0\left(\partial
_v\hat{f}_3\right)\right]\right\}
\eea
with the constant ${\cal N}$ given by 
\begin{equation}
\mathcal{N}=\frac{3}{\pi ^2} \overline{\lambda}^3 N_c M_{\text{KK}}^4.
\end{equation}
As shown in \cite{Bergman:2008qv,Preis:2010cq}, in order to arrive at a consistent definition for the charge density, the following additional boundary term 
\bea \label{eq:anomalyPRS}
\Delta S&=&\mathcal{N}\frac{b}{2}\int d^4x\int \text{d}v\left[\left(\partial _v\hat{f}_0\right)\hat{f}_3-\hat{f}_0\left(\partial
_v\hat{f}_3\right)\right] \, , 
\eea
is required and thus, the associated Hamiltonian takes the form 
\begin{equation} \label{Hamiltonian}
H =V\mathcal{N}\int _{v_0}^{\infty }\text{d}v\left\{v^{5/2}\sqrt{Q_0}\sqrt{Q_2}-
\frac{3}{2} b\left[\left(\partial _v\hat{f}_0\right)\hat{f}_3-\hat{f}_0\left(\partial
_v\hat{f}_3\right)\right]\right\} \, . 
\end{equation}


\subsection{The chirally broken phase}

In order to facilitate the understanding of the symmetries of the fields in the radial direction, we will introduce the coordinate $z$ defined by the relation 
\begin{equation}
v(z)=v_0\left(1+\frac{z^2}{v_0^2}\right)^{1/3} \, .
\end{equation}
In the chirally broken phase we look for a U-shape profile for $\hat \tau$, whereas the fields $\hat{f}_0$  and $\hat{f}_3$ are even and odd in the coordinate $z$, respectively. Then the boundary conditions at the tip of the brane $v=v_0$ are given by 
\bea \label{eq:BCv0}
\hat{\tau}'\left(v_0\right)=\infty \, , \quad  
\lim_{v \to v_0}\left[\sqrt{\frac{v}{v_0}-1} \, f_0'(v)\right]=0 \, , \quad
\hat{f}_3\left(v_0\right)=0 \, . 
\eea
Note from \eqref{eq:Tau} that the $\hat \tau$ boundary condition at the tip  implies that $Q_0/Q_2\vert_{v_0}=0$. 

The boundary conditions at $v=\infty$ are given by \footnote{The chemical potential appears with a negative sign to adapt the Lorentzian prescription to the thermodynamic relations in the Euclidean prescription.}
\begin{equation} \label{eq:BC}
 \hat{\tau }(\infty )=\frac{\ell}{2} \, , \quad 
 \hat{f}_0(\infty )=-\mu \, , \quad 
\hat{f}_3(\infty )= j \, , 
\end{equation}
where $\ell$ is the UV distance between the branes, $\mu$ is the chemical potential and $j$ is the supercurrent associated with the presence of a magnetic field \cite{Rebhan:2008ur}. Note that $\ell = \frac23 L/R$.

Using the fact that $\hat{f}_0$ ($\hat{f}_3$) is even (odd) in the coordinate $z$, we find the following expansions for $\hat{f}_3$ and $\widetilde{f}_0$ around the tip of the branes
\begin{equation} \label{eq:f3f0exp}
\hat{f}_3(v)=\alpha _0\sqrt{\frac{v}{v_0}-1} \left ( 1+\ldots \right ) 
\quad \mathrm{and} \quad 
\widetilde{f}_0(v)=\beta _0 + \beta _1 \left(\frac{v}{v_0}-1\right)+ \ldots 
\end{equation}
Using these expansions and the boundary condition $Q_0/Q_2\vert_{v_0}=0$ in  eq. (\ref{eq:f0}) we find that $\hat{c}=0$. On the other hand, if we take the ratio of \eqref{eq:f0} and \eqref{eq:f3} we find the equation 
\bea \label{eq:auxf0f3}
\widetilde{f}_0\partial _v\widetilde{f}_0=h(v)\widetilde{f}_3\partial _v\widetilde{f}_3
\eea
and we conclude that 
\bea
\beta _1=\left(3 b \right)^2\frac{\alpha _0^2}{2\beta _0}h\left(v_0\right).
\eea
Since $\hat{\tau}$ is an odd function in $z$, it immediately follows that it should be expanded as
\begin{equation}
\hat{\tau }(v)=\tau _0\sqrt{\frac{v}{v_0}-1}[\; 1+\ldots \;] \ .
\end{equation}
As suggested in \cite{Preis:2010cq}, it is convenient to define the integration constant 
\begin{equation}
\eta=v_0^{-3/2}\lim_{v\to v_0}\left[\frac{\hat{f}_3'(v)}{\hat{\tau }'(v)}\right]=v_0^{-3/2}\frac{\alpha _0}{\tau _0}.
\end{equation}
Using eq. \eqref{eq:Tau} we conclude that $\hat{k}$ takes the form 
\begin{equation} \label{eq:kcond1}
\hat{k}=\frac{v_0^4\sqrt{h\left(v_0\right)}\sqrt{Q_0\left(v_0\right)}}{\sqrt{1+\eta ^2}}.
\end{equation}
Note that if  $\mu $=0 or  $b=0$, the constant $\eta $ must vanish identically. 
Using  \eqref{eq:Q0overQ2v2} and some previous results, we find the interesting condition
\begin{equation} \label{eq:kcond2}
\hat{k}^2=v_0^8h\left(v_0\right)Q_0\left(v_0\right)-v_0^3\left[\widetilde{f}_0\left(v_0\right)\right]^2 \, .
\end{equation}
Finally using \eqref{eq:kcond1} and the condition \eqref{eq:kcond2}, we find  $\widetilde f_0(v_0)$ 
in terms of $\eta$ and $v_0$
\begin{equation} \label{eq:f0v0}
\widetilde{f}_0\left(v_0\right)=-v_0^{5/2}\sqrt{h\left(v_0\right)}\sqrt{Q_0\left(v_0\right)}\frac{\eta }{\sqrt{1+\eta ^2}}.
\end{equation}
Here, we are assuming that $\eta \geq  0 $ and that $\widetilde{f}_0$ is negative definite. 

In section \ref{sec:Numerics} we will use the results found in this subsection to obtain the numerical solution for the fields $\hat \tau(v)$, $\hat f_0(v)$ and $\hat f_3(v)$ describing the U-shaped profile for the chirally broken phase in the presence of a magnetic field $b$ and chemical potential $\mu$. 

\subsection{The chirally symmetric phase}

In the chirally symmetric phase, the $D8$-$\overline{D8}$ branes are separated by a distance $\ell$ and stretch from the boundary to the horizon $v=v_T$. Then we have a constant profile for $\hat{\tau}$ which immediately implies $\hat{k}=0$ . The boundary conditions for the fields $\hat{f}_0$ and $\hat{f}_3$ are 
\bea
\hat{f}_0\left(v_T\right)=0 \ , \quad  \hat{f}_0(\infty )=-\mu \ , \quad 
 \hat{f}_3(\infty )=0\ .
\eea
We will assume that $\partial _v \widetilde{f}_3$ does not diverge at the horizon $v=v_T$. Thus, from \eqref{eq:Q0overQ2} with $\hat k=0$, we see that $\frac{Q_0}{Q_2}\left(v_T\right)$ is finite 
(there are no real solutions for the case $\frac{Q_0}{Q_2}\left(v_T\right)=0$). On the other hand, from \eqref{eq:Q0overQ2v2} with $\hat k=0$, we find that, in order to get a real $\frac{Q_0}{Q_2}\left(v_T\right)$, we have to impose $\widetilde{f}_0 =\beta _0 \left(v-v_T\right)^{r_0}$, with $r_0 \geq 1/2 $ .  

Using eq. \eqref{eq:f3}, the fact that $\frac{Q_0}{Q_2}\left(v_T\right)$ is finite, and the auxiliary eq. \eqref{eq:auxf0f3}, we find that $r_0=1$ and $\widetilde{f}_{0,3}$ admit the following expansions 
\begin{equation} \label{eq:asympexp}
\widetilde{f}_3(v)=\alpha_0+\alpha _1\left(v-v_T\right)+ \ldots  \ , \quad \widetilde{f}_0(v)=\beta_0\left(v-v_T\right)+ \ldots \, . 
\end{equation}
Plugging these results into eq. \eqref{eq:Q0overQ2} yields 
\begin{equation}
\frac{Q_0}{Q_2}\left(v_T\right)=Q_0\left(v_T\right)+\frac{\alpha _0^2}{v_T^5}.
\end{equation}
Using eq.  \eqref{eq:f0}, a relation between $\beta _0$ and $\alpha _0$ can be established, i.e., 
\begin{equation}
\beta _0=-\frac{\left(3 b\right)\alpha _0}{v_T^{5/2}\sqrt{Q_0\left(v_T\right)+\frac{\alpha _0^2}{v_T^5}}}=-\frac{\left( 3b \right)\alpha
_0}{\sqrt{\alpha _0^2+ b^2 v_T^2+v_T^5}}.
\end{equation}
The auxiliary eq. \eqref{eq:auxf0f3} then can be used to determine $\alpha _1$ in terms of $\alpha _0$ and $\beta _0$, namely
\begin{equation}
\alpha _1=\frac{\beta _0^2}{\alpha _0h'\left(v_T\right)}=\frac{\beta _0^2}{3\alpha _0}v_T.
\end{equation}
Note that the boundary condition $\hat{f}_0\left(v_T\right)=0$ and the asymptotics \eqref{eq:asympexp} imply that $\hat{d}=0$. 


\section{Solving the DBI-CS equations and the chiral transition}

\label{sec:Numerics} 

In this section, we present the numerical calculations for the chirally broken and chirally symmetric profiles. 
Evaluating the corresponding Hamiltonians, we find the phase diagram for the chiral transition in the $(b, \mu)$  
plane for fixed $T$ and the $(T,b)$ plane for fixed $\mu$. 

For the UV distance between the $D8$ and $\overline{D8}$ branes, we set $\ell=1$. As pointed out in \cite{Preis:2010cq}, results for other values of $\ell$ can be obtained from the $\ell=1$ results noticing that the DBI-CS equations actually depend on the quantities $v_0 \ell^2$, $\mu  \ell^2$,  $v_T \ell^2$, $b \ell^3$
and can thus be rescaled accordingly.

\subsection{The chirally broken phase}

\label{NumXB}

The non-trivial parameters in the problem are $\alpha _0$, $\eta$ and $v_0$. Our strategy to numerically find the chirally broken profiles is the following: we integrate numerically the second order differential eq. \eqref{eq:Master} for $\widetilde {f}_3 = 3 b \hat f_3$, with $\sqrt{Q_0/Q_2}$ given by \eqref{eq:Q0overQ2}, from the  tip to the boundary, using as initial condition the $\widetilde {f}_3$ expansion given in \eqref{eq:f3f0exp}. Then, by utilising eq. \eqref{eq:f3}, we can extract $\widetilde{f}_0$. Subsequently, integrating eq. \eqref{eq:Tau} we obtain $\hat{\tau}$, i.e., 
\begin{equation}
\hat{\tau }(v)=\int_{v_0}^v \frac{\hat{k}}{\bar{v}^{11/2}h\left(\bar{v}\right)} \,
\sqrt{\frac{Q_2}{Q_0} (\bar v)} \, 
d\bar{v} \ ,
\end{equation}
where we used the condition $\hat \tau (v_0) = 0$. For fixed $v_0$ we will impose the boundary conditions 
\begin{equation} \label{eq:conds}
\hat{\tau }(\infty )=\frac{\ell}{2} \ ,\quad \widetilde{f}_0\left(v_0\right)=-v_0^{5/2}\sqrt{h\left(v_0\right)}\sqrt{Q_0\left(v_0\right)}\frac{\eta}{\sqrt{1+\eta^2}}.
\end{equation}
The second condition in \eqref{eq:conds} was obtained in \eqref{eq:f0v0}. The two conditions \eqref{eq:conds} fix $\left(\alpha_0,\eta \right)$ for a given value of $v_0$. This is a 2D shooting method that can be solved in Mathematica combining {\fontfamily{cmtt}\selectfont ContourPlot} and {\fontfamily{cmtt}\selectfont FindRoot}. We solve for the profiles of the chirally broken phase in the range $0 \le v_T \le 0.4 $ (for the temperature) and $0 < b \le 0.5$ (for the magnetic field). Note that the profiles do not depend at all on the value of $\mu$. The latter appears only in the Hamiltonian and will determine the ground state for the chirally broken phase.

\begin{figure}[H] 
   \centering
   \includegraphics[width=4.6cm]{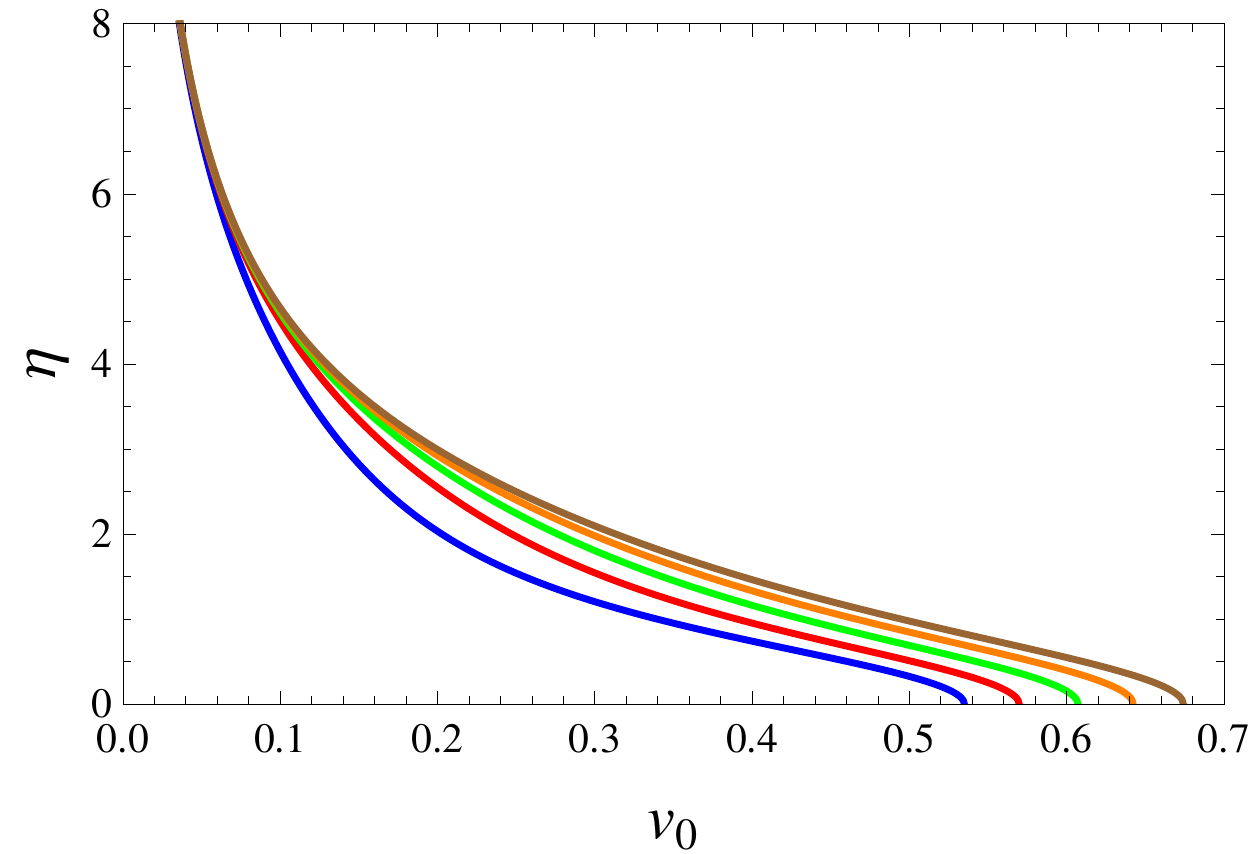} 
   \hspace{0.2cm}
   \includegraphics[width=4.6cm]{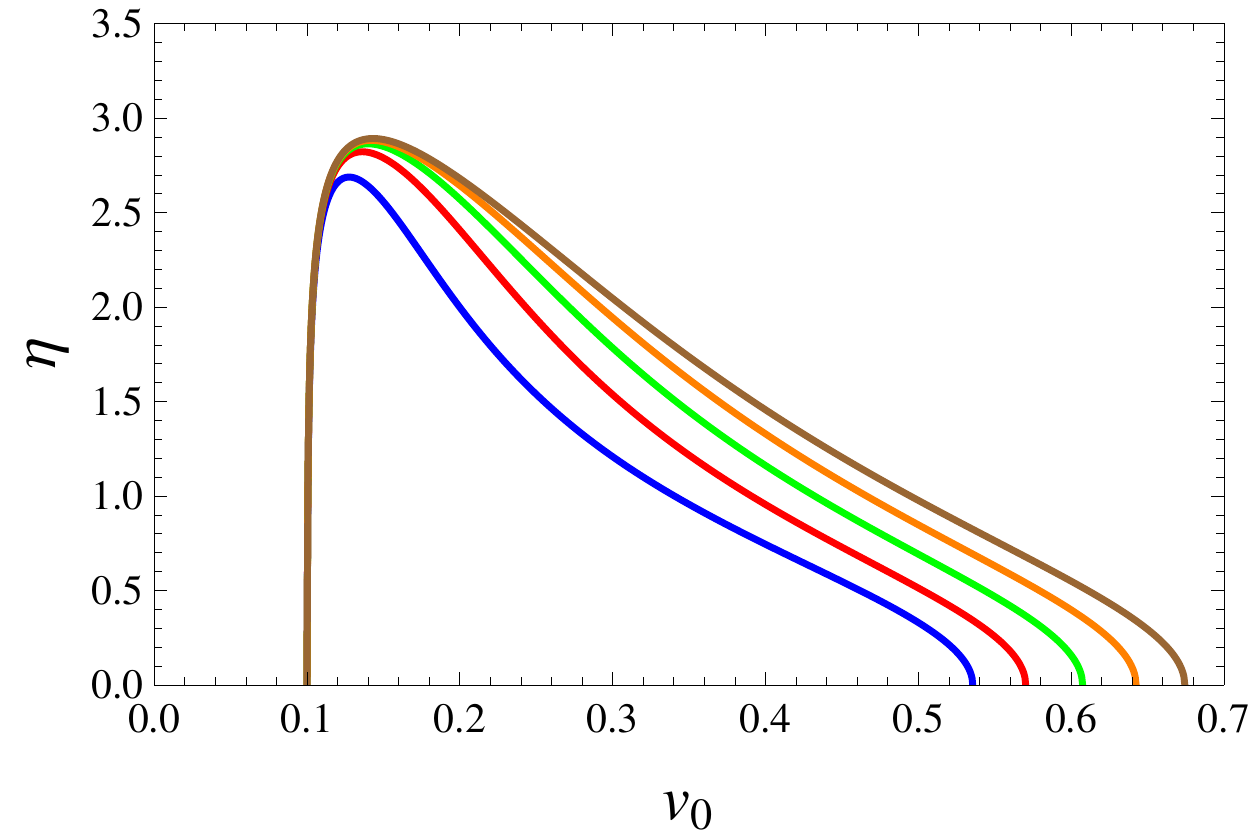}
   \hspace{0.2cm}
   \includegraphics[width=4.6cm]{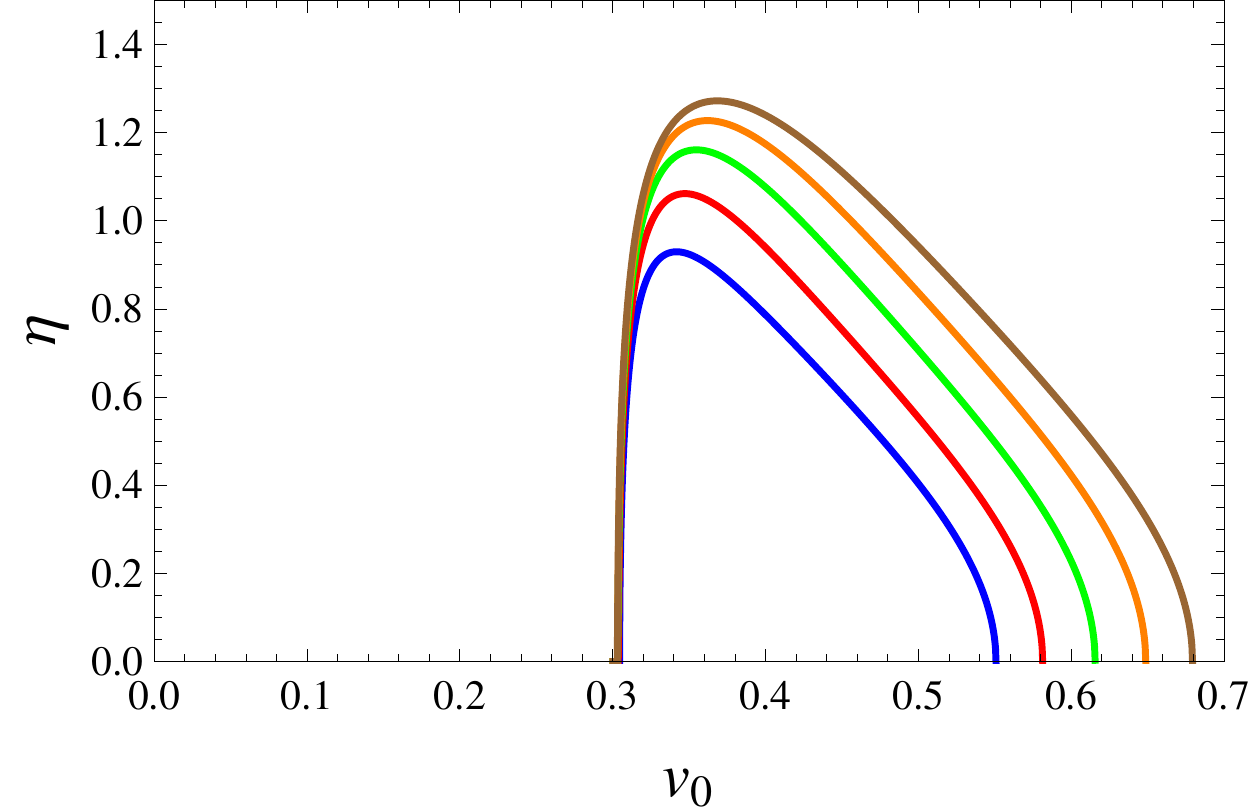} 
   \caption{\small Chirally broken profiles characterized by $\eta(v_0)$ for different values of the magnetic field $b=0.1$ (blue), $b=0.2$ (red), $b=0.3$ (green), $b=0.4$ (orange) and $b=0.5$ (brown).  The left, center and right panels correspond to $v_T=0$, $v_T=0.1$  and  $v_T=0.3$, respectively. }
  \label{fig:etavsv0}
\end{figure}

In Fig. \ref{fig:etavsv0} we show three solutions for  $\eta(v_0)$ corresponding to $v_T=(0,0.1,0.3)$. 
The different lines correspond to different values of $b$. 
While for zero temperature the minimum value for $v_0$ is zero (with $\eta \rightarrow \infty$) for finite temperature
it is given by the horizon radius (with $\eta \rightarrow 0$).
Note that the maximum value of $v_0$ increases with the magnetic field. 

\begin{figure}[H] 
   \centering
   \includegraphics[width=4.6cm]{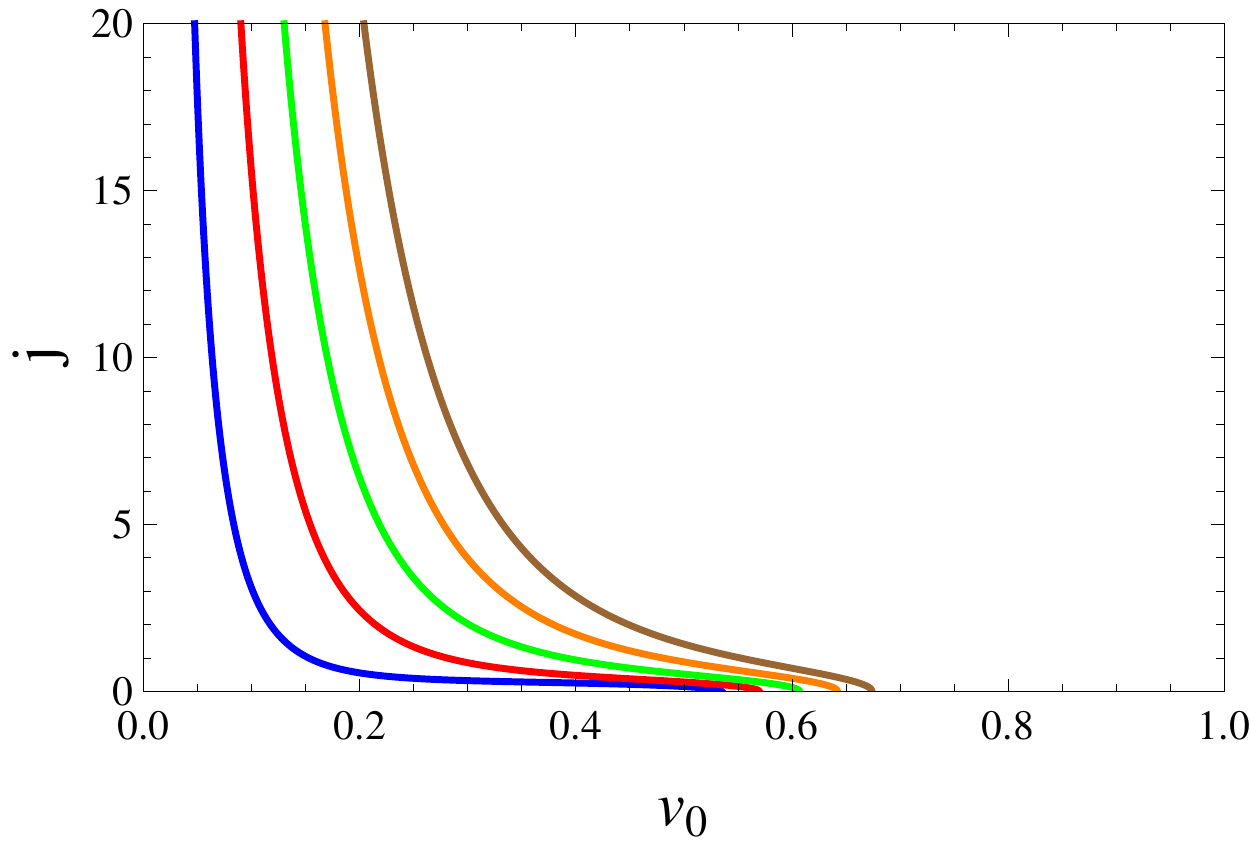} 
   \hspace{0.2cm}
   \includegraphics[width=4.6cm]{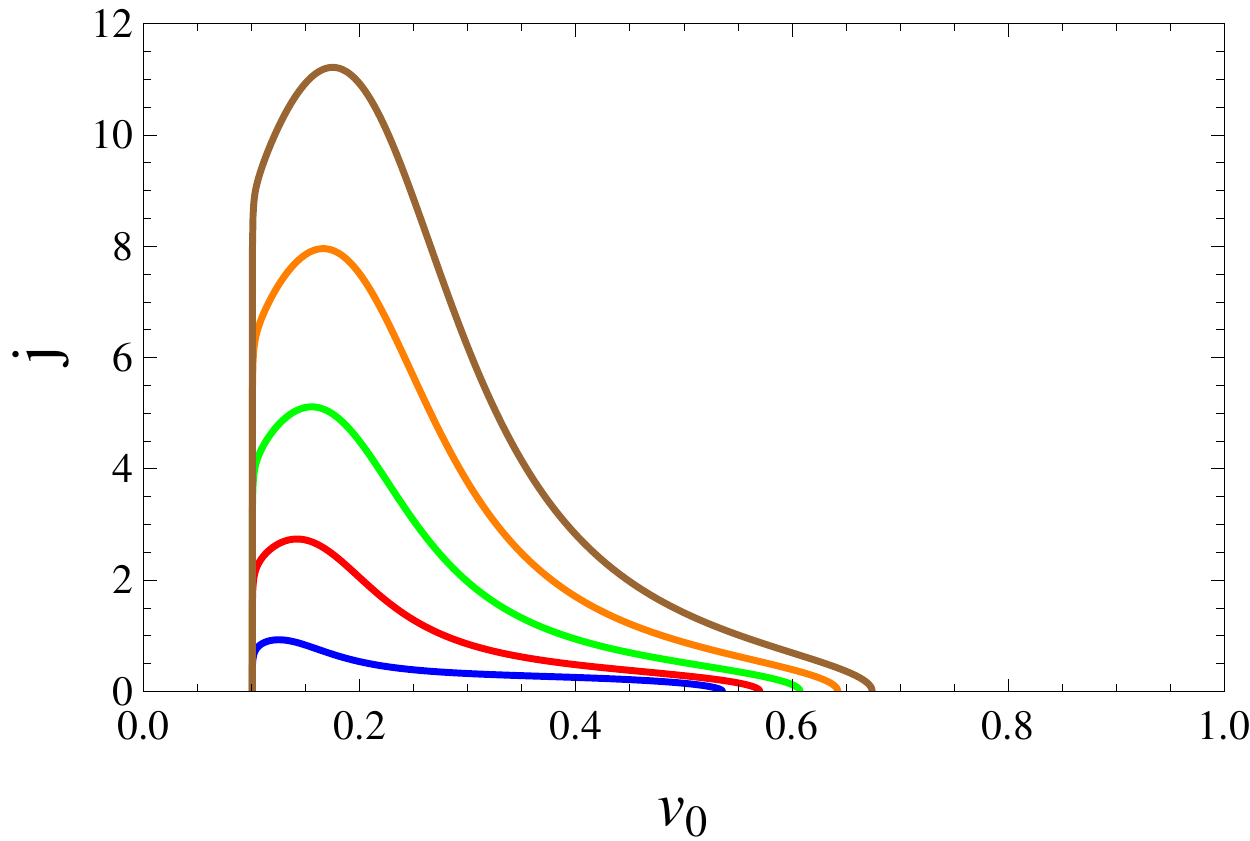}
   \hspace{0.2cm}
   \includegraphics[width=4.6cm]{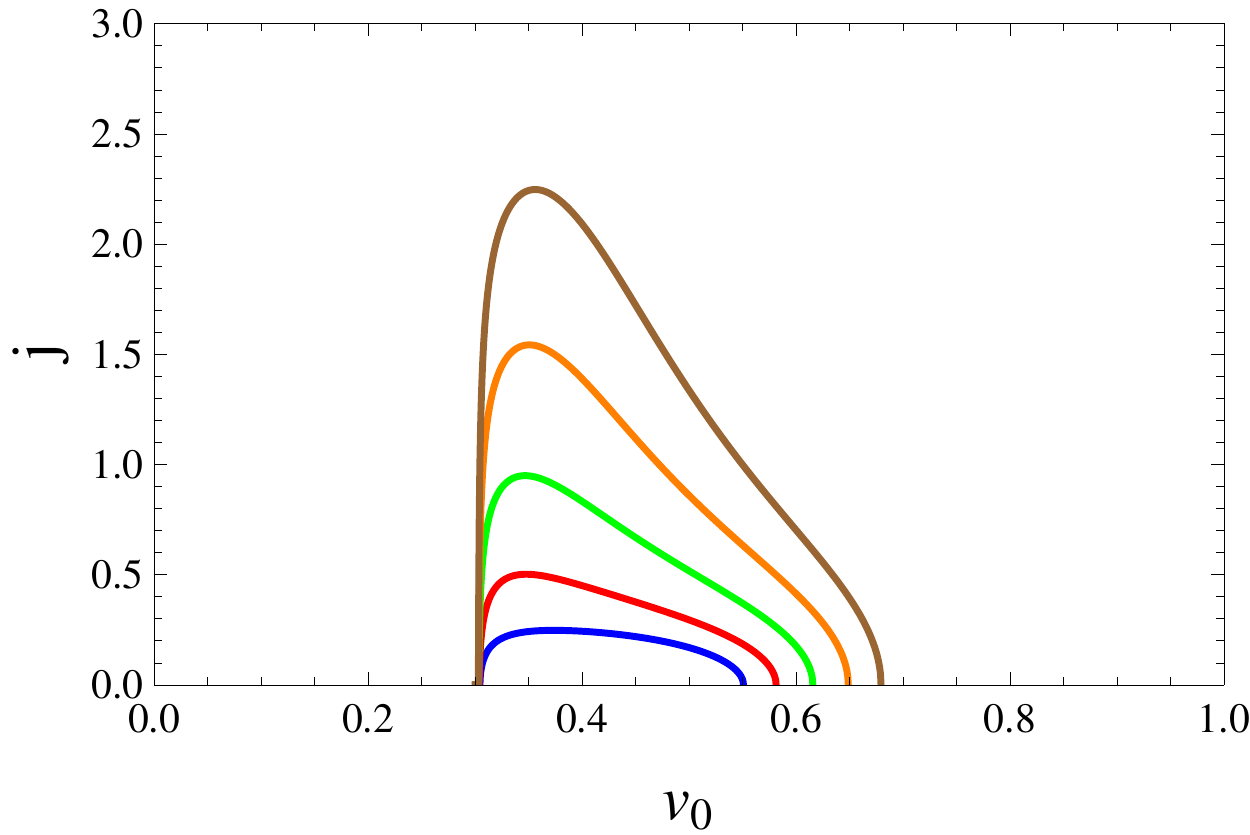}
   \caption{\small The supercurrent $ j$ as a function of $v_0$ for five different values of the magnetic field $b=0.1$ (blue), $b=0.2$ (red), $b=0.3$ (green), $b=0.4$ (orange) and $b=0.5$ (brown). The left, center and right panels correspond to $v_T=0$, $v_T=0.1$  and  $v_T=0.3$, respectively. }
  \label{fig:jvsv0}
\end{figure}

For each profile characterised by $v_0$ we can evaluate the supercurrent $j = \hat f_3 (\infty)$. From the Hamiltonian definition in (\ref{Hamiltonian}) it is not difficult to see that the quark density $\rho$ will be proportional to $j$. We show in Fig. \ref{fig:jvsv0} the results for $j(v_0)$ for $v_T=(0,0.1,0.3)$ and five different values of the magnetic field $b$.  Note that, for fixed $j$, there are either one or two profiles, so in some cases there may be a transition between two chirally broken phases (corresponding to different profiles) in the $b$ vs $\mu$ phase diagram. 

\begin{figure}[H] 
   \centering
   \includegraphics[width=4.6cm]{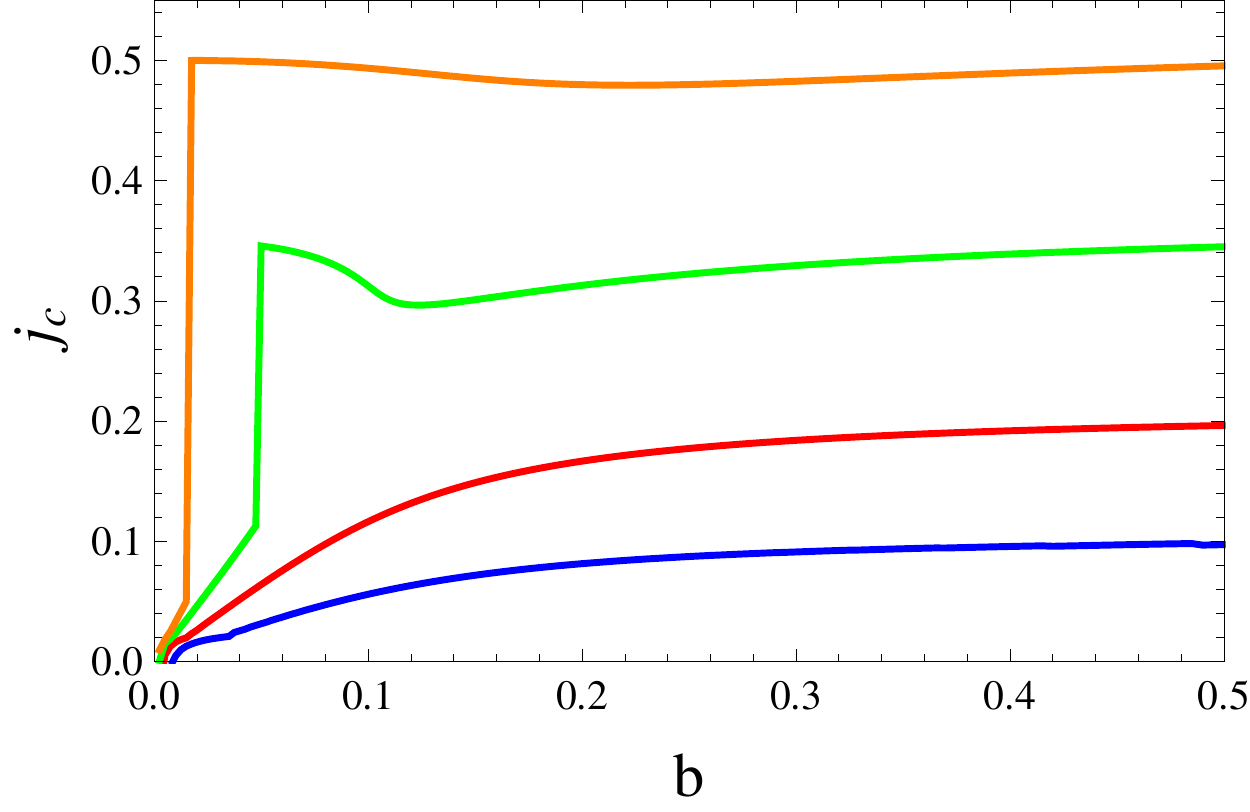} 
   \hspace{0.2cm}
   \includegraphics[width=4.6cm]{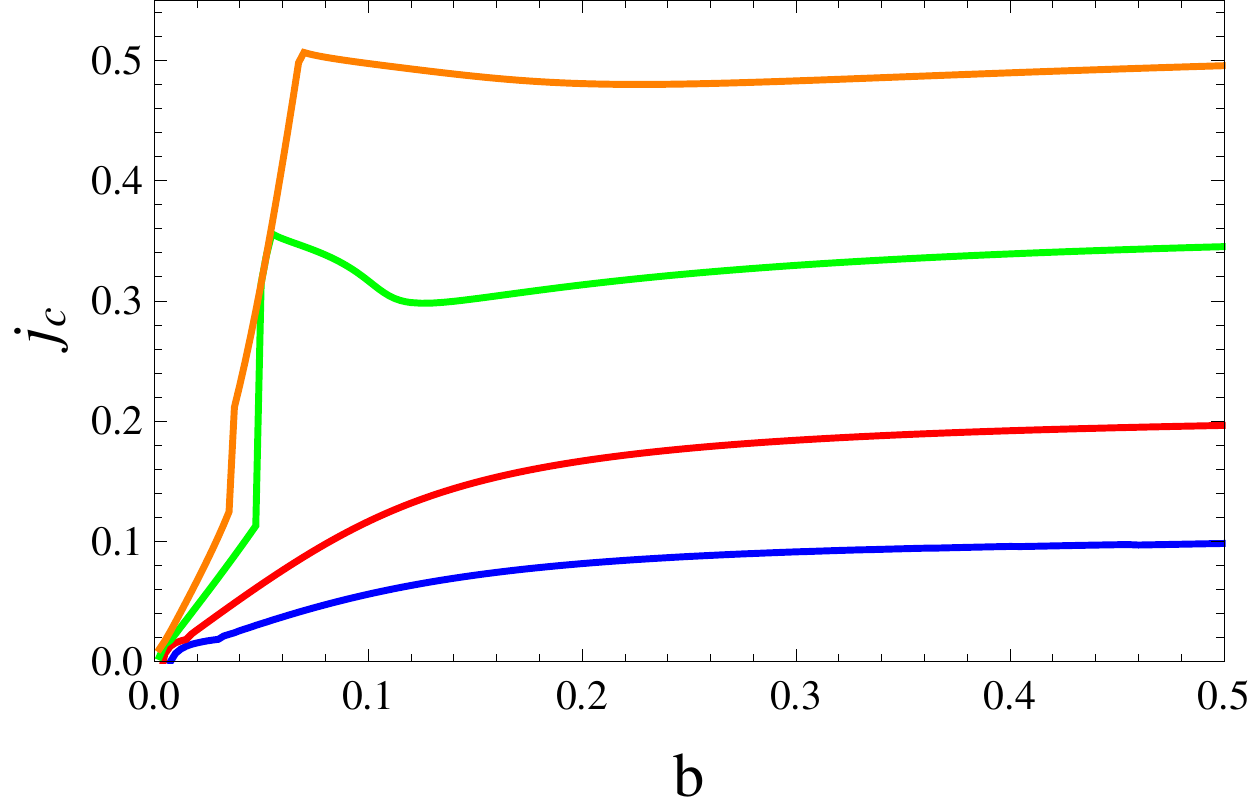}
   \hspace{0.2cm}
   \includegraphics[width=4.6cm]{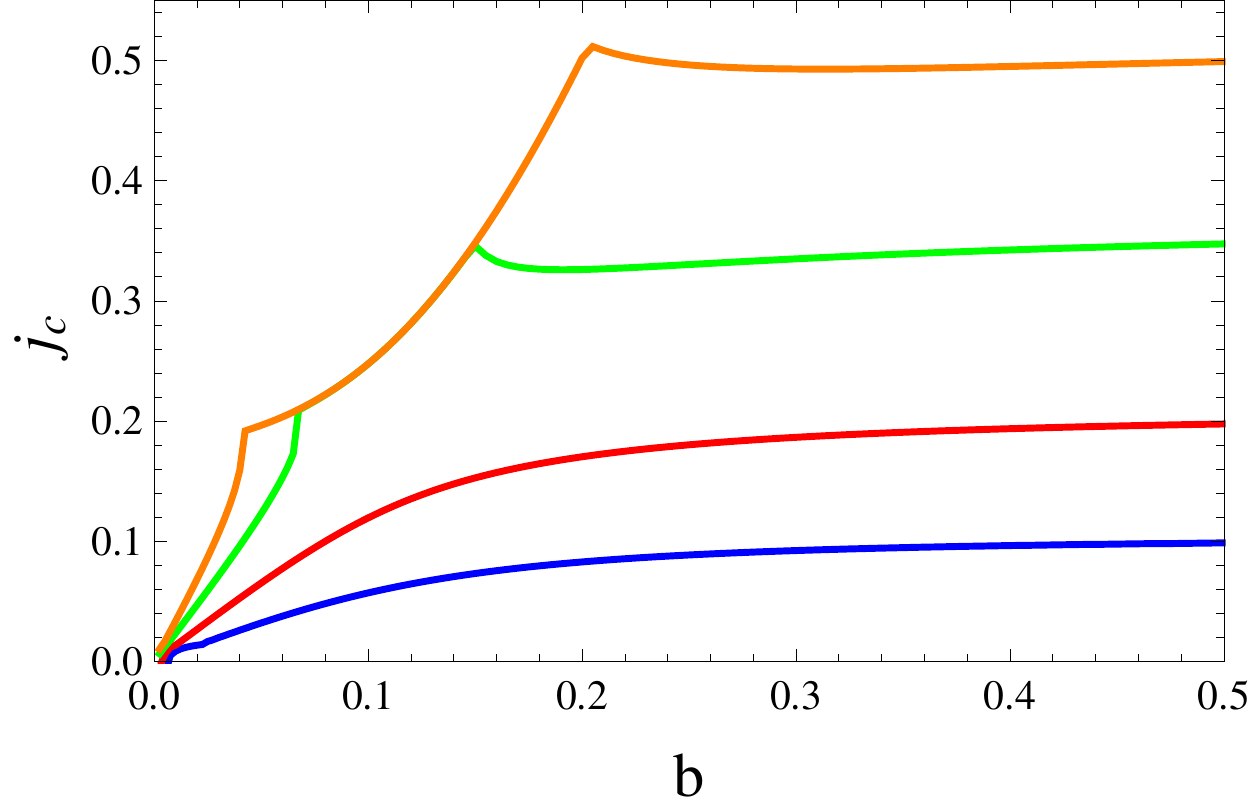}
   \caption{\small  The parameter $j (v_0^c) = j_c$ characterising the ground state as a function of the magnetic field $b$ for four different values of the chemical potential $ \mu = 0.2$ (blue),  $\mu = 0.4$ (red),  $\mu = 0.7$ (green) and  $\mu = 1$ (orange). The left, center and right panels correspond to $v_T=0$, $v_T=0.1$  and  $v_T=0.3$, respectively. }
  \label{fig:v0cvsb}
\end{figure}

For each profile, we evaluate the Hamiltonian using the formula (\ref{Hamiltonian}). The Hamiltonian then depends on the parameter $v_0$ and there is a value $v_0^c$ where the Hamiltonian has a minimum that corresponds to the ground state. In Fig. \ref{fig:v0cvsb} we show the value of $j$ at $v_0^c$ as a function of the magnetic field $b$ for $v_T=(0,0.1,0.3)$ and four different values of $\mu$.  

As $\mu$ increases, at fixed temperature, a transition between two chirally broken phases takes place. This transition is first order at low temperatures, as shown in the first plot of Fig. \ref{fig:v0cvsb} because there is a jump in the density (a jump in $j$). As the temperature increases this transition becomes second order (a jump in the derivative of $j$). Note in the last plot of Fig. \ref{fig:v0cvsb} that whereas the first cusp is truly a second order phase transition, under numerical scrutiny the second apparent corner seems not to be a jump in the derivative and therefore there is no second order phase transition there.

Closing this subsection, we should mention that an approximate analytic solution of the equations of 
motion \eqref{eq:Tau}, \eqref{eq:f0} and \eqref{eq:f3} for small values of the magnetic field and the temperature
is presented in Appendix \ref{AppCBP}. The zero temperature limit of that solution can be obtained from a small $b$
expansion of the semi-analytic PRS solution that will be summarised in the next subsection.


\subsection{The PRS approximation}

\label{PRSapprox}

The PRS approximation was proposed in \cite{Preis:2010cq} as a simple method to find a semi-analytic 
solution for the chirally broken phase. The approximation is $h(v) \to 1,$ which simplifies the analysis of the corresponding equations of motion enormously. The approximation is justified when either  
\begin{itemize}
\item the temperatures are very small: $h(v) \sim 1$ implies $v_T \ll v_0 < v$. Since $v_T = ( 4 \pi T/3)^2$ this implies the limit of very small temperatures. Note, however, that since we are interested in the deconfined phase of the S-S model, where $T > T_c = \frac{M_{\text{KK}}}{2 \pi}$, we necessarily have to take a very small $M_{\text{KK}} = 1/R \to 0$ simultaneously. This is the decompactification limit ($R \to \infty$) of the S-S model. 
\item the UV separation of the $D8-\overline {D8}$ branes is very small: very small values of the distance $\ell$ has the effect of increasing $v_0$ which in turn satisfies the condition $v_T \ll v_0$ at fixed $v_T$. This again corresponds to a decompactification limit, in a sense that we keep the radius $R$ fixed while decreasing $L = \frac32 \ell R$, which is equivalent to keeping $L$ fixed and increasing $R$.
\end{itemize}
The net effect of both limits is that the quark constituent mass becomes large compared to the temperature and we arrive at a situation where the gluon dynamics is (almost) decoupled from the flavour physics and the field theory dual approaches a non-local version of the Nambu-Jona-Lasinio model.

Here we briefly review the PRS approximation since we are going to compare those results with the results arising from the full numerical solutions (described in the previous subsection). 

\noindent Using the PRS approximation  the  ratio of \eqref{eq:f0} and \eqref{eq:f3} takes the form 
\bea
-\frac{\partial _v\hat{f}_0}{\partial _v\hat{f}_3}=\frac{3 b\hat{f}_3}{d-3 b\hat{f}_0} \quad \to \quad 
\tilde{f}_0\partial \text{}_v\tilde{f}_0=\tilde{f}_3\partial \text{}_v\tilde{f}_3.
\eea

\noindent Integrating this equation we find the relation
\bea
\tilde{f}_0^2-\tilde{f}_3^2=\text{const.}=\left[\tilde{f}_0\left(v_0\right)\right]{}^2=v_0^5Q_0\left(v_0\right)\frac{\eta ^2}{1+\eta ^2},
\eea
where we have used the b.c.  $\hat f_3\left(v_0\right)=0$ and the condition \eqref{eq:f0v0} with $h(v)=1$. Another key ingredient in the PRS procedure is that, for $h(v)=1$, the equations \eqref{eq:f0} and \eqref{eq:f3} can be rewritten as
\bea \label{eq:PRSsystem}
-\partial _y\tilde{f}_0=\tilde{f}_3 \quad \mathrm{and} \quad \partial _y\tilde{f}_3=-\tilde{f}_0,
\eea
where the new variable $y$ is related to $v$ through the differential equation 
\bea
\frac{\text{d}y}{\text{d}v}=\frac{3b}{v^{5/2}\sqrt{\frac{Q_0}{Q_2}}}=\frac{3b}{v^{5/2}\sqrt{Q_0(v)-\frac{k^2}{v^8}-\frac{v_0^5}{v^5}Q_0\left(v_0\right)\frac{\eta ^2}{1+\eta ^2}}}.
\eea
Solving the system \eqref{eq:PRSsystem} one finds the solutions
\begin{equation}
\tilde{f}_3(y) = \tilde{c}_1\sinh y+\tilde{c}_2 \cosh y 
\quad \mathrm{and} \quad 
\tilde{f}_0(y) = - \tilde{c}_1\cosh y-\tilde{c}_2\sinh y \, .
\end{equation}
The b.c. $\hat f_3\left(v_0\right)=0$ becomes $f_3(y=0)=0$
and implies that $\tilde{c}_2 =0$. Moreover, the b.c. $ \hat f_3\left(y_{\infty }\right)= j$ implies that
\be
\tilde{c}_1=\frac{3 b  j}{\sinh y_{\infty}} \, .
\ee
Evaluating $\tilde{f}_0$ at $v_0$, corresponding to $y=0$, and using \eqref{eq:f0v0}, one finds the first PRS condition
\bea
-\frac{3 b j}{\sinh y_{\infty }}=-v_0^{5/2}\sqrt{Q_0\left(v_0\right)}\frac{\eta }{\sqrt{1+\eta ^2}}. \label{eq:PRScond1}
\eea
Finally, integrating $\hat \tau $ in \eqref{eq:Tau} for $h(v)=1$, we find
\be
\hat \tau (v) = \int_{v_0}^v \frac{\hat k}{\tilde{v}^{\frac{11}{2}}\sqrt{Q_0\left(\tilde{v}\right)-\frac{\hat k^2}{\tilde{v}^8}-\frac{v_0^5}{\tilde{v}^5}Q_0\left(v_0\right)\frac{\eta ^2}{1+\eta ^2}}} \, d\tilde{v}
\quad {\rm where} \quad 
\hat k = \frac{v_0^4\sqrt{Q_0\left(v_0\right)}}{\sqrt{1+\eta ^2}} \,.
\ee
Imposing the condition \eqref{eq:PRScond1} and $\hat \tau (\infty )=\frac{\ell}{2}$, one finds $j$ and $\eta$ for a given $v_0$. At the end of this section we will show that the PRS approximation can be used to describe the chiral transition very well at small temperatures. However, as the temperature increases, it must be abandoned in favour of the full numerical solution, since the departure turns out to be quite significant. 

\subsection{The chirally symmetric phase}

The strategy for finding the chirally symmetric profile is a bit different from the one introduced above: We will numerically integrate eq. \eqref{eq:Master} w.r.t. $\widetilde{f}_3$, with $\sqrt{Q_0/Q_2}$ given by \eqref{eq:Q0overQ2}, and $\hat k = 0$, from the horizon to the boundary. In the numerical integration we use as the initial conditions the expansion \eqref{eq:asympexp}. 

\begin{figure}[H] 
   \centering
   \includegraphics[width=4.6cm]{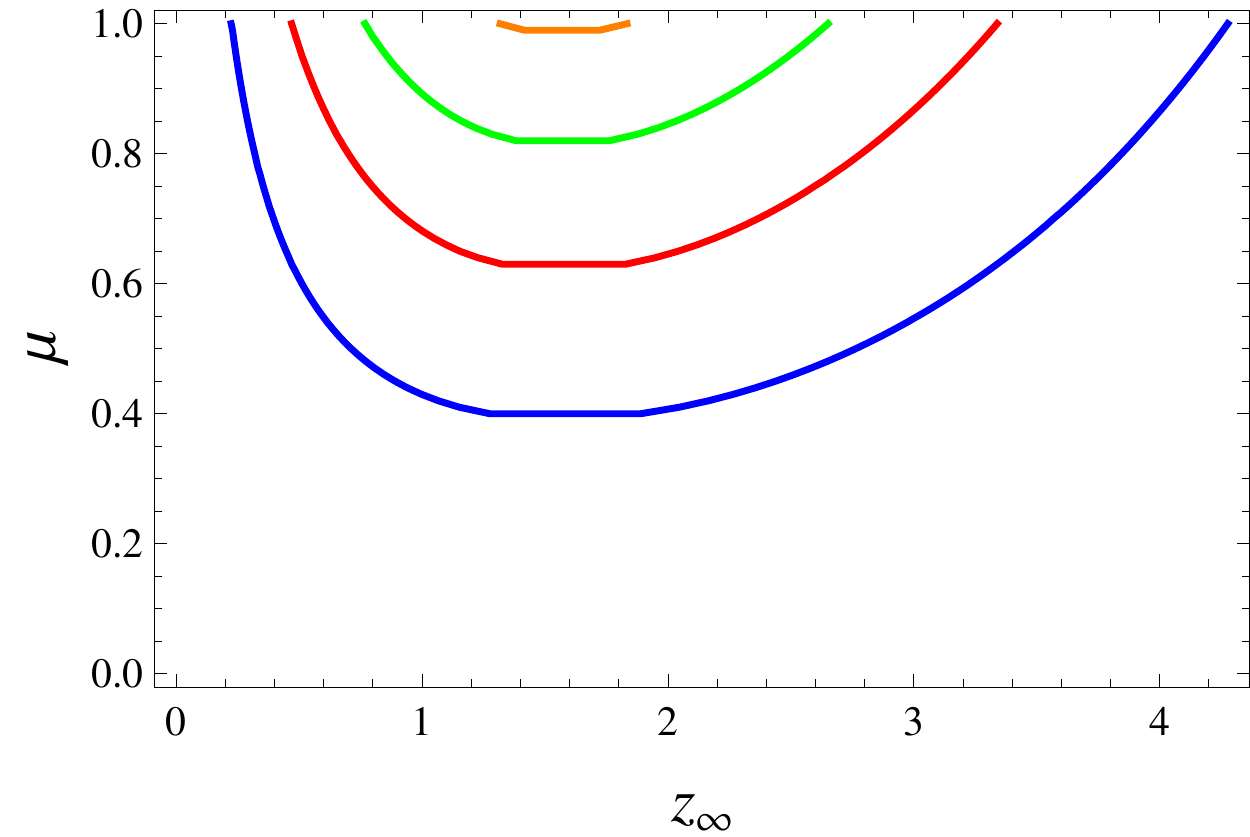} 
   \hspace{0.2cm}
   \includegraphics[width=4.6cm]{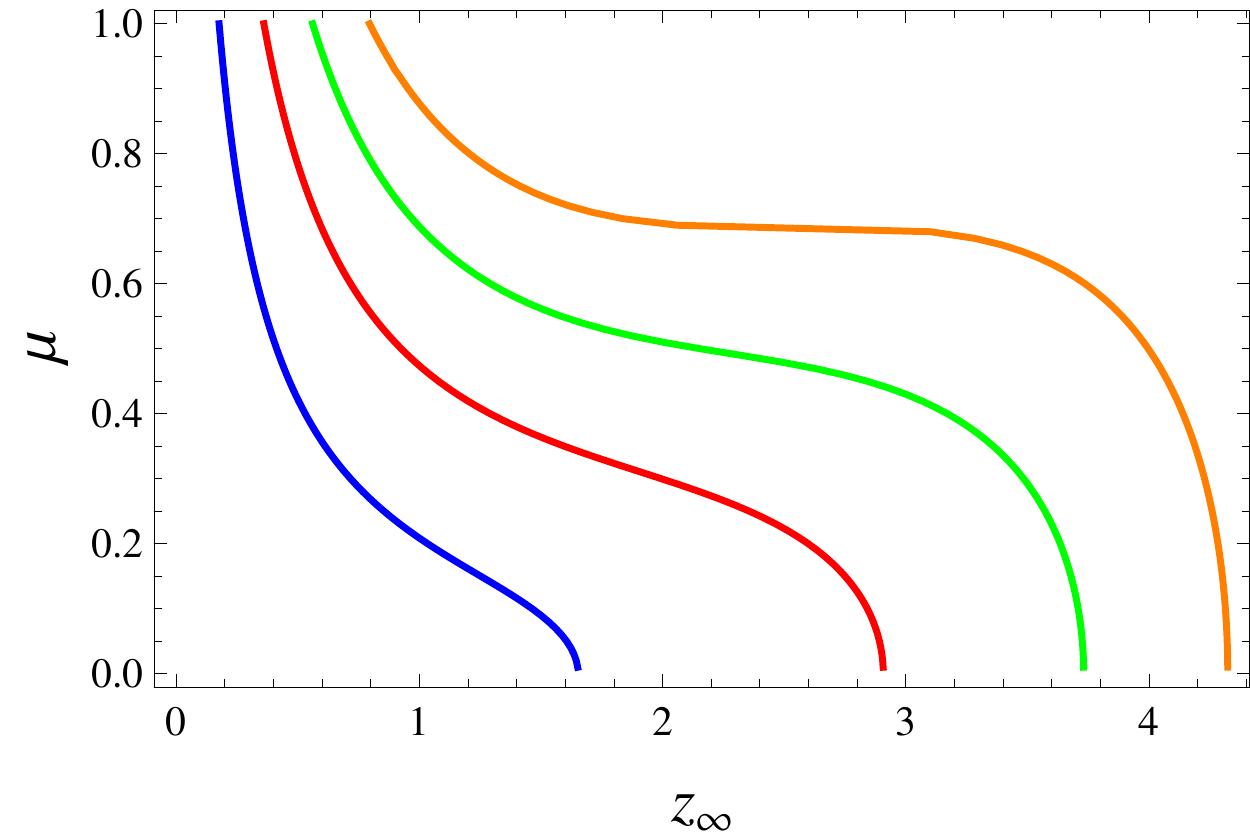}
   \hspace{0.2cm}
   \includegraphics[width=4.6cm]{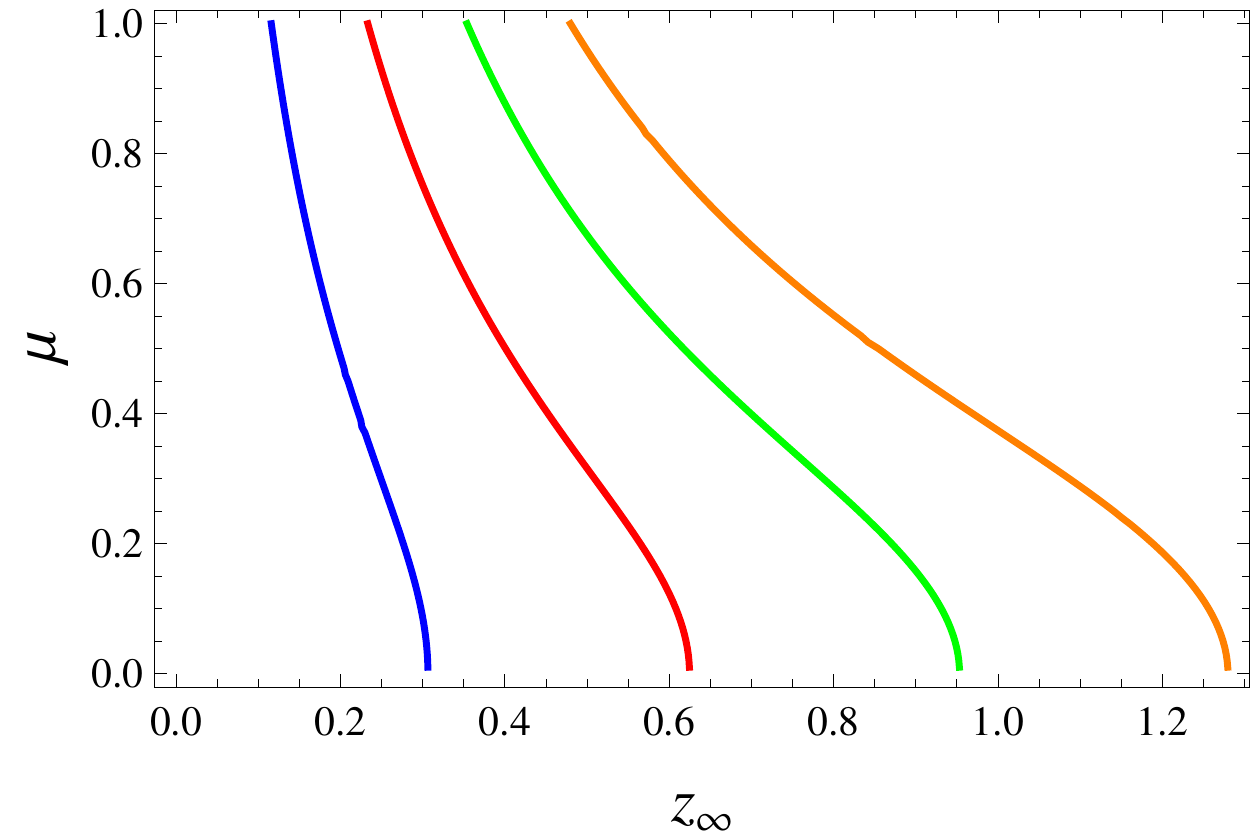}
      \caption{\small  Chirally symmetric profiles characterized by $z_{\infty}$ for given $\mu$. Different lines correspond to different values of the magnetic field $b=0.025$ (blue), $b=0.05$ (red), $b=0.075$ (green) and $b=0.1$ (orange). The left, center and right panels correspond to $v_T=0$, $v_T=0.1$  and  $v_T=0.3$  respectively. Note that the apparent perfectly straight sections at the bottom of the curves in the left figure are a numerical artefact. }
  \label{fig:yinfvsmu}
\end{figure}

Then using eq. \eqref{eq:f3}, we can determine $\widetilde{f}_0$. The parameter $\alpha_0$ in the expansion \eqref{eq:asympexp} is  obtained from the boundary condition 
\bea
\widetilde{f}_0(\infty )= - 3 b \mu \ , \label{eq:BCf0}
\eea
where we have used the result $\hat d=0$, that simplifies the relation between $\tilde f_0$ and $\hat f_0$ for the chirally symmetric profiles. Notice that in the chirally symmetric case it is the chemical potential $\mu$ that characterises the different profiles (in contrast to the chirally broken case where the relevant parameter was $v_0$ or $j$). For a given $\mu$ we need to find the value of $\alpha_0$ using the condition \eqref{eq:BCf0}. 
This is a 1-d shooting method that again is solved combining {\fontfamily{cmtt}\selectfont ContourPlot} and 
{\fontfamily{cmtt}\selectfont FindRoot}. The challenge now is that for a given $ \mu$ we may find more than one value of $\alpha_0$ which is the problem of finding multiple roots. 
Since the values of $\alpha_0$ in general are very small, it turns out to be convenient to introduce the parameter $z_{\infty}$ related to $\alpha_0$ through the equation
\bea
\alpha_0 (z_{\infty}) = \frac{ 3 b  \mu}{\sinh(z_{\infty})} \,. 
\eea
We solve numerically the chirally symmetric profiles in the range $0 \le v_T < 0.4$ for the temperature, $0<b<0.5$ for the magnetic field and $0 < \mu < 1$ for the chemical potential.  In Fig. \ref{fig:yinfvsmu} we show some profiles, characterised by $z_{\infty}$ for a given $ \mu$,  for  $v_T=(0,0.1,0.3)$ and four different values of the magnetic field $b$.  Note that there is more than one value for $z_{\infty}$ for fixed $\mu$, so again there may be a transition between different chirally symmetric profiles.

At zero temperature there is an extra profile not shown in Fig. \ref{fig:yinfvsmu} that can only be obtained analytically. This is the case where $z_{\infty}=\infty$; it can be interpreted as the lowest Landau level \cite{Preis:2010cq}. As the temperature increases, the lowest Landau level becomes the highest value of $z_{\infty}$ whereas higher Landau levels correspond to lower values of $z_{\infty}$. At high enough temperatures there is only one solution for $z_{\infty}$ so there is no distinction between the lowest and the higher Landau levels. In the thermodynamic analysis for the chiral transition, we always take the ground state for the chirally symmetric phase, i.e. the profile that minimises the Hamiltonian. 
 
Similarly to the chirally broken phase, an approximate analytic solution of the equations of 
motion \eqref{eq:f0} and \eqref{eq:f3} for small values of the magnetic field and the temperature
is presented in Appendix \ref{AppCSP}.

\subsection{The phase diagram}

Evaluating the difference of the Hamiltonians associated with the chirally broken and chirally symmetric ground states, i.e. 
\bea
\Delta H = H_{\chi S} - H_{\chi B} \, , \label{DeltaHam}
\eea
we are able to find the phase diagram associated with the chiral transition. 

From equations \eqref{eq:Q0Q2} and \eqref{Hamiltonian} one finds that in each phase the Hamiltonian diverges as $\frac27 v_{max}^{7/2} + b^2 v_{max}^{1/2}$  with $v_{max} \to \infty$. As expected, both divergences cancel in the Hamiltonian difference \eqref{DeltaHam}. In the numerical calculations we consider a finite but large value for $v_{max}$. Since we will calculate other thermodynamic quantities such as the magnetisation, density and entropy for each phase, we actually substract the divergences independently.

\begin{figure}[H] 
   \centering
   \includegraphics[width=4.6cm]{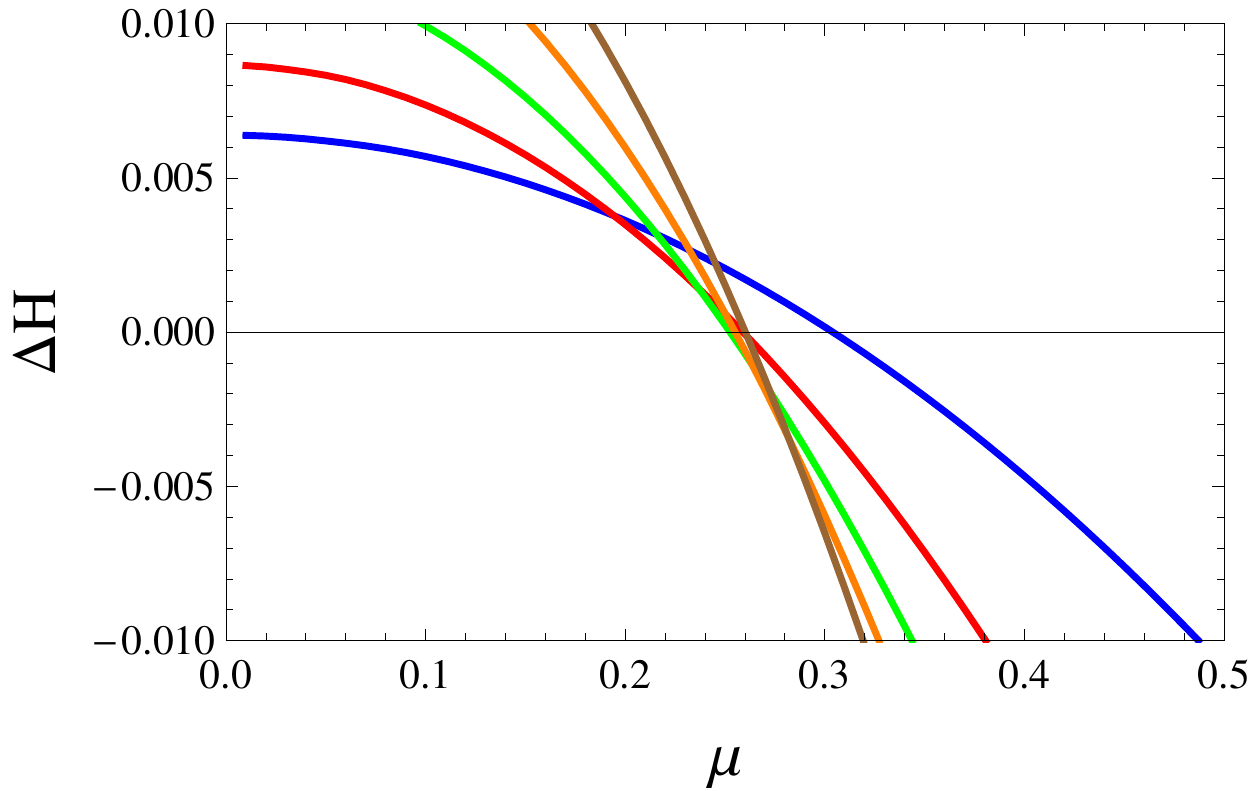} 
   \hspace{0.2cm}
   \includegraphics[width=4.6cm]{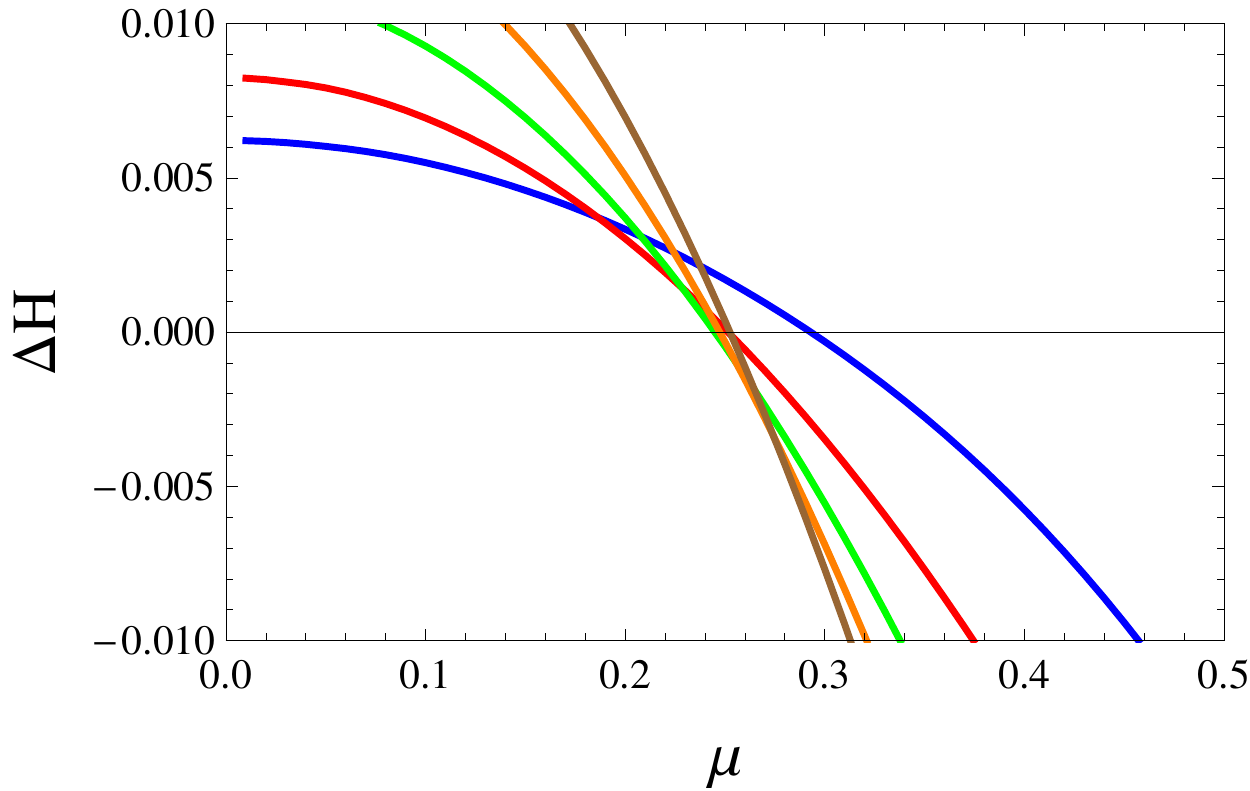}
   \hspace{0.2cm}
   \includegraphics[width=4.6cm]{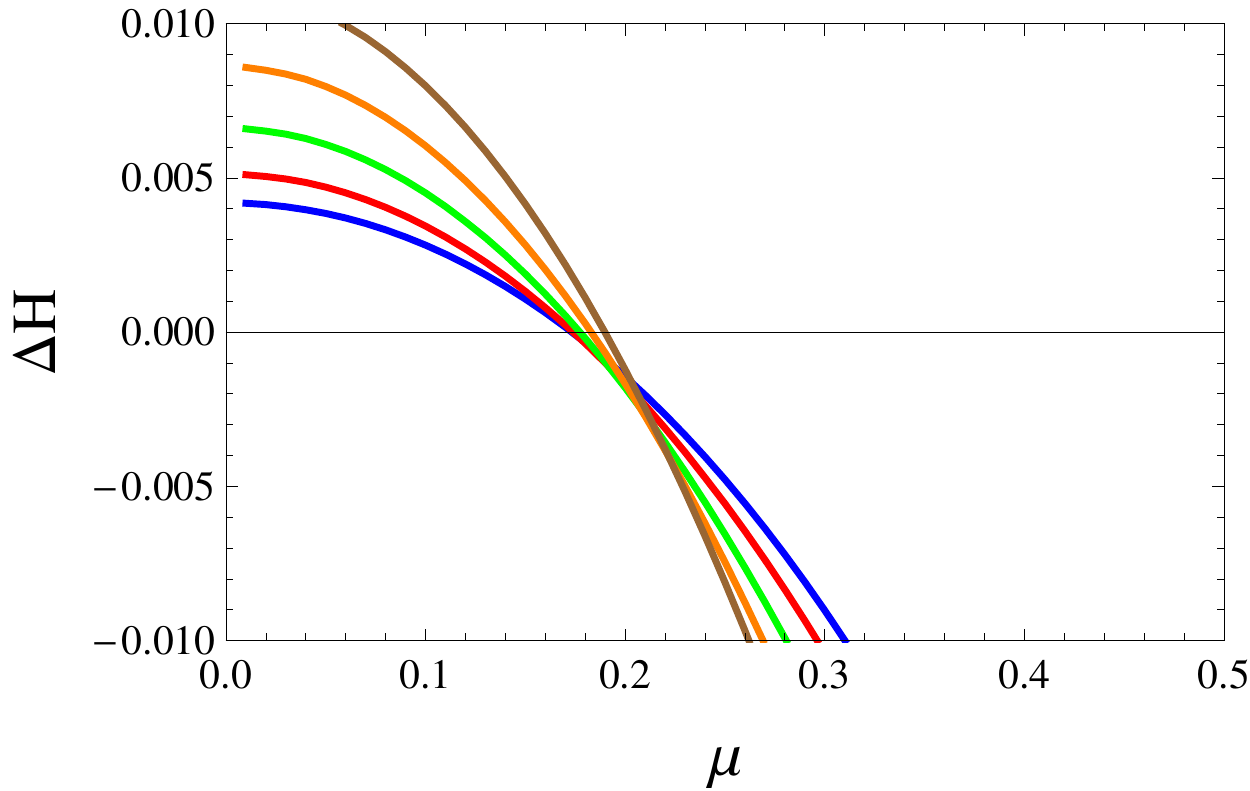}
     \caption{\small  Difference of ground state Hamiltonians $\Delta H$ as a function of $\mu$ and different values of the magnetic field $b=0.05$ (blue), $b=0.1$ (red), $b=0.15$ (green), $b=0.2$ (orange) and $b=0.25$ (brown).  The left, center and right panels correspond to $v_T=0$, $v_T=0.1$  and  $v_T=0.3$, respectively. }
  \label{fig:DeltaHvsmu}
\end{figure}

\begin{figure}[H] 
   \centering
   \includegraphics[width=7cm]{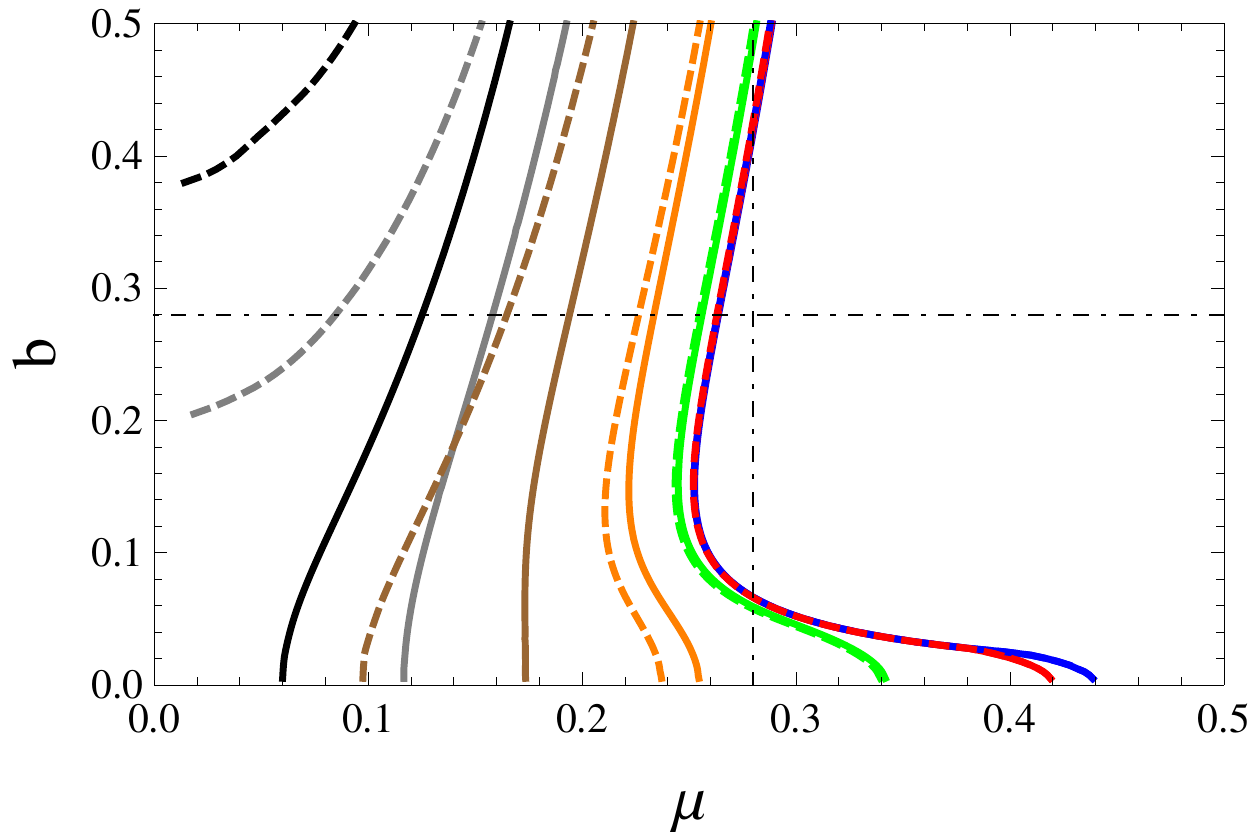} 
      \caption{\small  Phase diagram for the chiral transition in the $(b, \mu)$ plane.  Each critical line divides the plane in two sides corresponding to the chirally broken (left) and chirally symmetric (right) phases. Different colors correspond to different temperatures $v_T=0$ (blue), $v_T=0.02$ (red), $v_T=0.1$ (green), $v_T=0.2$ (orange), $v_T=0.3$ (brown), $v_T=0.36$ (gray) and $v_T=0.4$ (black). The solid lines are the results using the full numerical solutions for the chirally broken phase. The dashed lines are the results using the PRS approximation. The dot-dashed horizontal and vertical lines correspond to the cases $b={\rm const}$ and $\mu={\rm const}$ respectively. The intersection between those lines and the solid lines allows us to extract the critical values $\mu_c(b)$ and $b_c(\mu)$ respectively.  }
  \label{fig:bvsmuc}
\end{figure}

For each temperature $T$, we find a critical line in the $(b, \mu)$  plane. In Fig. \ref{fig:DeltaHvsmu} we show our results for $\Delta H$ as a function of $\mu$ for the temperatures $v_T=(0,0.1,0.3)$ and five different values of the magnetic field $b$. The intersection points with the horizontal axis correspond to the critical values of the chemical potential $\mu_c$ where the chiral transition takes place. 
Collecting the results for $\mu_c$ for each different value of $b$, at fixed $T$, we obtain the phase diagram for the chiral transition in the $(b, \mu)$ plane (at fixed temperature). In Fig. \ref{fig:bvsmuc}, we present our results for the chiral transition using the full numerical procedure described in subsection \ref{NumXB} (solid lines) compared to those obtained using the PRS approximation (dashed lines), described in subsection \ref{PRSapprox}. Different colors represent seven different temperatures from $v_T=0$ (blue) to $v_T=0.4$ (black). It is clear from Fig. \ref{fig:bvsmuc} that the validity of the 
the semi-analytic approximation of the PRS breaks well before the system reaches the critical temperature for which the effect of IMC disappears.
\begin{figure}[H] 
   \centering
   \includegraphics[width=7cm]{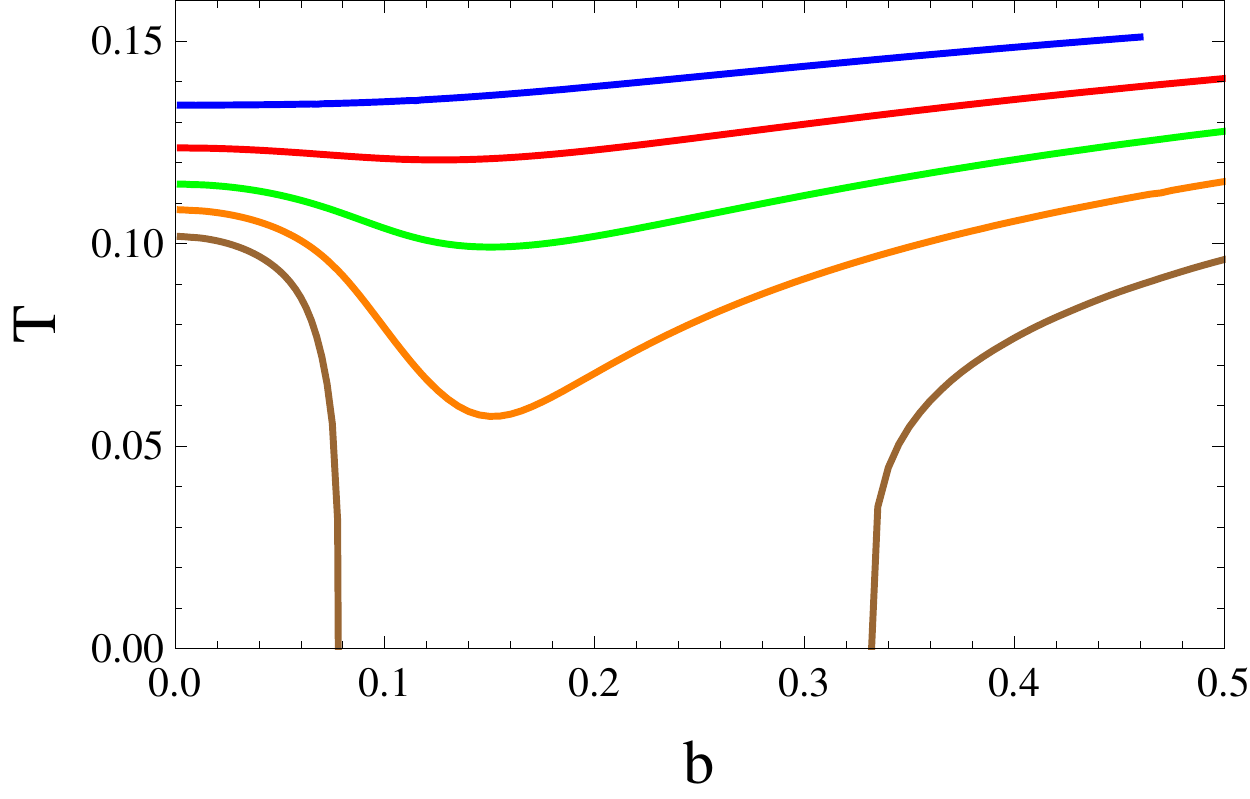} 
      \caption{\small Phase diagram in the $(T,b)$ plane for four different values of the chemical potential: $\mu = 0.16$ (blue) , $\mu = 0.2$ (red), $\mu = 0.23$ (green),   $\mu = 0.25$ (orange) and $\mu = 0.27$ (brown). The blue, red, green and orange lines describe the transition from the chirally broken phase (below the line) to the chirally symmetric phase. The two brown lines describe two consecutive transitions. The first transition is between the chirally broken and chirally symmetric phases (left to center) whereas the second transition turns the chirally symmetric phase into a chirally broken phase (center to right). }
  \label{fig:Tcvsbc}
\end{figure}

Note that the critical value for $\mu$ at fixed $b$, i.e. $\mu_c(b)$ is obtained in Fig. \ref{fig:bvsmuc} from the intersection of the critical lines and horizontal lines $b={\rm const}$. Alternatively, critical values for $b$ at fixed $\mu$, i.e. $b_c(\mu)$ can be obtained in Fig. \ref{fig:bvsmuc} by intersecting the critical lines with vertical lines $\mu={\rm const}$. In this way  we obtain the phase diagram for the chiral transition in the $(T,b)$ plane, at fixed $\mu$. In Fig. \ref{fig:Tcvsbc}, we plot some critical lines in that plane. Different colors correspond to different values of the chemical potential $\mu$. This plot shows that as the chemical potential $\mu$ decreases the IMC effect disappears at some critical $b$ and becomes MC. In particular, at $\mu=0$ IMC has disappeared completely. This is related to the probe approximation used in our model, where backreaction effects are neglected. Incorporating those effects, as in \cite{Ballon-Bayona:2013cta,Mamo:2015dea,Rougemont:2015oea,Dudal:2015wfn,Fang:2016cnt,Evans:2016jzo, Gursoy:2016ofp}, IMC appears again due to the interplay between the deconfinement and chiral transitions.   

\begin{figure}[H] 
   \centering
   \includegraphics[width=7cm]{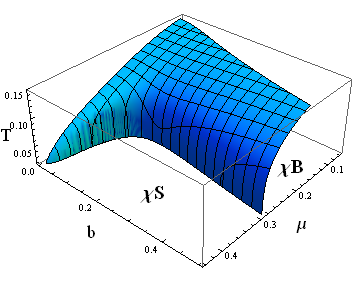} 
      \caption{\small The 3D phase diagram in the parameter space $(b,\mu,T)$. Projections onto the $(b,\mu)$ and $(T,b)$ planes, shown in figures \ref{fig:bvsmuc} and \ref{fig:Tcvsbc}, correspond to fixing $T$ or $\mu$ respectively. The projection onto the $(T,\mu)$ plane corresponds to fixing $b$.}
  \label{fig:PhaseDiag3D}
\end{figure}

The full 3D phase diagram in the parameter space $(b,\mu,T)$ is shown in Fig. \ref{fig:PhaseDiag3D}. In addition to the projections onto the $(b,\mu)$ and $(T,b)$ planes, already shown in figures \ref{fig:bvsmuc} and \ref{fig:Tcvsbc}, from Fig. \ref{fig:PhaseDiag3D} one can obtain the projection onto the $(T,\mu)$ plane. 

Our main results for the  chiral transition, displayed in figures \ref{fig:bvsmuc}, \ref{fig:Tcvsbc} and \ref{fig:PhaseDiag3D}, clearly indicate that the finite density deconfined Sakai-Sugimoto model allows IMC and this effect typically occurs at small $b$. At large $b$ the IMC effect disappears and the traditional MC becomes the dominant effect. A detailed analysis of the transition from IMC to MC will be developed in the next section, in terms of a universal order parameter. Here we provide a physical interpretation of this transition, following \cite{Preis:2010cq,Preis:2012fh}. Since we work in the probe approximation, where backreaction effects are neglected, in the chirally broken phase the chiral condensate always increases with $b$. This can be seen in Fig. \ref{fig:etavsv0}, where the parameter $v_0$, characterising the constituent quark mass, increases with $b$. However, as explained in the introduction, at finite density the magnetic field $b$ also contributes to the energy cost of creating that condensate. The Hamiltonian difference, defined in \eqref{DeltaHam}, then has two contributions, a negative term associated with the energy gain of having a chiral condensate and a positive term associated with the energy cost of creating it. At small $b$ the energy cost is bigger than the energy gain and as a consequence IMC is the dominant effect.

As described at the beginning of this section, our results for the chiral transition correspond to the case $\ell=1$ where $\ell$, defined below (\ref{eq:BC}), is the dimensionless descendent of $L$; the separation between the $D8$ and $\overline{D8}$ branes. We want to stress that the DBI-CS equations actually depend on the quantities $v_0 \ell^2$, $\mu  \ell^2$,  $v_T \ell^2$, $b \ell^3$, and therefore the results for a different value of $\ell$ can be extracted from the $\ell=1$ results by replacing $\mu$, $v_T$ and $b$ by $\mu  \ell^2$,  $v_T \ell^2$ and $b \ell^3$, respectively.

We would like to finish this section pointing out that when constructing the phase diagram for the chiral transition, we have ignored the presence of baryonic matter. If baryonic matter were to be included, the phase diagram would change dramatically, as can be seen in \cite{Preis:2011sp}, using the PRS approximation.  

\section{Magnetisation as an order parameter for IMC} \label{sec:magnetisation}

In this section we will present a detailed study of the magnetisation, emphasising its role as an order 
parameter of  IMC.  The magnetisation can be obtained from the Hamiltonian through the formula 
\begin{equation}
M = - \frac{\partial H}{\partial b}\Big{|}_{T,\mu} \, ,
\label{eq:magnetisation}
\end{equation}
with temperature $T$ and chemical potential $\mu$ held fixed.

\subsection{Magnetisations and charge densities near the critical line}

\begin{figure}[H] 
   \centering
   \includegraphics[width=4.5cm]{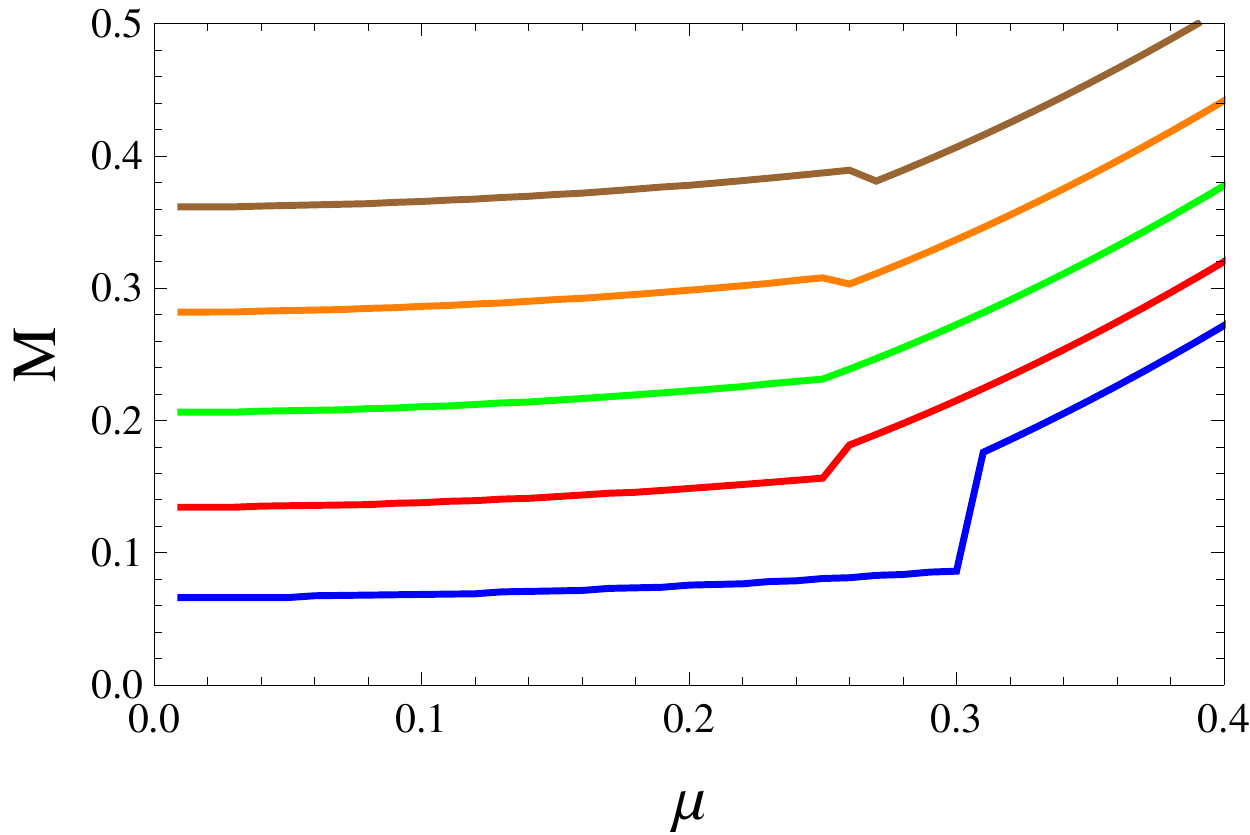} 
   \hspace{0.2cm}
   \includegraphics[width=4.5cm]{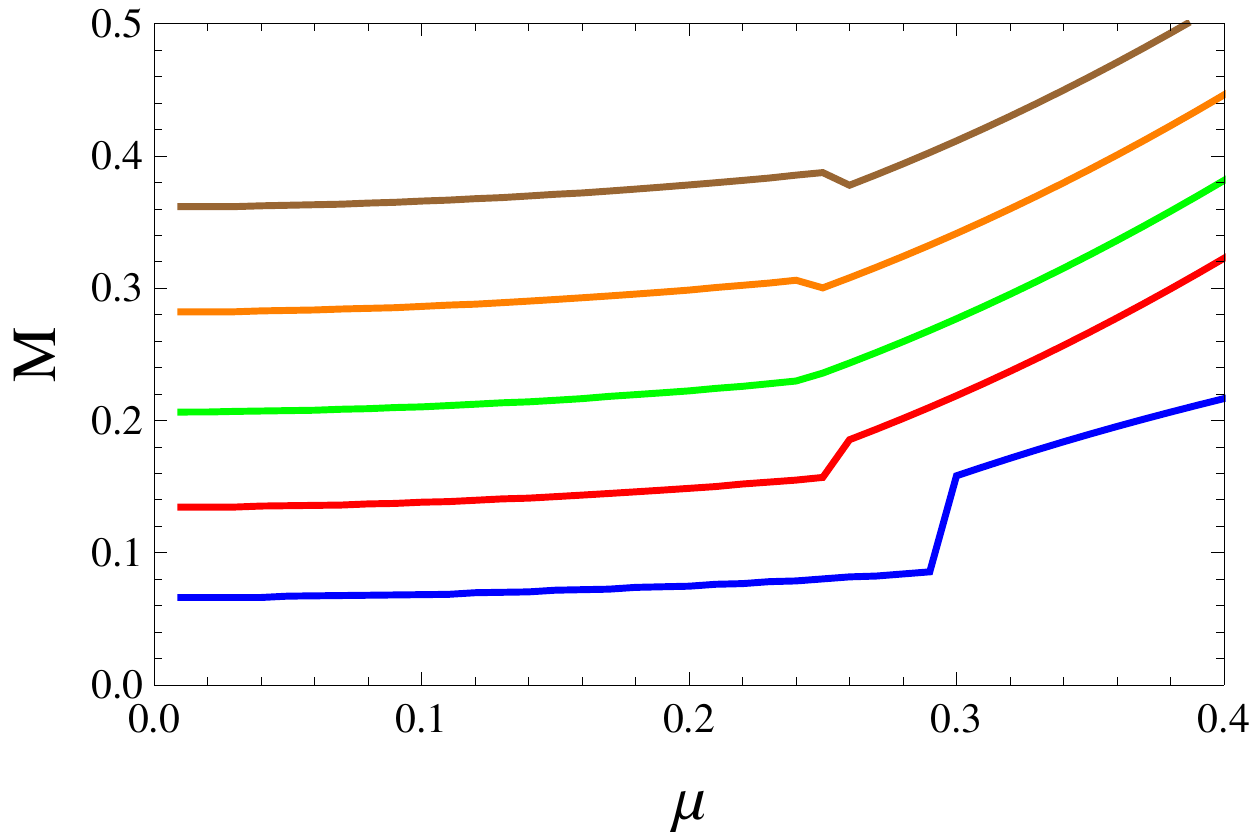}
   \hspace{0.2cm}
   \includegraphics[width=4.5cm]{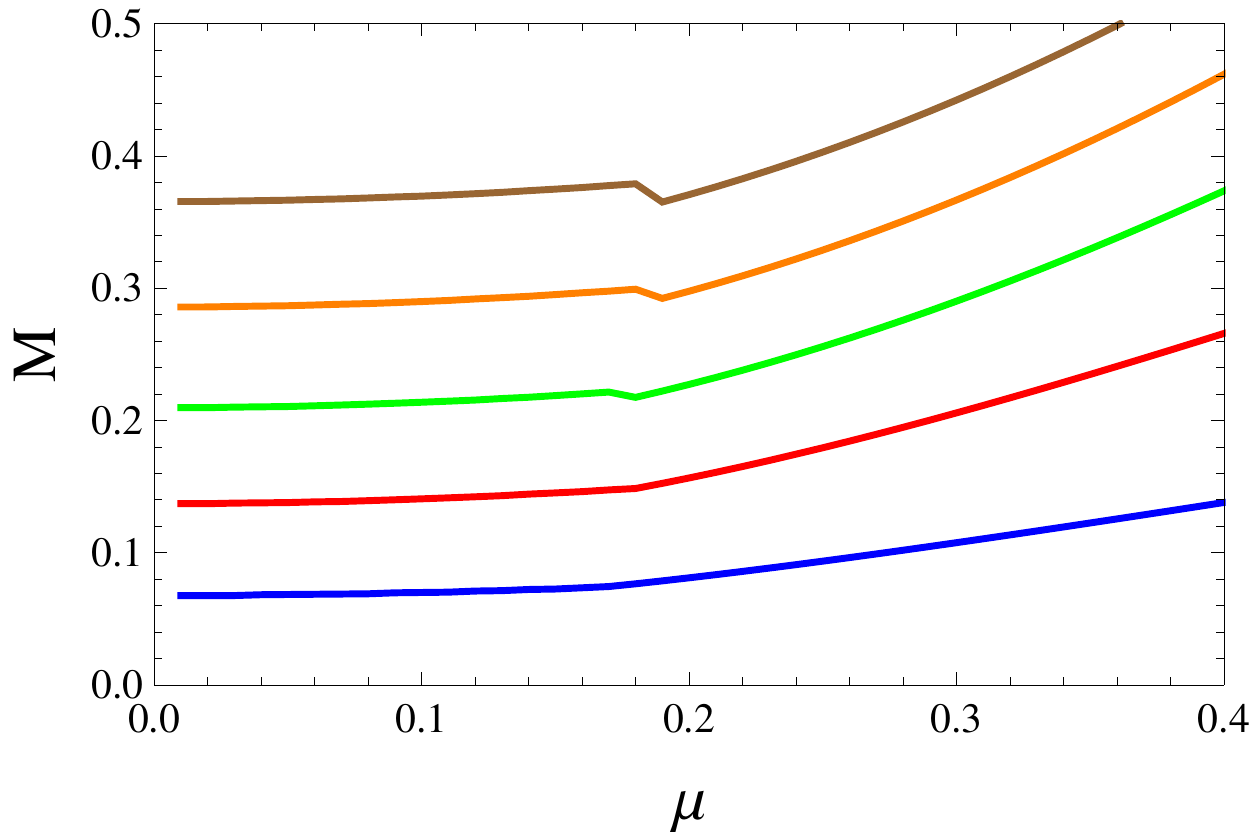}
     \caption{\small  The magnetisation $M$ as a function of the chemical potential $\mu$ for five different values of the magnetic field $b=0.05$ (blue), $b=0.1$ (red), $b=0.15$ (green), $b=0.2$ (orange) and $b=0.25$ (brown). The left, center and right panels correspond to $v_T=0$, $v_T=0.1$  and  $v_T=0.3$  respectively. }
  \label{fig:Magvsmu}
\end{figure}

It turns out that across the first order phase transition between the chirally broken and chirally symmetric phases, the magnetisation is discontinuous, and more specifically, the  magnetisation variation $\Delta M$ along the chiral transition has a specific sign that distinguishes the MC regime from the IMC regime. This is shown in Fig. \ref{fig:Magvsmu} where we plot the magnetisation $M$ as a function of the chemical potential $\mu$ for $v_T=(0,0.1,0.3)$ and five different values of the magnetic field $b$. For fixed $v_T$ we see that the discontinuity on the magnetisation occurs at the same $\mu_c(b)$ found in the phase diagram shown in Fig. \ref{fig:bvsmuc}. The magnetisation variation $\Delta M$ at $\mu_c(b)$ changes from positive to negative when going from the IMC regime (small $b$) to the MC regime (large $b$). The plot in the right panel of Fig. \ref{fig:Magvsmu} shows how the IMC regime disappears as the temperature gets high enough.  From this analysis we conclude that the magnetisation behaves as an order parameter that distinguishes IMC from MC. 

\begin{figure}[H] 
   \centering
   \includegraphics[width=4.6cm]{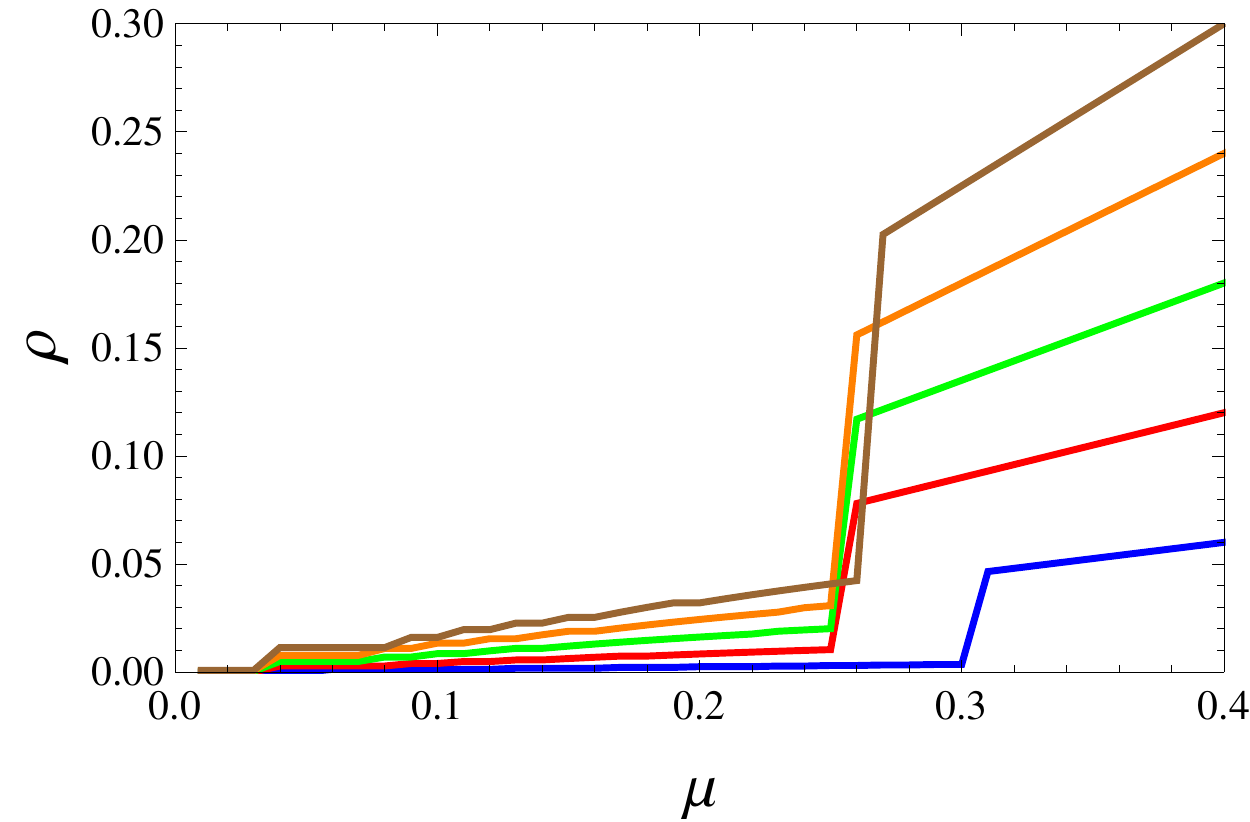} 
   \hspace{0.2cm}
   \includegraphics[width=4.6cm]{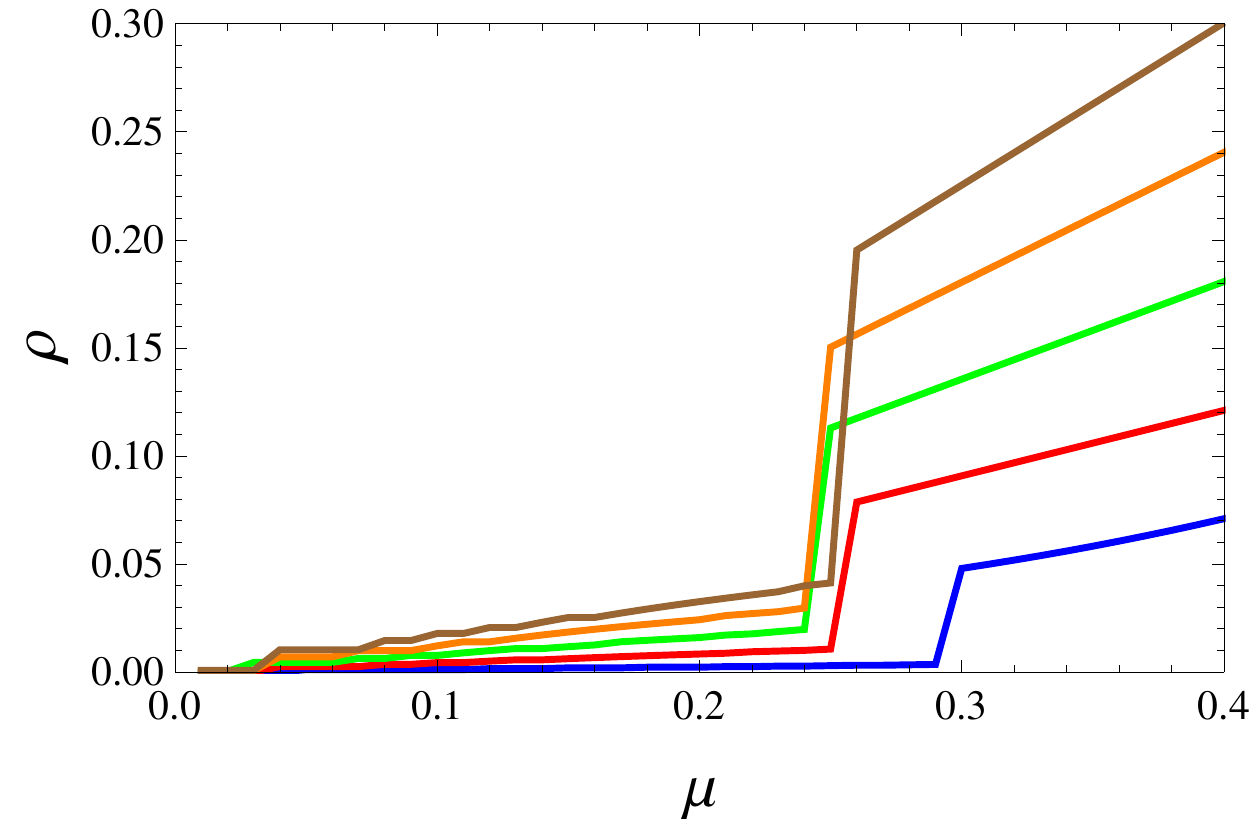}
   \hspace{0.2cm}
   \includegraphics[width=4.6cm]{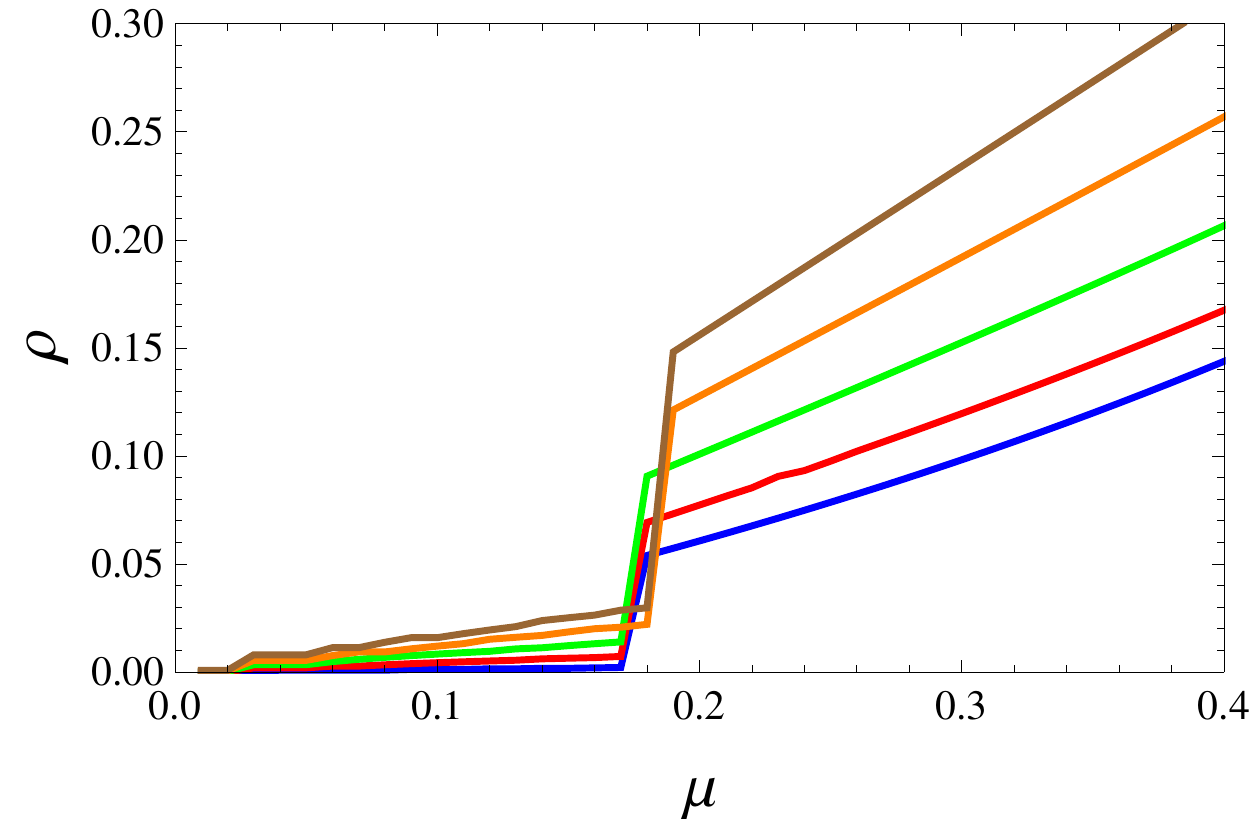}
     \caption{\small  The charge density $\rho$ as a function of the chemical potential $\mu$ for five different values of the magnetic field $b=0.05$ (blue), $b=0.1$ (red), $b=0.15$ (green), $b=0.2$ (orange) and $b=0.25$ (brown). The left, center and right panels correspond to $v_T=0$, $v_T=0.1$  and  $v_T=0.3$, respectively. }
  \label{fig:rhovsmu}
\end{figure}

Another interesting observable across the chiral transition is the charge density, defined in terms of the Hamiltonian by 
\begin{equation}
\rho = -\frac{\partial H}{\partial \mu} \Big{|}_{T,b} \ , 
\label{eq:chargedensity}
\end{equation}
with temperature $T$ and magnetic field $b$ held fixed. In the chirally broken phase the charge density reduces to $\rho = (3/2) b j$, where $j$ is the supercurrent defined in (\ref{eq:BC}). As with the magnetisation, at fixed $T$ the charge density also shows a discontinuity at the critical line $\mu_c(b)$ where the chiral transition takes place. This is shown in Fig. \ref{fig:rhovsmu} where we plot the charge density $\rho$ as a function of the chemical potential $\mu$ for $v_T=(0.0.1,0.3)$ and five different values of the magnetic field $b$. We note, however, that the variation of the charge density $\Delta \rho$ at $\mu_c$ always remains positive, thus not distinguishing between the IMC and MC regimes.  

\subsection{The critical line $\mu_c(b)$ from $\Delta M$}

The analysis presented in the previous section led us to suspect the existence of a relation, at fixed temperature $T$, between the evolution of the critical line $ \mu_c (b)$, that separates the chirally broken and chirally symmetric phases, and the behaviour of the magnetisation variation at the chiral transition, $\Delta M$. It seems that whenever $\Delta M$ is positive at some fixed magnetic field $b$ the critical chemical potential $\mu_c$ decreases at the next value of the magnetic field $b + \Delta b$, which is what traditionally characterizes the IMC regime. Similarly, when $\Delta M$ is negative at fixed $b$ the critical chemical potential $\mu_c$ increases at $b + \Delta b$ (MC regime). 

\begin{figure}[t] 
   \centering
   \includegraphics[width=6.5cm]{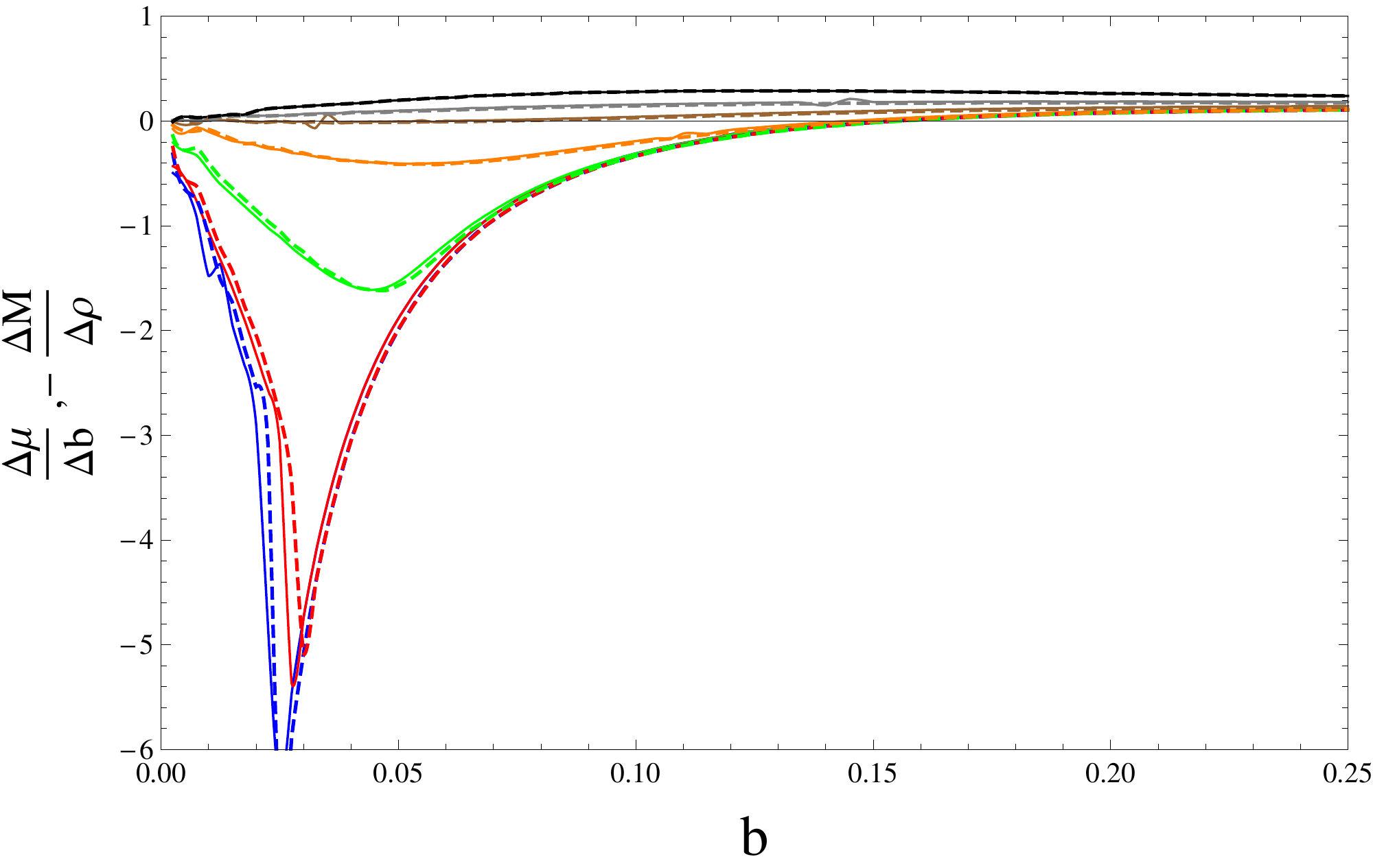}
   \hspace{0.75cm} 
   \includegraphics[width=6cm]{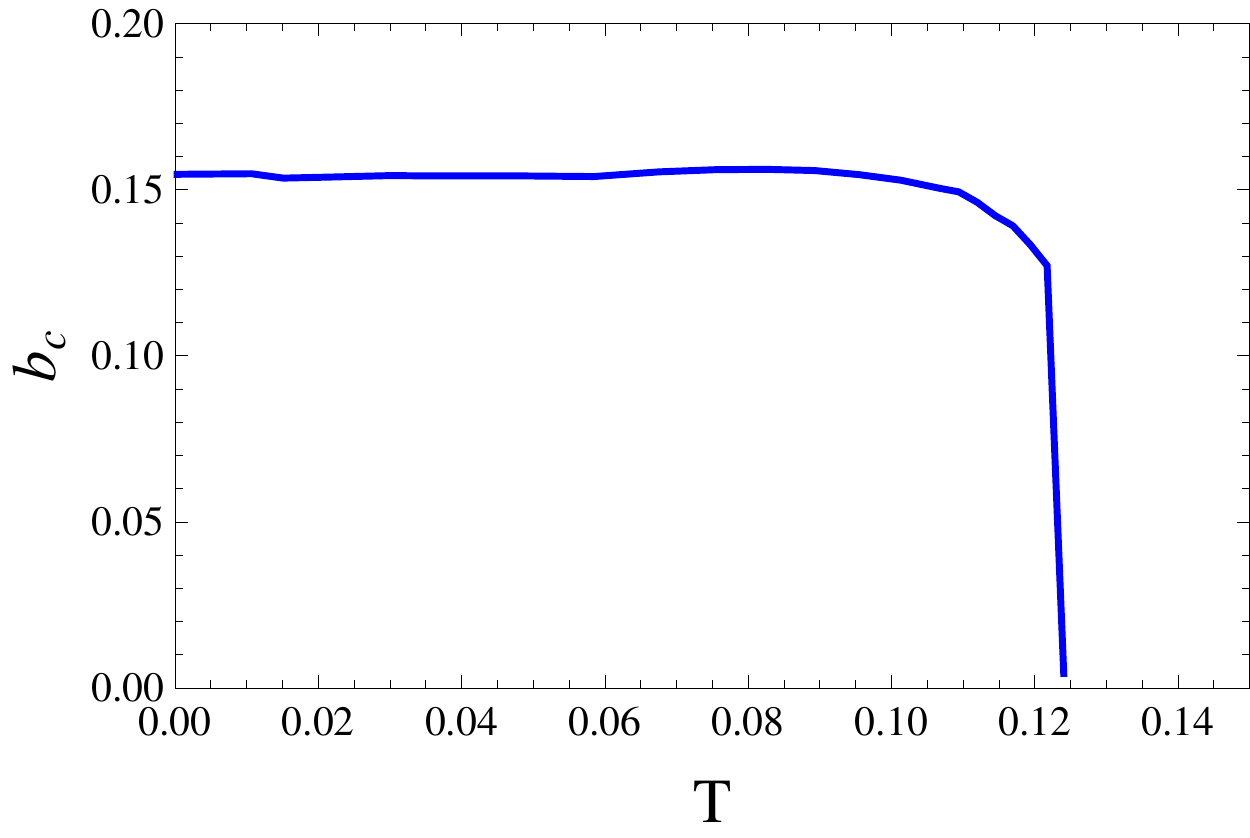}
      \caption{\small {Left panel:} The ratios $\Delta \mu / \Delta b$ (solid lines) and $-\Delta M / \Delta \rho$ (dashed lines) plotted as functions of the magnetic field $b$ for seven different values of the temperature $v_T=0$ (blue), $v_T=0.02$ (red), $v_T=0.1$ (green), $v_T=0.2$ (orange), $v_T=0.3$ (brown), $v_T=0.36$ (gray) and $v_T=0.4$ (black). {Right panel:} Critical magnetic field $b_c$ where IMC becomes MC as a function of the temperature $T$. Below $b_c$ the chemical potential $\mu$ decreases with $b$ (IMC) whereas above $b_c$ one finds the opposite behaviour (MC).}
  \label{fig:DeltamucvsDeltab}
\end{figure}

Motivated by this observation we consider, at fixed $T$, a perturbative expansion for the Hamiltonian along the critical line $\mu_c (b)$. When the magnetic field evolves from $b$ to $b + \Delta b$ the chemical potential evolves from $\mu_c$ to $\Delta \mu_c$ and the Hamiltonian evolves as 
\bea
H(\mu_c + \Delta \mu_c, b + \Delta b) = H(\mu_c , b) - \rho(\mu_c,b) \Delta \mu_c  - M(\mu_c,b) \Delta b \,, \label{eq:Hamexp}
\eea
where we have used the definitions (\ref{eq:magnetisation}) and (\ref{eq:chargedensity}). The expansion (\ref{eq:Hamexp}) holds for both the chirally broken and chirally symmetric phases. Since $(b, \mu_c)$ and $(b + \Delta b , \mu_c + \Delta \mu_c)$ are points along the critical line for the chiral transition, the Hamiltonian at these points satisfies the relations 
\bea
H_{\chi S} (\mu_c , b) =  H_{\chi B} (\mu_c , b) \ , \quad 
H_{\chi S} (\mu_c + \Delta \mu_c , b + \Delta b) =  H_{\chi B} (\mu_c + \Delta \mu_c, b + \Delta b) \ , \label{eq:Hamrels}
\eea
where $\chi B$ and $\chi S$ refer to the chirally broken and chirally symmetric phases, respectively. From (\ref{eq:Hamexp}) and (\ref{eq:Hamrels}), we find  the interesting relation 
\bea
\frac{\Delta \mu_c }{\Delta b} = 
- \frac{M_{\chi S} - M_{\chi B}}{ \rho_{\chi S} - \rho_{\chi B}} = - \frac{\Delta M}{\Delta \rho} \,. \label{eq:univrel}
\eea
The relation (\ref{eq:univrel}) is universal, i.e. it does not depend on the  model used to describe the chiral transition. This relation demonstates the role of the discontinuity of the magnetisation as an order parameter to distinguish IMC from MC, and also provides a method to reconstruct the critical line $\mu_c (b)$ from the discontinuities $\Delta M$ and $\Delta \rho$. We check numerically the validity of the formula (\ref{eq:univrel}) in our set-up and find a very good match of the ratios  $\Delta \mu_c / \Delta b$ and $- \Delta M / \Delta \rho$, within the limits of numerical errors. The results are shown in Fig. \ref{fig:DeltamucvsDeltab}, where we plot the ratios $\Delta \mu_c / \Delta b$ and $- \Delta M / \Delta \rho$ as functions of the magnetic field $b$ for five different values of the temperature $T$. In the same figure, we exhibit the behaviour of the critical magnetic field $b_c$ vs. $T$ which separates the regimes of  IMC and MC. More concretely, when the magnetic field is in the regime $b<b_c$ IMC takes place whereas in the regime $b>b_c$ MC is recovered. 

Two interesting observations can be drawn from the second plot in in Fig. \ref{fig:DeltamucvsDeltab}: The critical magnetic field $b_c$ is almost constant (cf. also Fig. \ref{fig:bvsmuc}) for a range of temperatures and then drops rapidly, reaching $b_c=0$ at $T_e \approx 0.124$. Above this temperature, the effect of inverse magnetic catalysis disappears completely and only the normal effect of magnetic catalysis can be observed. The numerical computation of $T_e$ is a novel result of this paper.

\subsection{The critical line $T_c(B)$ at fixed $\mu$}

The analysis in the previous subsection was done along the critical line $\mu_c(b)$ at fixed $T$. 
Alternatively, we can fix the chemical potential $\mu$ and analyse the Hamiltonian along the critical line $T_c(b)$, cf. e.g. the critical lines in Fig. \ref{fig:Tcvsbc}.
 
\begin{figure}[H] 
   \centering
   \includegraphics[width=4.5cm]{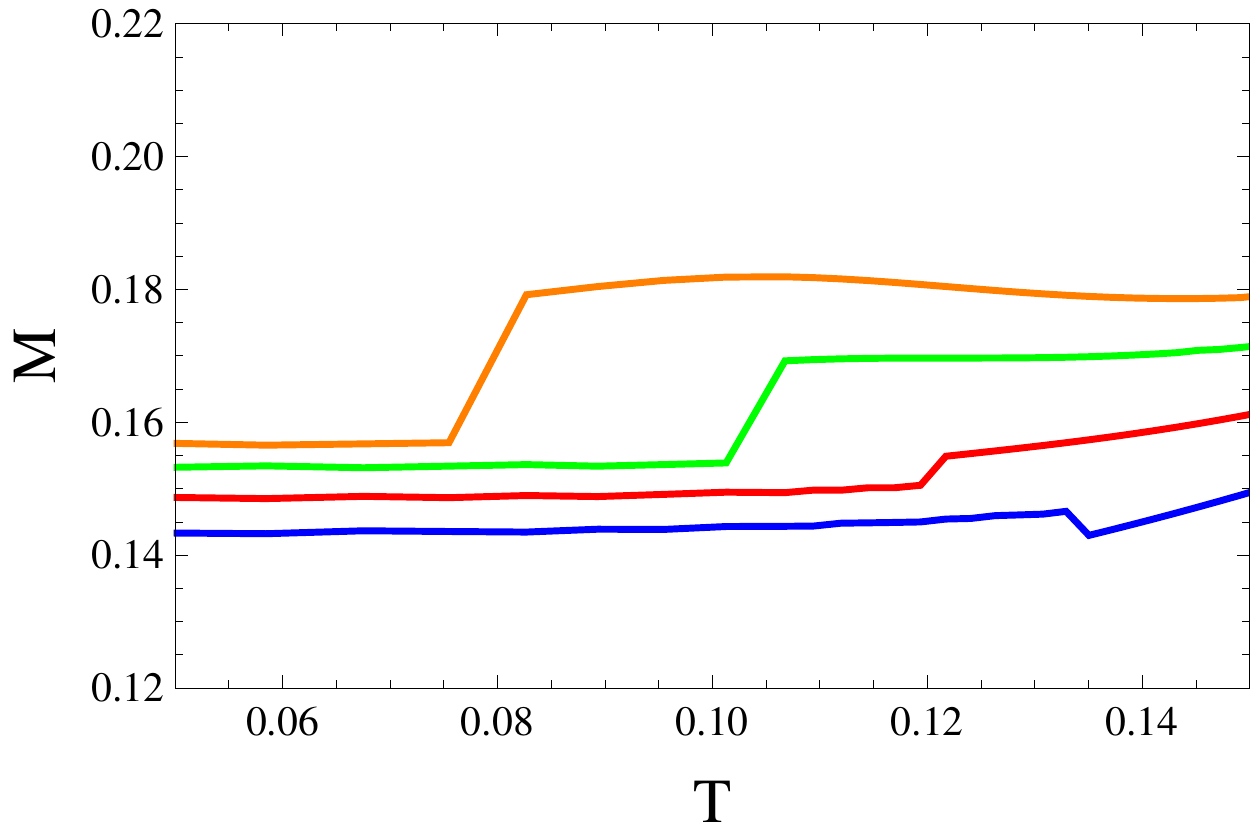} 
   \hspace{0.2cm}
   \includegraphics[width=4.5cm]{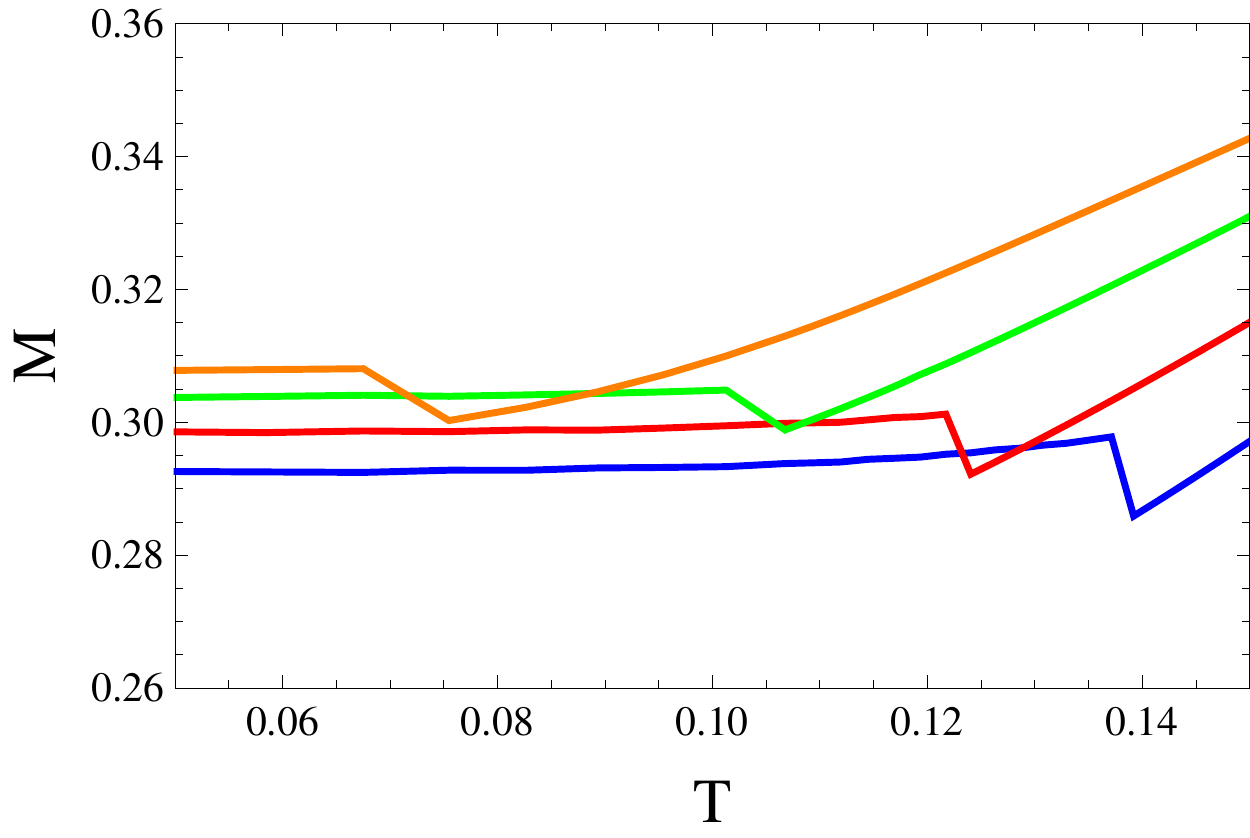}
   \hspace{0.2cm}
   \includegraphics[width=4.5cm]{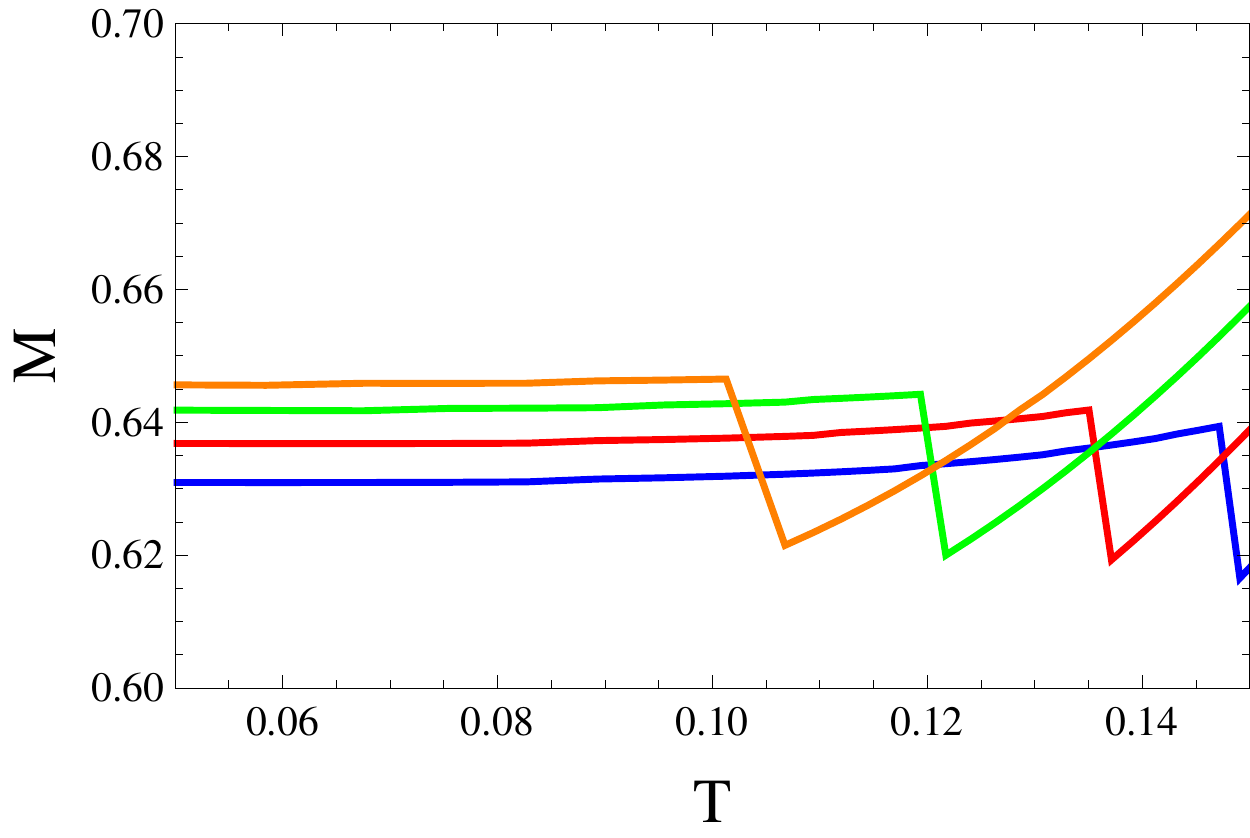}
     \caption{\small  The magnetisation $M$ as a function of the temperature $T$ for four different values of the chemical potential $\mu=0.16$ (blue), $\mu=0.2$ (red), $\mu=0.23$ (green) and $\mu=0.25$ (orange). The left, center and right panels correspond to $b=0.1$, $b=0.2$  and  $b=0.4$, respectively. }
  \label{fig:MagvsT}
\end{figure}

\begin{figure}[H] 
   \centering
   \includegraphics[width=4.5cm]{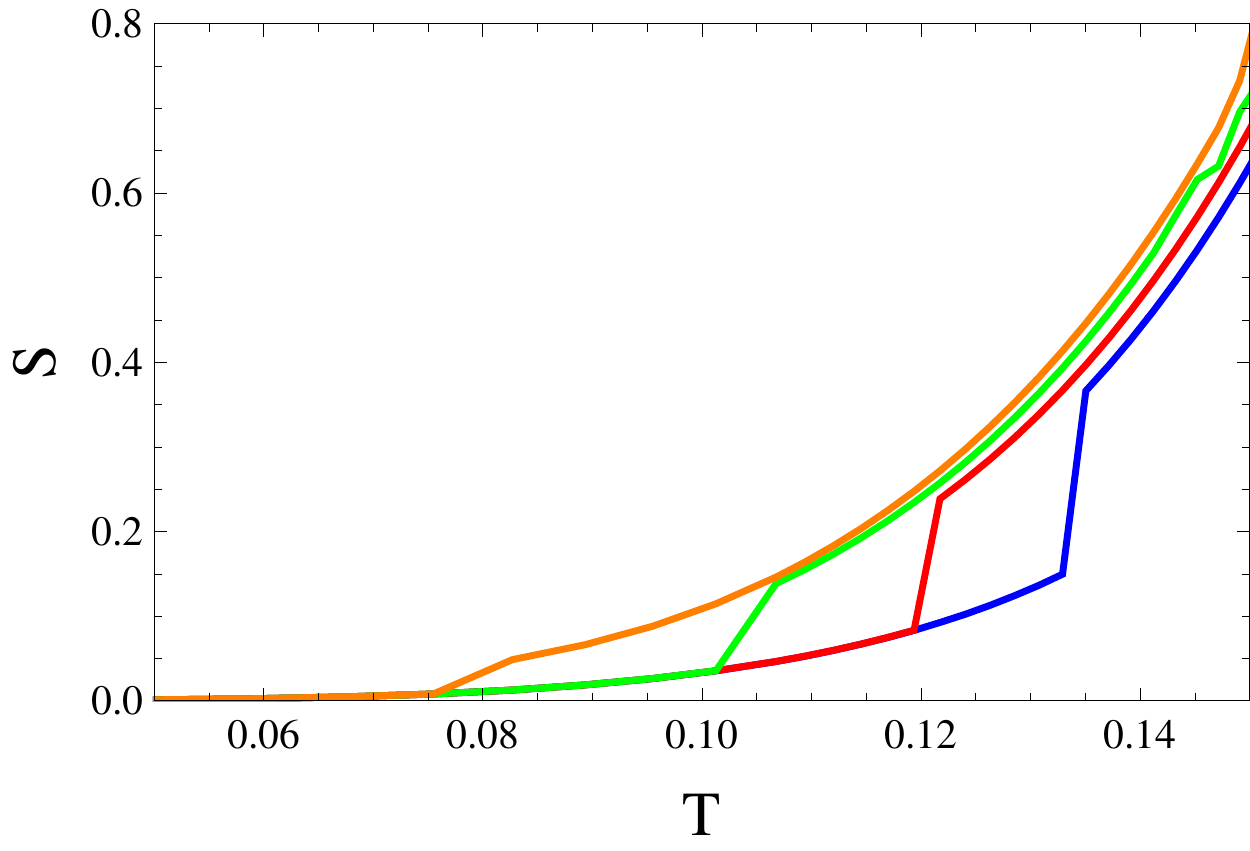} 
   \hspace{0.2cm}
   \includegraphics[width=4.5cm]{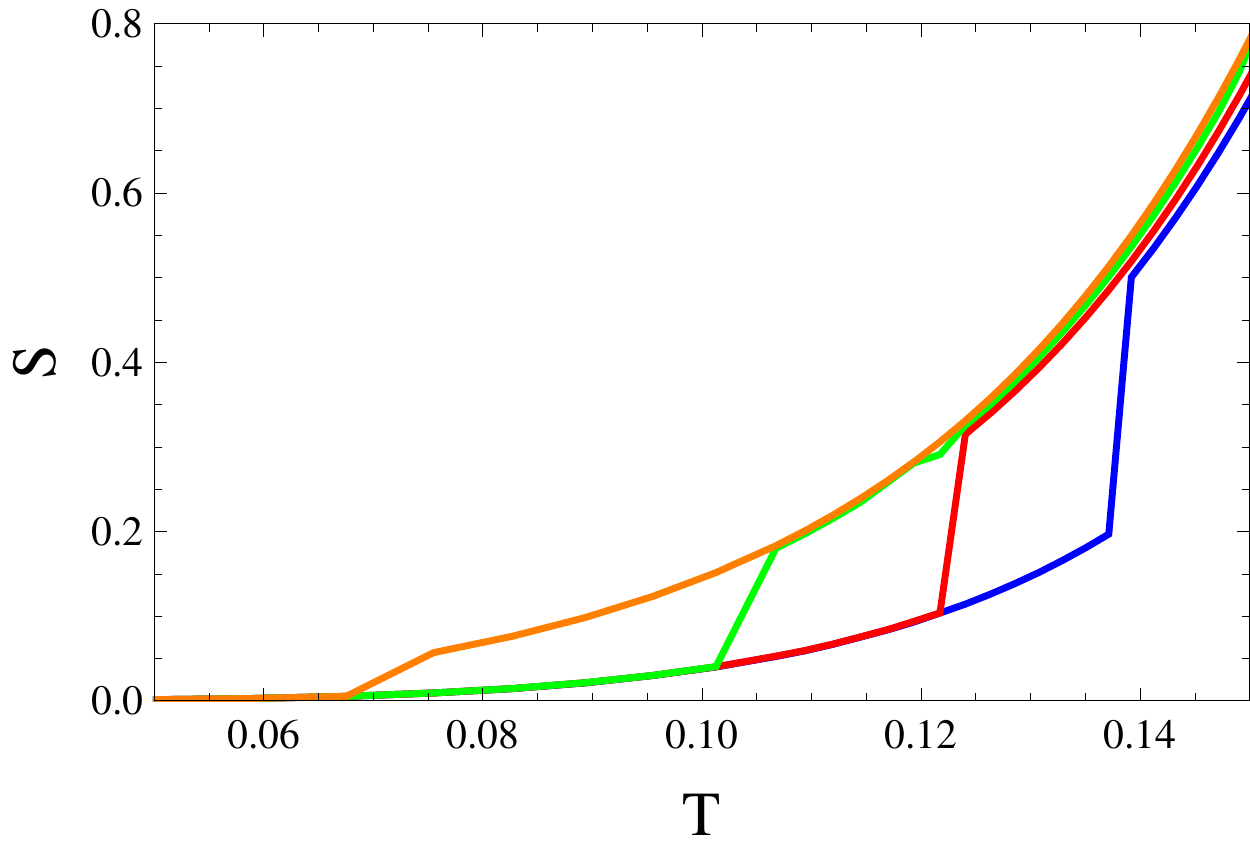}
   \hspace{0.2cm}
   \includegraphics[width=4.5cm]{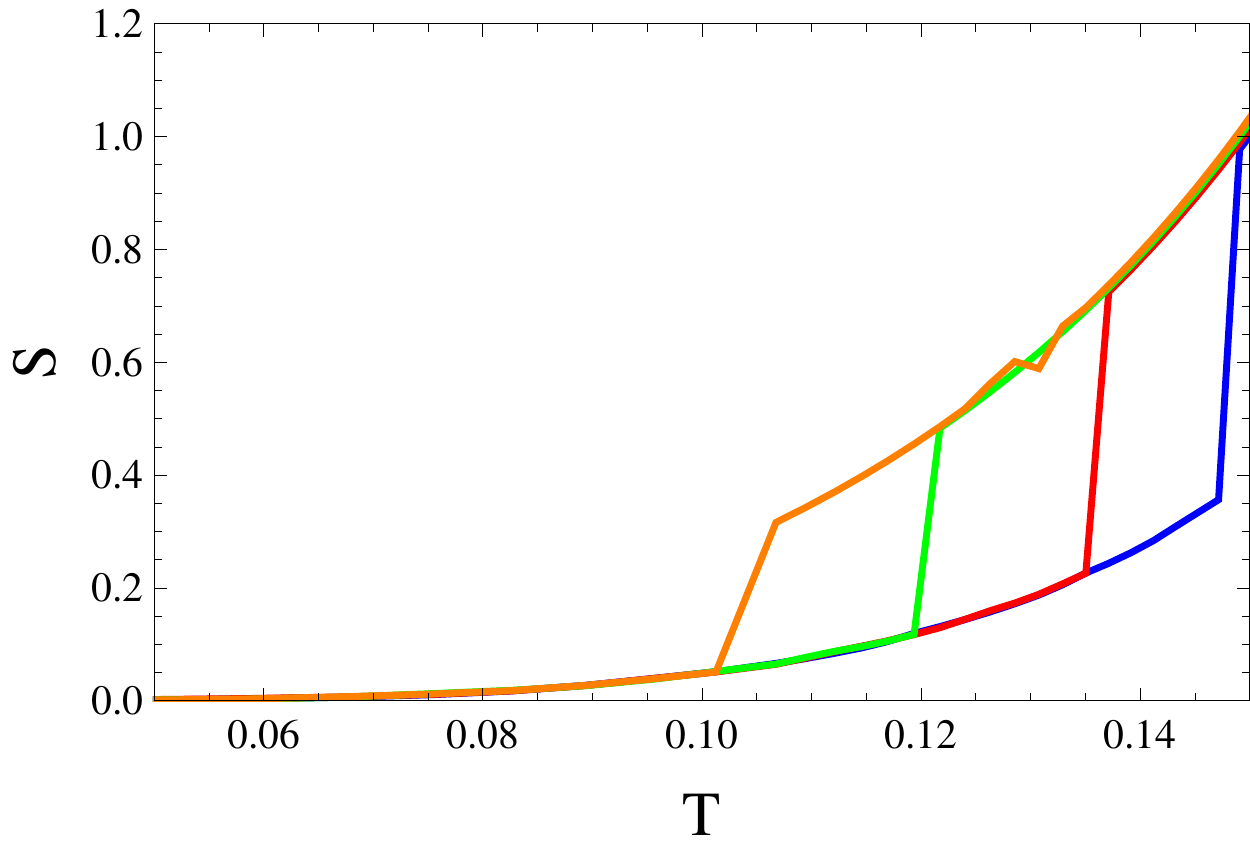}
     \caption{\small  The entropy ${\cal S}$ as a function of the temperature $T$ for four different values of the chemical potential $\mu=0.16$ (blue), $\mu=0.2$ (red), $\mu=0.23$ (green) and $\mu=0.25$ (orange). The left, center and right panels correspond to $b=0.1$, $b=0.2$  and  $b=0.4$  respectively. }
  \label{fig:EntropyvsT}
\end{figure}

In analogy with \eqref{eq:Hamexp}, when the magnetic field increases from $b$ to $b + \Delta b$, the critical temperature evolves from $T_c$ to $T_c + \Delta T_c$, and we find a perturbative expansion for the Hamiltonian   
\begin{equation}
H(T_c + \Delta T_c, b + \Delta b) = H(T_c , b) - {\cal S} (T_c,b) \Delta T_c - M(T_c,b) \Delta b \,, \label{eq:Hamexpv2}
\end{equation}
where 
\begin{equation}
{\cal S} = -\frac{\partial H}{\partial T} \Big{|}_{\mu,b} \, , 
\label{eq:entropy}
\end{equation}
is the entropy, for fixed magnetic field $b$ and chemical potential $\mu$. In analogy with (\ref{eq:Hamrels}) we have the relations 
\begin{equation}
H_{\chi S} (T_c , b) =  H_{\chi B} (T_c , b) \ , \quad 
H_{\chi S} (T_c + \Delta T_c, b + \Delta b) =  H_{\chi B} (T_c + \Delta T_c, b + \Delta b) \, ,  \label{eq:Hamrelsv2}
\end{equation}
and from (\ref{eq:Hamexpv2}) and (\ref{eq:Hamrelsv2}) we find the relation 
\begin{equation}
\frac{\Delta T_c}{\Delta b} = -
\frac{M_{\chi S} - M_{\chi B}}{ {\cal S}_{\chi S} - {\cal S}_{\chi B}} = - \frac{\Delta M}{\Delta {\cal S}} \,. \label{eq:univrelv2}
\end{equation}
On general grounds one always expects a positive jump of the entropy at the chiral transition, i.e. $\Delta S >0$, so again the sign of the magnetisation jump $\Delta M$ will distinguish the regime of magnetic catalysis ($T_c$ increasing with $b$) from the regime of inverse magnetic catalysis ($T_c$ decreasing with $b$). In figures \ref{fig:MagvsT} and \ref{fig:EntropyvsT}, we show the magnetisation and entropy, respectively, as a function of the temperature $T$ for $b=0.1$, $b=0.2$ and $b=0.4$ and four different values of the chemical potential, corresponding to different colours. Comparing the plots in figure \ref{fig:MagvsT} with the plot in figure \ref{fig:Tcvsbc}, we see that the magnetisation jump is (positive) negative  in the regime of (inverse) magnetic catalysis. On the other hand, from the plots in figure \ref{fig:EntropyvsT}, we find that the entropy jump is always positive irrespective of the regime. These results are consistent with our formula (\ref{eq:univrelv2}). 

\begin{figure}[H] 
   \centering
   \includegraphics[width=6.5cm]{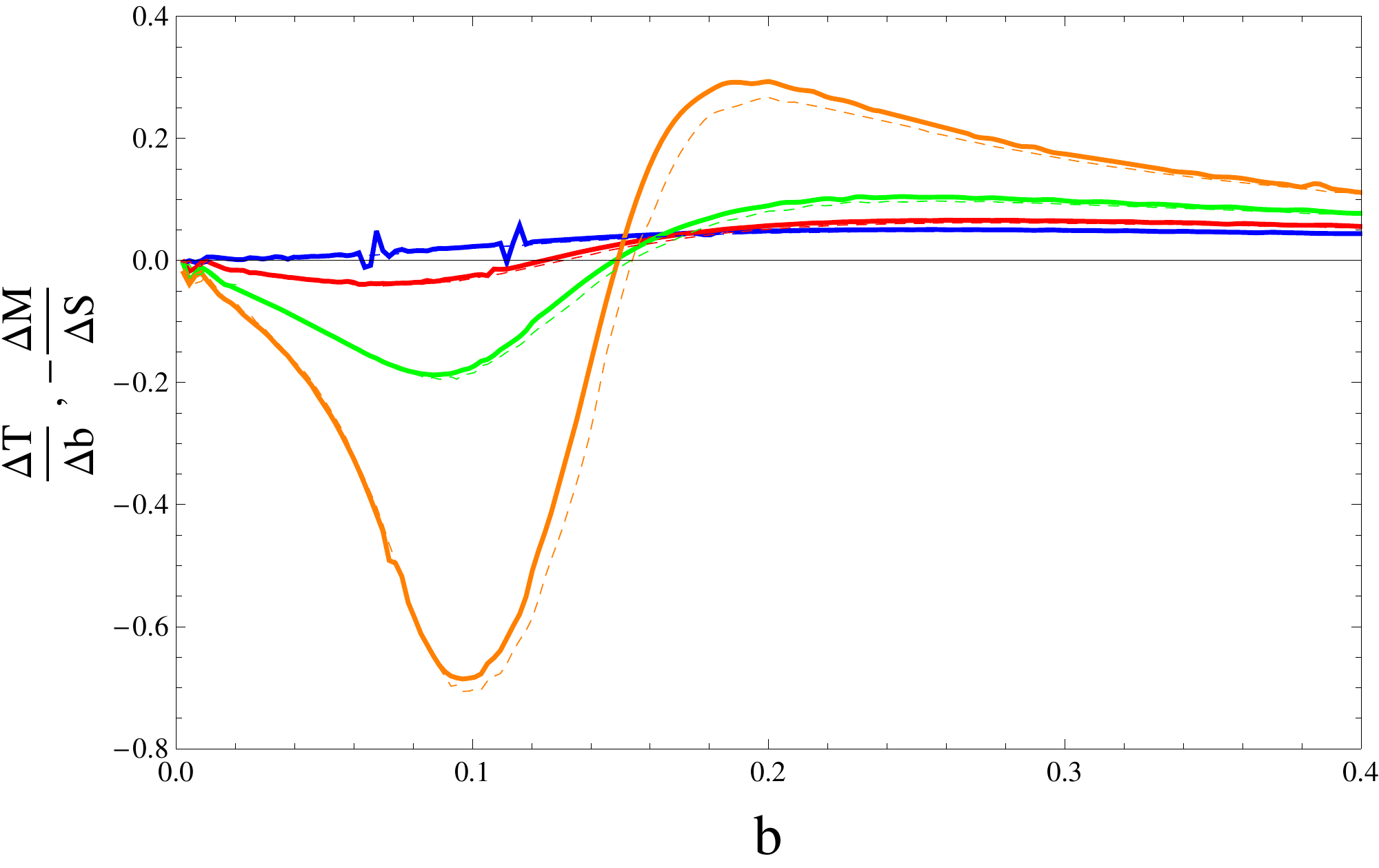}
   \hspace{0.75cm} 
   \includegraphics[width=6cm]{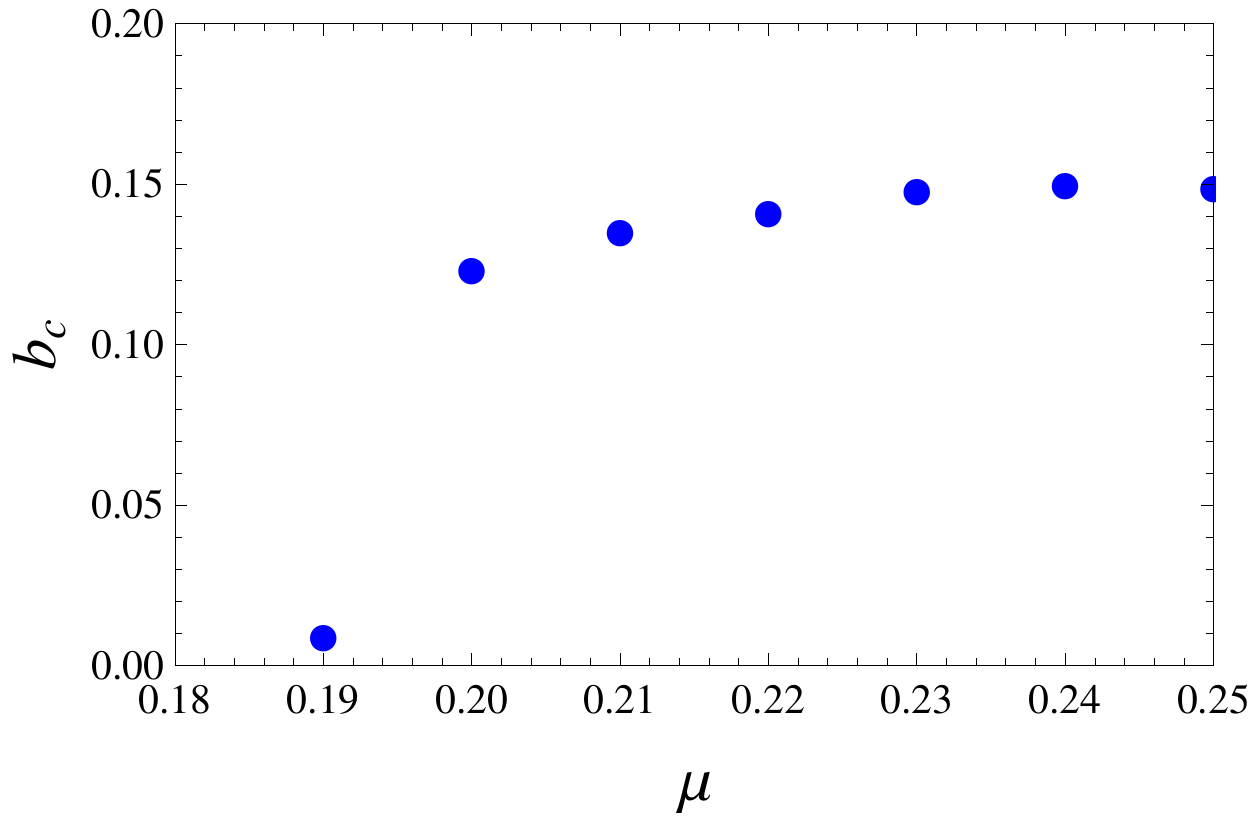}
      \caption{\small {Left panel:} The ratios $\Delta T / \Delta b$ (solid lines) and $- \Delta M / \Delta {\cal S}$ (dashed lines) plotted as functions of the magnetic field $b$ for four different values of the chemical potential $\mu=0.16$ (blue), $\mu=0.2$ (red), $\mu=0.23$ (green) and $\mu=0.25$ (orange). {Right panel:} Critical magnetic field $b_c$ where IMC becomes MC as a function of the chemical potential $\mu$. Note that the trigger of IMC occurs approximately at $\mu\approx0.19$ where $b_c$ starts increasing from zero. }
  \label{fig:DeltaTcvsDeltab}
\end{figure}

In figure \ref{fig:DeltaTcvsDeltab}, we compare the two sides of our identity (\ref{eq:univrelv2}) and find a good match within our available numerical precision. We use these results to estimate the critical value of the magnetic field $b_c$ (where IMC becomes MC) as a function of the chemical potential. We find that IMC first occurs approximately at $\mu_e \approx 0.19$ where $b_c$ starts growing. For values of the chemical potential $\mu$ lower than $\mu_e $, only MC occurs. As explained previously, the fact that IMC disappears in the case of $\mu=0$ is an artifact of the probe approximation considered in this model.

We would like to stress that the equality (\ref{eq:univrelv2}), although motivated by a particular holographic model, is universal and should be very useful for tracking inverse magnetic catalysis in different scenarios.  In particular, it can be used in the regime where the chemical potential is small or even zero, which is the regime accessible to lattice QCD computations. Moreover, although we have used (\ref{eq:univrelv2}) in a model where the chiral transition is first order, it should also be useful when the chiral transition is second or higher order. In those cases the magnetisation does not jump near the transition and the variation $\Delta M$ becomes infinitesimal ($dM$). But in any case, the sign of $\Delta M$ (or $d M$) distinguishes IMC from MC. 

Moreover, our formula (\ref{eq:univrelv2}) does not depend on the particular physical mechanism behind IMC which, in our framework, is related to including or not backreaction effects. Although we did not find IMC at $\mu=0$ in this particular model, due to the absence of backreaction, we expect that models incorporating backreaction effects, such as \cite{Ballon-Bayona:2013cta,Mamo:2015dea,Rougemont:2015oea,Dudal:2015wfn,Fang:2016cnt,Evans:2016jzo, Gursoy:2016ofp}, will exhibit a positive (negative) magnetisation variation whenever IMC (MC) appears. Interestingly, the authors of \cite{Gursoy:2016ofp} arrived at a formula equivalent to (\ref{eq:univrelv2}) for the deconfinement transition in a model where $\Delta M$ is positive and the magnetic field favours deconfinement\footnote{We thank the authors of \cite{Gursoy:2016ofp} for explaining the details of their analysis.}. As a matter of fact, although we propose the use of (\ref{eq:univrel}) and (\ref{eq:univrelv2}) as a criteria for distinguishing IMC from MC, both of them should be useful when investigating any phase transition in the $(T,b,\mu)$ phase diagram. The reason is that (\ref{eq:univrel}) and (\ref{eq:univrelv2}) were derived from perturbative expansions that can be interpreted as particular cases for the thermodynamic evolution of the grand canonical potential, i.e.  $d \Omega = - S dT - \rho d \mu - M dB$. The formula (\ref{eq:univrel}) is useful when the phase transition takes place in the $(b, \mu)$ plane (fixed $T$) whereas the formula (\ref{eq:univrelv2}) is useful for transitions taking place in the $(b,T)$ plane (fixed $\mu$). In particular, we would like to encourage further exploration of the relation (\ref{eq:univrelv2}) in non-perturbative models for the chiral or deconfinement transition (holographic or non-holographic) at finite or zero chemical potential and also in lattice QCD computations.

\section{Conclusions}
\label{sec:Conclusions}
In this paper we have explored the phase space of the chiral transition in a holographic model for QCD with a focus on the effect of (inverse) magnetic catalysis. Namely, we have investigated the deconfined finite temperature phases of the Sakai-Sugimoto model at non-vanishing magnetic field and chemical potential. We provided a full numerical solution to the field equations, 
building on and extending the previous semi-analytic approximation of \cite{Preis:2010cq}. As a consistency check for our numerical results, approximate analytic results at small values of the magnetic field and temperature were also obtained and are described in appendix \ref{AppB}.

We remind the reader that in the confined phase of the Sakai-Sugimoto model the quarks are always in the chirally broken phase. It is only after the deconfinement transition takes place that we have access to the chiral transition. Investigating the effect of a nonzero magnetic field on the deconfinement transition requires including backreaction effects. This is an important question because magnetic inhibition of confinement provides a plausible explanation of IMC at zero density. In this work we were mainly interested in the IMC effect at finite density which, as described in the introduction, has a physical origin very different from the case of zero density. For this reason we opted to work in the probe approximation, neglecting backreaction effects. Including those effects can be done, at least perturbatively, following the progress made in  \cite{Ballon-Bayona:2013cta} and \cite{Bigazzi:2014qsa}. In that scenario, it would also be interesting to study the inclusion of baryonic matter along the lines of \cite{Bergman:2007wp,Rozali:2007rx} and more recently \cite{Preis:2011sp}. Such a calculation is beyond the scope of the current investigation.

The main results are comprised of a detailed discussion of the chirally symmetric and chirally broken phases 
and the effect of (inverse) magnetic catalysis on the chiral phase transition between those phases. 
We identified and discussed a universal order parameter that distinguishes between magnetic catalysis (MC) and inverse magnetic catalysis (MC). This parameter is the magnetisation, which exhibits a jump $\Delta M$ across 
the critical line from the chirally broken to the chirally symmetric phase. We found that, for a given temperature $T$, a positive (negative) magnetisation jump signifies IMC (MC) in the sense that $\mu_c(b)$ is a decreasing (growing) function of $b$. Moreover, considering a perturbative expansion for the Hamiltonian along the chiral transition we arrived at the universal relation (\ref{eq:univrel}) that allowed us to track the critical line $\mu_c(b)$ from the evolution of the magnetisation and density. We used our criteria to find the value of the magnetic field $b_c$ for which IMC disappears. We found that, as the temperature increases, this critical magnetic field $b_c(T)$ remains almost constant and after $T\approx 0.1$ drops abruptly until it vanishes at $T_e \approx 0.124$. 

We also provided the universal relation (\ref{eq:univrelv2}) for the chiral transition in the phase diagram $T$ vs. $b$ (temperature vs. magnetic field), at fixed chemical potential.
Again, we find that the sign of the magnetisation jump $\Delta M$ at the chiral transition distinguishes the regime where $T$ increases with $b$ (MC) from the regime where $T$ decreases with $b$ (IMC). Using these results we found the critical magnetic field $b_c$ where IMC becomes MC as a function of the chemical potential $\mu$. We observed that IMC is triggered at $\mu_e \approx 0.19$ where $b_c$ starts increasing from zero. As previously remarked, our formulas (\ref{eq:univrel}) and (\ref{eq:univrelv2}) are universal and in particular do not rely on the physical mechanism behind inverse magnetic catalysis. Interestingly, the formula (\ref{eq:univrelv2}), although utilised  in the finite density regime, can actually be used at small or even zero chemical potential, where lattice QCD computations are performed. In fact, as explained in the previous section, the use of (\ref{eq:univrel}) and (\ref{eq:univrelv2}) is not restricted to the chiral transition but applies to any phase transition in the $(T,b,\mu)$ phase diagram. 

As observed in \cite{Preis:2010cq,Preis:2012fh}, for the case of small temperatures, the way IMC occurs in the deconfined Sakai-Sugimoto model bears a strong resemblance with the Nambu-Jona-Lasinio model. This suggests a universality of the effect of IMC in models where confinement is absent. It is important to remark, however, that the specific region in parameter space where IMC occurs is model dependent and thus the existence or non-existence of IMC has to be checked on a case by case basis. In some cases it may be that the parameter space in which IMC occurs vanishes completely. In any case, it would be interesting to find a general constraint in holographic QCD backgrounds that exhibit IMC. Since our formulas (\ref{eq:univrel}) and (\ref{eq:univrelv2}) provide universal criteria for IMC in terms of the magnetisation, the natural strategy would be to calculate the magnetisation for a general class of backgrounds and look for a general constraint that leads to a positive jump for the magnetisation at the chiral transition.

There are some interesting scenarios in holographic QCD where further information could be gained from considering the magnetisation as an order parameter for IMC and the use of identities (\ref{eq:univrel}) and (\ref{eq:univrelv2}). For instance, in the recent bottom-up proposals arising from five dimensional dilaton-gravity \cite{Rougemont:2015oea,Gursoy:2016ofp} it should be possible to find a relation between the magnetisation and the beta function for each phase and test a possible connection between (inverse) magnetic catalysis and the response of the beta function to a non-zero magnetic field, as suggested in \cite{Farias:2014eca}. However, there is an important caveat: Although the ad hoc beta function considered in \cite{Farias:2014eca} for the NJL model fits the lattice data well, its origin is not clear. One usually derives the associated beta function for each effective model separately; however, even after considering $b$-dependent parameters, it is not possible to reproduce IMC in a sustained way \cite{Fraga:2013ova}. 

We also intend to further study the effect of IMC in top-down holographic models of QCD that are similar to the S-S model. We are particularly interested in: 
(i) The non-critical $AdS_6$ background that can be lifted to massive type IIA with a Romans mass, cf. e.g.~\cite{Cvetic:1999un, Brandhuber:1999np, Klebanov:1998yya, Kuperstein:2004yk, Elander:2013jqa}; (ii) The (Dymarsky-) Kuperstein-Sonnenschein models of chiral symmetry breaking in the Klebanov-Witten and Klebanov-Strassler backgrounds, cf.~\cite{Kuperstein:2008cq, Dymarsky:2009cm, Ihl:2010zg}, and their generalisation to the Veneziano limit $N_f \sim N_c$ \cite{Ihl:2012bm, Alam:2013cia}. Another interesting framework for investigating IMC is the holographic QCD model proposed in \cite{Iatrakis:2010zf,Iatrakis:2010jb} that combines features of the bottom-up and top-down approaches.

Another interesting direction that our results suggest is the possible connection between the behaviour of the magnetisation and the chiral condensate near the chiral transition, since both act as order parameters that distinguish IMC from MC. We suggest to investigate this in holographic models such as  \cite{Gursoy:2016ofp}, where the chiral condensate is well defined.
In other non-perturbative approaches, e.g. \cite{Mueller:2015fka}, it would be interesting to find a relation between the magnetisation and the gap equation at non-zero magnetic field. 

{\bf Note added:} While revising this paper, Ref. \cite{Gursoy:2017wzz} appeared that investigates the deconfinement and chiral transitions in the $(T,b,\mu)$ phase diagram using a bottom-up holographic QCD model. The authors of Ref. \cite{Gursoy:2017wzz} also analyse the magnetisation as a criterion for distinguishing IMC from MC and their conclusions agree with ours. 


\section*{Acknowledgments}
The authors would like to acknowledge useful conversations and correspondence with Eduardo Fraga, Umut G\"{u}rsoy, Matti J\"{a}rvinen, Luis Mamani, Carlisson Miller, Govert Nijs, Jorge Noronha, Florian Preis, Anton Rebhan and Andreas Schmitt. 
The work of A.B-B is funded by S\~ao Paulo Research Foundation (FAPESP) under the grant 2015/17609-3. M.~I. was funded by the FCT fellowship SFRH/BI/52188/2013 up until the final stages of this work. The Centro de F\'isica do Porto is partially funded by FCT through the project CERN/FIS-NUC/0045/2015.
M.I. and J.S. would like to thank the African Institute for Mathematical Sciences (AIMS) in Muizenberg, South Africa for hospitality and support in form of a short-term visit grant "Research in pairs" in November 2015, and for an ongoing long-term research visit (M.I.).  

\appendix 

\section{Identities for DBI-CS equations}
\label{AppA}

In this appendix we show some identities that are very useful when solving the DBI-CS equations (for more details cf. \cite{BallonBayona:2013gx}). First of all, we recall that the DBI action can be written in terms of $\sqrt{-E}$ where $E$ is the determinant of the tensor 
\bea
E_{mn} = g_{mn} + \beta F_{mn} \,. 
\eea

The first identity is related to the expansion of $\sqrt{-E}$ in five dimensions,  
\bea \label{eq:detexp}
\sqrt{-E}&=&\sqrt{-g}\sqrt{Q} \, , \cr
Q &=& 1+\frac{\beta ^2}{2}F^{mn}F_{mn}+\frac{\beta ^4}{4!} F^{mnpq} F_{mnpq} \ , 
\eea
where
\bea
F_{mnpq} &=& F_{mn}F_{pq}-F_{mp}F_{nq}+F_{mq}F_{np}  \, ,
\eea
is a totally antisymmetric 4-tensor. The next two identities are useful in the DBI-CS equations (\ref{eq:DBICS}),
\bea
\sqrt{-E}E^{<ml>}&=&-\beta
\frac{\sqrt{-g}}{\sqrt{Q}}\left[F^{ml}+\frac{\beta ^2}{2}F_{pq}F^{mlpq}\right] \, , \cr
\frac{\partial E}{\partial \left( \partial_u \tau \right)}&=& 2 E_0 \, g_{\text{xx}} \partial_u \tau  \, ,
\eea
where we have defined the four dimensional determinant $E_0 =\det \left(E_{\mu \nu }\right)$, which admits the expansion
\bea 
E_0 &=& g_0 Q_0 \quad , \quad 
g_0 = \det \left [g_{\mu \nu }\right ] \, , \cr 
Q_0 &:=& 1+\frac{\beta ^2}{2}F^{\mu \nu }F_{\mu \nu }+\frac{\beta ^4}{4!}F^{\mu \nu \rho \sigma }F_{\mu \nu \rho \sigma} \ . 
\eea



\section{Analytic results for the free energy at small magnetic field and temperature}
\label{AppB}

Here we will solve perturbatively the equations of motion \eqref{eq:Tau}, \eqref{eq:f0} and \eqref{eq:f3}   
 for both the chirally broken and the chirally symmetric phase in order to arrive at analytic expressions for the free energy. This analysis complements the numerical analysis we have 
presented in section \ref{sec:Numerics}.


\subsection{Chirally broken phase}
\label{AppCBP}

The equations of motion of the three functions $\hat{\tau}$, $\hat{f_0}$ and $\hat{f_3}$ that appear in the 
chirally broken phase are
\begin{eqnarray}\label{XB}
\frac{v^3\,\hat{\tau}' h \, \sqrt{u^5+b^2 u^2}}{\sqrt{1+h \hat{f}_3'^2 - \hat{f}_0'^2 + v^3 h \, \tau'^2}} \, = \,  \hat{k}\, , \quad 
\frac{\hat{f}_0'\sqrt{u^5 + b^2u^2}}{\sqrt{1+h \hat{f}_3'^2 - \hat{f}_0'^2 + v^3 h \, \tau'^2}} \, = \, - 3 b \hat{f}_3 \, ,
\nonumber \\[1.5ex]
\frac{\,\hat{f}_3' h \sqrt{u^5 + b^2 u^2}}{\sqrt{1+h \hat{f}_3'^2 - \hat{f}_0'^2 + v^3 h \, \tau'^2}} \, =
 \, -  3 b \hat{f}_0 + \hat{d} \, ,
\end{eqnarray}
with integration constants $\hat{d}$ and $\hat{k}$.\footnote{The value of $\hat{d}$ is determined by extremising the free energy, 
as in the zero temperature analysis of \cite{Preis:2010cq}, 
so we have $\hat{d} = \frac{3}{2} b \, \mu$.} The boundary conditions we will use are  
\begin{equation}
\hat{f}_0(\infty) = - \mu \, , \quad \hat{f}_3(v_0) = 0 \, \quad 
\hat{\tau}'(v_0) = \infty \, , \quad \mathrm{and} \quad \frac{1}{2}=\int_{v_0}^{\infty}dv\, \hat{\tau}' \, .\label{XBBC}
\end{equation}
In order to solve the system of equations \eqref{XB} perturbatively for small $T$ and $b$, we consider 
expansions of the following form\footnote{Note that the first index is related to the magnetic field
and the second to the temperature.}
\begin{equation} \label{expansion.ansatz}
W[v,\mu] \, = \, \sum_{{i,j}=0}W^{ij}[v,\mu] \, b^i \, v_{T}^j \, , 
\end{equation}
for every one of the functions and constants (namely $\hat{f}_0$, $\hat{f}_3$, $\hat{\tau}$, 
$\hat{k}$ and $v_0$) that appear in \eqref{XB}. 
Substituting \eqref{expansion.ansatz} into \eqref{XB}, expanding in $T$ and $b$, and using the boundary conditions
\eqref{XBBC} in every step of the expansion, we arrive at the following expressions for the different functions and 
constants\footnote{We restrict the analysis to the first non-zero temperature correction that will 
affect the calculation of the free energy. 
It is possible to continue the perturbative analysis for higher values of the exponent of $v_T$. Note 
that the zero temperature result can be obtained by expanding the semi analytic 
solution of appendix A of \cite{Preis:2010cq}, for small values of the magnetic field.}.
We start from the expansion of $v_0$
\begin{equation}
v_0 = \left(v_{00}+v_T^3 v_{03}\right) \, + \, b^{2} \left(v_{20}+v_T^3 v_{23}\right)
\end{equation}
with\footnote{
$P_1 \equiv \frac{2\sqrt{\pi}\,\Gamma\left(\frac{9}{16}\right)}{\Gamma\left(\frac{1}{16}\right)}$
and
$P_2 \equiv \frac{\sqrt{\pi}\, \Gamma\left(\frac{3}{16}\right)}{8\Gamma\left(\frac{11}{16}\right)}$.}
\begin{eqnarray}
&& v_{00} = \frac{16 \pi  \Gamma \left(\frac{9}{16}\right)^2}{\Gamma \left(\frac{1}{16}\right)^2} \, ,  \quad
v_{20} = \frac{1}{8v_{00}^2} \left[\cot\frac{\pi}{16}-1-
\left(\frac{3\mu}{2v_{00}}\right)^2\left(\frac{3P_2}{P_1}-1\right)\right]  \, , 
\nonumber \\
&& v_{03} = \frac{1}{56 v_{00}^2} \, \left[8+\sqrt{2}+\sqrt{2 \left(2+\sqrt{2}\right)}\right] \, , \quad
v_{23} =-16.6821 +54.9847 \, \mu ^2 \, . 
\end{eqnarray}
For the expansion of $ \hat{k}$, we have
\begin{equation}
\hat{k} = \left(v_{00}^4+v_T^3 \hat{k}_{03}\right) \, + \, b^{2} \left(\hat{k}_{20}+v_T^3 \hat{k}_{23}\right)
\end{equation}
with
\begin{eqnarray}
&&\hat{k}_{20} = \frac{1}{2} v_{00} \cot \left(\frac{\pi }{16}\right) - 
\frac{18 \, \mu ^2 \cos \left(\frac{\pi }{16}\right) \Gamma\left(\frac{7}{16}\right) \Gamma \left(\frac{17}{16}\right) \Gamma 
\left(\frac{19}{16}\right)}{\pi \, v_{00} \,\Gamma \left(\frac{11}{16}\right)} \, , 
\nonumber \\
&& \hat{k}_{03} = 4 \, v_{00}^3 \,v_{03}-\frac{1}{2}\, v_{00} \, , \quad
\hat{k}_{23} =-6.36931 - 17.4008 \, \mu^2 \, . 
\end{eqnarray}
For the functions $\hat{f}_0$ and $\hat{f}_3$, we have
\begin{eqnarray} \label{XBexpansions}
& \hat{f}_0[v]  \, = \, - \, \mu \, - \, \left[\hat{f}_0^{20}[v] \, + \, \hat{f}_0^{23}[v] \, v_T^3\right] b^2 \, ,   &
\\[5pt]
& \hat{f}_3[v]  \, = \left[\hat{f}_3^{10}[v] \, + \, \hat{f}_3^{13}[v] \, v_T^3 \right] b 
\quad \& \quad
\hat{\tau}  \, = \, \left[\hat{\tau}^{00} \, + \, \hat{\tau}^{03}\, v_T^3 \right]  \, + \, 
\left[\hat{\tau}^{20} \, + \, \hat{\tau}^{23}\, v_T^3 \right] b^2 \, .  & 
\nonumber
\end{eqnarray}
All the functions in the \eqref{XBexpansions} expansion are analytic (the majority has complicated, non-illuminating expressions)
except for $ \hat{f}_0^{23}$ that can only be calculated numerically (solving a simple integral).
Here we list the analytic expressions for three of the eight functions, namely $\hat{f}_0^{20}$, $\hat{f}_3^{10}$  
and $\hat{\tau}^{00}$,
\begin{equation}
\hat{f}_0^{20}[v,\mu] = \frac{\mu}{v^3}  \, _2F_1\left(\frac{3}{16},\frac{1}{2};\frac{19}{16};\frac{v_{00}^8}{v^8}\right) \left[\,
_2F_1\left(\frac{3}{16},\frac{1}{2};\frac{19}{16};\frac{v_{00}^8}{v^8}\right)-\frac{2 \sqrt{\pi } v^{3/2}
\Gamma \left(\frac{19}{16}\right)}{v_{00}^{3/2} \Gamma \left(\frac{11}{16}\right)}\right] \, ,
\end{equation}
\begin{equation}
\hat{f}_3^{10}[v,\mu] = \mu \left[\frac{\sqrt{\pi } \, \Gamma \left(\frac{19}{16}\right)}{v_{00}^{3/2} \Gamma \left(\frac{11}{16}\right)} 
-\frac{1}{v^{3/2}} \, _2F_1\left(\frac{3}{16},\frac{1}{2};\frac{19}{16};\frac{v_{00}^8}{v^8}\right) \right] \, ,
 \end{equation}
\begin{equation}
\hat{\tau}^{00} [v] = \frac{2}{15} \,\frac{\sqrt{v^8-v_{00}^8}}{v_{00}^{12} v^{1/2}} \left[7 v^8 \,
   _2F_1\left(1,\frac{23}{16};\frac{31}{16};\frac{v^8}{v_{00}^8}\right)+15 v_{00}^8\right]
   +\frac{14 i \sqrt{\pi }\,  \Gamma \, \left(\frac{31}{16}\right)}{15 \sqrt{v_{00}} \, \Gamma
   \left(\frac{23}{16}\right)} \, , 
\end{equation}
while for the rest we present plots in Fig. \ref{CSapproxPlots} for $\mu=1$. 
\begin{figure}[h] 
   \centering
   \includegraphics[width=4.7cm]{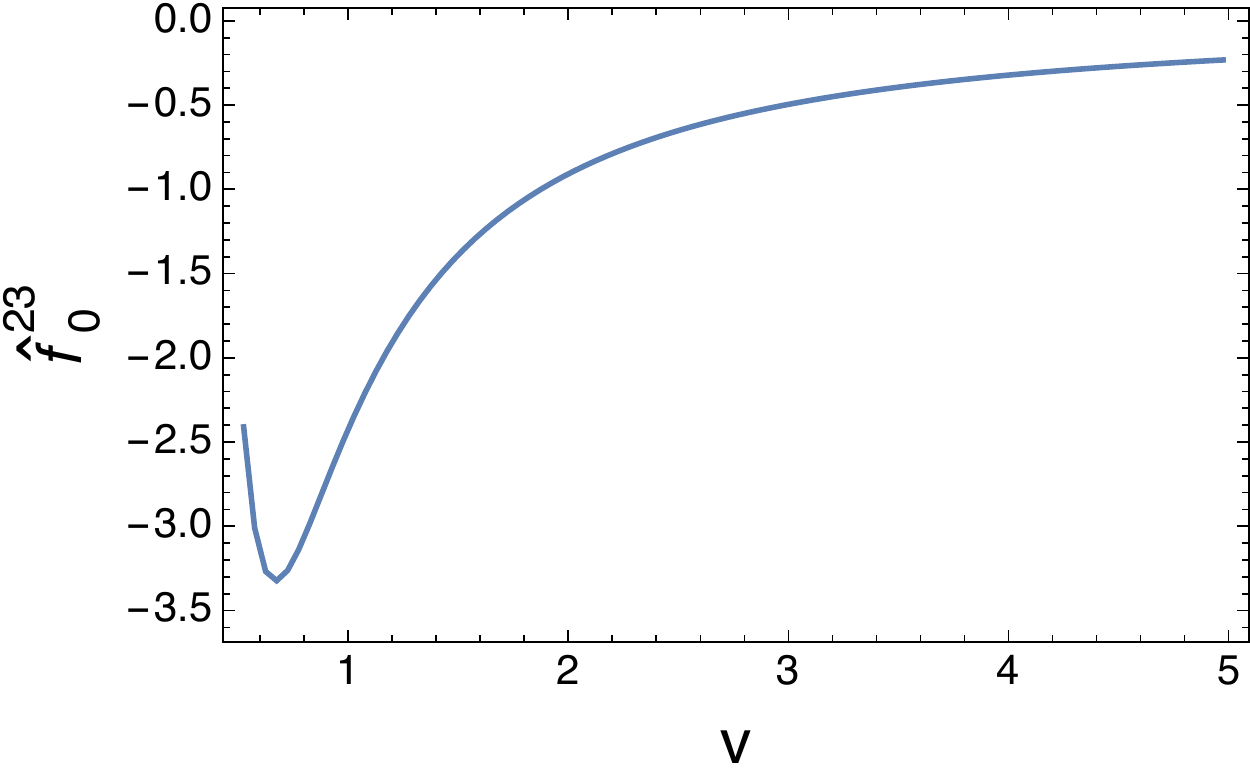}
   \hspace{0.2cm}
    \includegraphics[width=4.7cm]{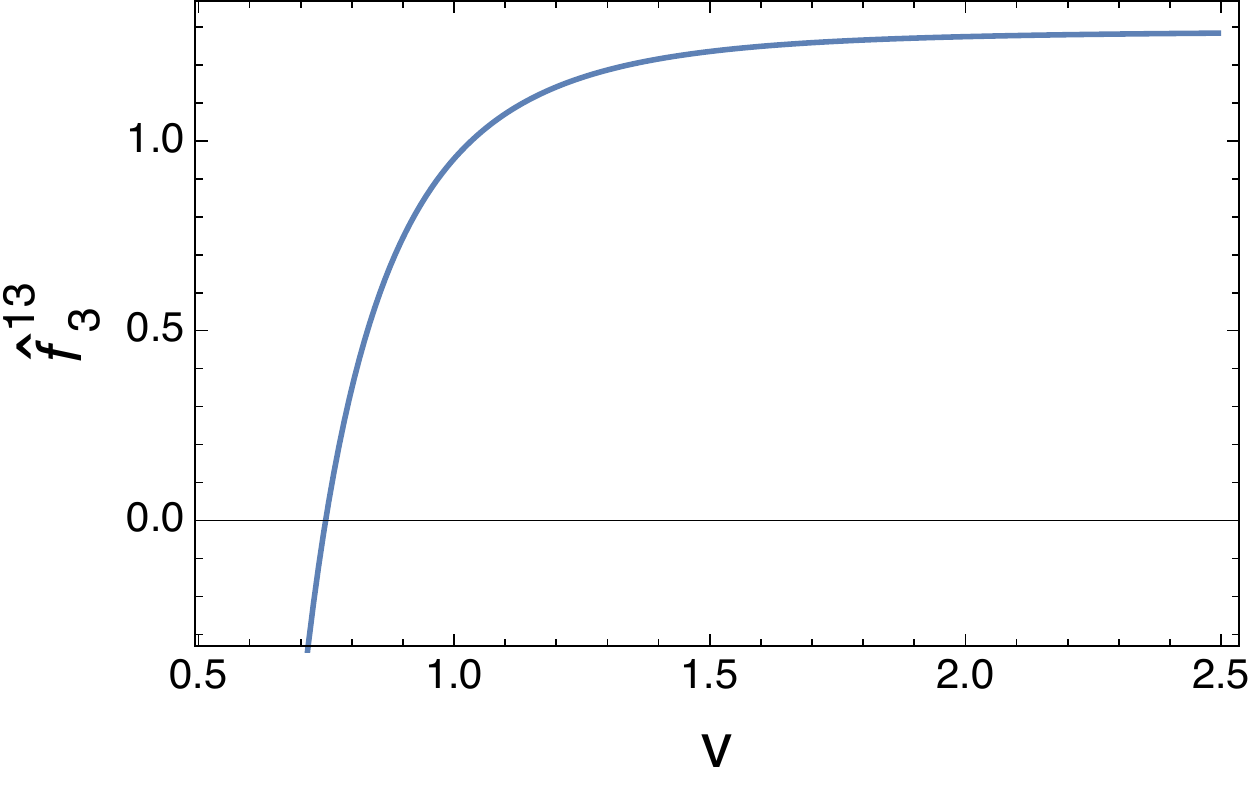}
    \hspace{0.2cm}
    \includegraphics[width=4.7cm]{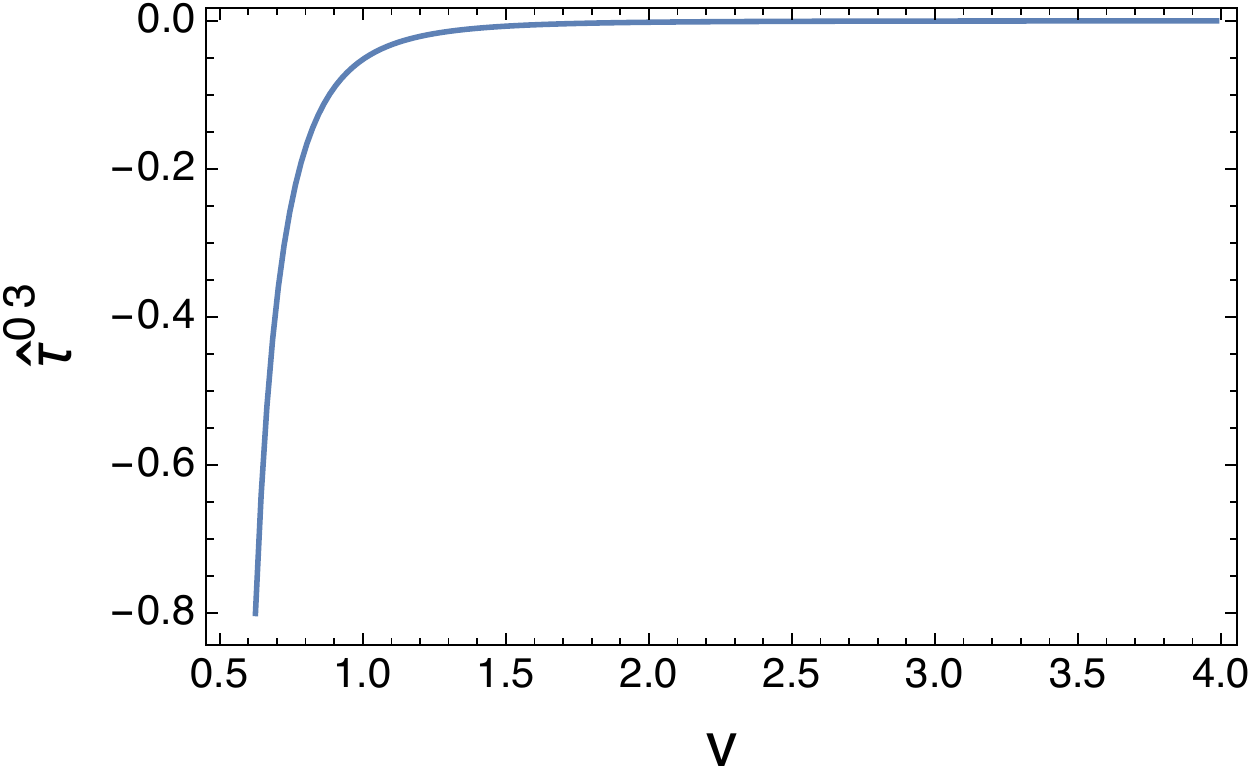}
    \hspace{0.2cm}
    \includegraphics[width=4.7cm]{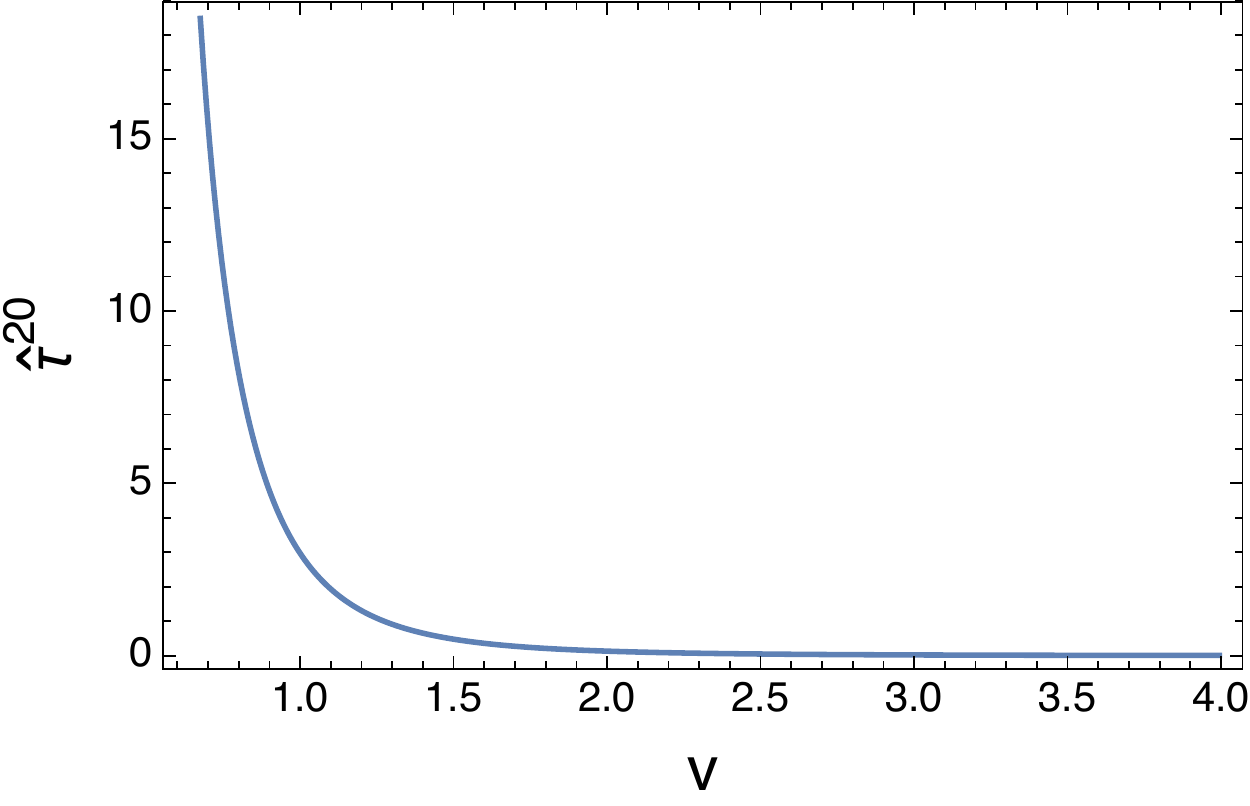}
    \hspace{0.2cm}
    \includegraphics[width=4.8cm]{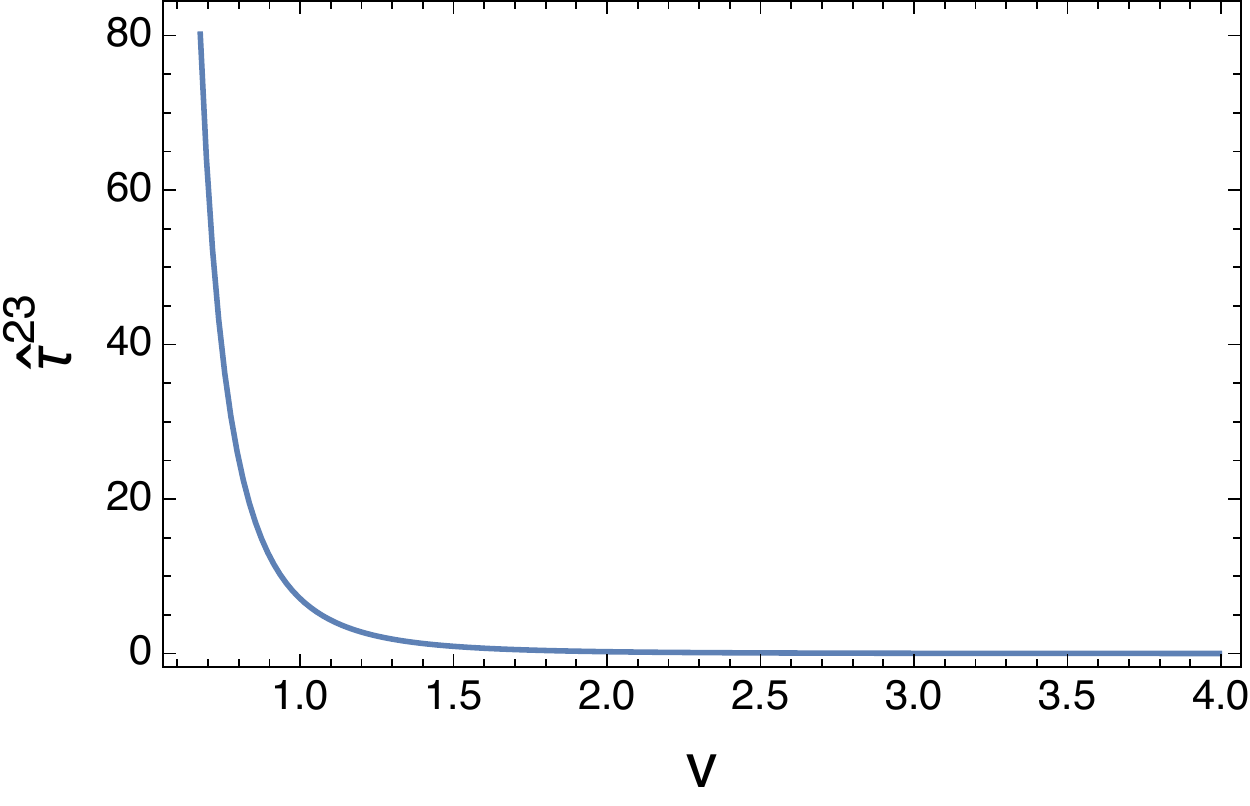}
       \caption{Plots for the functions $\hat{f}_0^{23}$, $\hat{f}_3^{13}$, $\hat{\tau}^{03}$, $\hat{\tau}^{20}$ and 
       $\hat{\tau}^{23}$ that appear in the expansion
       \eqref{XBexpansions}. Note that with the exception of the function $\hat{f}_0^{23}$, all the others that appear in this figure 
       have a singularity as $v$ approaches $v_{00}$. However,  since the brane bents before reaching $v_{00}$, the singularity
       is never approached and the function is perfectly smooth at the tip of the brane. We plot from $v_{00}$ but the real 
       solution starts on the right of that point, where none of the functions is infinite.}
   \label{CSapproxPlots}
\end{figure}
The function $\hat{f}_3^{13}$, in the limit $v \rightarrow \infty$, approaches the value
\begin{equation}
\hat{f}_3^{13}[\infty,\mu] \approx \frac{\mu \sqrt{\pi}}{80 \, v_{00}^{9/2} } 
   \left[\frac{9 \cot \left(\frac{3 \pi }{16}\right) \Gamma \left(\frac{21}{16}\right)
   \left(1-8 \,v_{00}^2 \, v_{03}\right)}{\Gamma \left(\frac{13}{16}\right)}+\frac{49 \tan \left(\frac{\pi
   }{16}\right) \Gamma \left(\frac{31}{16}\right)}{\Gamma \left(\frac{23}{16}\right)}\right] = 1.28895 \mu \, . 
\end{equation}
Putting all these ingredients together in the equation for the Hamiltonian \eqref{Hamiltonian}, 
it is possible to obtain the expression for the free energy. 
To obtain a finite result we subtract the vacuum contribution, expand up to order 
${\cal O}(b^2)$ and ${\cal O}(v_T^3)$, and finally obtain the following result
\begin{eqnarray} \label{FEUapprox}
\frac{\Omega_\cup}{\mathcal{N}} &\simeq &  - \frac{2}{7}\frac{P_1 \, v_{00}^{7/2}}{2} \,
\left(1+\frac{v_T^3}{2 \, v_{00}^3} \, \cot \frac{\pi}{16} \right)
-b^2\,\left(\frac{P_1v_{00}^{1/2}}{2} \cot\frac{\pi}{16}+\frac{9\mu^2P_2}{8 v_{00}^{3/2}}\right) 
\nonumber \\ [4pt]
&& \, -  b^2 v_T^3 \left(2.293 \, + \, 6.0386 \,\mu^2 \right) \, . 
\end{eqnarray}
In order to make the comparison with the zero temperature result (see eq (D.2) of \cite{Preis:2010cq}), we have used 
the notation of that paper. 


\subsection{Chirally symmetric phase}
\label{AppCSP}

In the case of disconnected $D8$ flavour branes, we will write the system of equations of motion 
and boundary conditions after performing the following change of variables and a redefinition of the constants
\begin{equation} \label{XSredefinitions}
V = \frac{v}{\mu}\, , \quad 
V_{T} = \frac{v_T}{\mu}\, , \quad
\epsilon = \frac{b}{\mu^{3/2}} \, , \quad
\hat{c} = \hat{C} \, \mu^{5/2} \, , \quad
\hat{F}_3 = \frac{\hat{f}_3}{\mu} \quad \mathrm{and} \quad
\hat{F}_0 = \frac{\hat{f}_0}{\mu} \, . 
\end{equation}
In this way the equations of motion are unchanged ($b$ has to be traded for $\epsilon$) 
and only the boundary condition for $\hat{f}_0$ changes (now the 
value at infinity is not  $\mu$ anymore but $1$ instead). To avoid clutter,
we keep the same notation as before but set the value of 
$\mu$ equal to 1. 
The equations of motion then become
\begin{equation}
\frac{\hat{f}_0' \, \sqrt{v^5 \, + \, b^2 \, v^2}}{\sqrt{1\, + \, h\, \hat{f}_3'^2 \, - \, \hat{f}_0'^2}}  = 
- \, 3 \, b \, \hat{f}_3 - \hat{c} 
\quad \mathrm{and} \quad
\frac{h \, \hat{f}_3' \, \sqrt{v^5 \, + \, b^2 \, v^2}}{\sqrt{1 \, + \, h \, \hat{f}_3'^2 \, - \, \hat{f}_0'^2}} = 
 - \, 3 \, b \, \hat{f}_0 \, , \label{XS}
\end{equation}
with integration constant $\hat{c}$, while the boundary conditions for the fields $\hat{f}_0$ and $\hat{f}_3$ are 
\begin{equation} \label{XSBC}
\hat{f}_0\left(v_T\right) = 0 \, , \quad  
\hat{f}_0(\infty ) = - 1 \quad \mathrm{and} \quad
\hat{f}_3(\infty ) = 0 \ .
\end{equation}
Considering expansions of the form \eqref{expansion.ansatz} for every quantity that appears in  \eqref{XS}, 
expanding in $T$ and $b$, and using the boundary conditions
\eqref{XSBC} in every step of the expansion,
we arrive at the following expressions for the different functions and constants,
\begin{eqnarray} \label{XSexpansions}
& \hat{f}_0[v]  \, = \, - \, \left[\hat{f}_0^{00}[v] \, + \, \hat{f}_0^{01}[v] \, v_T\right] \, - \, 
\left[\hat{f}_0^{20}[v] \, + \, \hat{f}_0^{21}[v] \, v_T\right] b^2 \, ,   &
\\[5pt]
& \hat{f}_3[v]  \, = \, - \, \left[\hat{f}_3^{10}[v] \, + \, \hat{f}_3^{11}[v] \, v_T\right] b 
\quad \mathrm{and} \quad
\hat{c}  \, = \left[\hat{c}_{00} \, + \, \hat{c}_{01} \, v_T\right] \, + \, 
\left[\hat{c}_{20} \, + \, \hat{c}_{21}\, v_T\right] b^2 \, .  & 
\nonumber
\end{eqnarray}
For the expansion of the constant $\hat{c}$, we get the following 
\begin{align} 
\hat{c}_{00} \, = \, \left[\frac{\sqrt{\pi }}{\Gamma \left(\frac{3}{10}\right) \Gamma \left(\frac{6}{5}\right)}\right]^{5/2} \, , 
\quad 
\hat{c}_{01} \, = \, \frac{3 \pi ^{5/4}}{\Gamma \left(\frac{3}{10}\right)^{5/2} 
\Gamma \left(\frac{6}{5}\right)^{3/2} \Gamma \left(\frac{11}{5}\right)}
\nonumber \\[5pt]
\hat{c}_{20} \, = \, \frac{9 \Gamma \left(\frac{3}{10}\right)^3 \Gamma \left(\frac{6}{5}\right)^3-\pi  \Gamma 
\left(-\frac{1}{10}\right) \Gamma \left(\frac{8}{5}\right)}{12 \pi ^{5/4}
\sqrt{\Gamma \left(\frac{3}{10}\right) \Gamma \left(\frac{6}{5}\right)}} \, , 
\quad 
\hat{c}_{21} \, \approx \, 5.56359 \, . 
\end{align}
For the functions $\hat{f}_0^{00}$, $\hat{f}_3^{10}$ and $\hat{f}_0^{01}$, there are simple analytic expressions, namely
\begin{equation} 
\hat{f}_0^{00}[v] \, = \, v \, _2F_1\left(\frac{1}{5},\frac{1}{2};\frac{6}{5};-\frac{v^5}{\hat{c}_{00}^2}\right)
\end{equation}
\begin{equation} 
\hat{f}_3^{10}[v] \, = \, -\frac{3}{2 \, \hat{c}_{00} }
\left[1-\hat{f}_0^{00}[v]^2 \right]
\end{equation}
\begin{equation} 
\hat{f}_0^{01}[v] \, = \,\frac{2 \hat{c}_{01}}{5 \hat{c}_{00}} \, \hat{f}_0^{00}[v] -
\frac{2 \hat{c}_{01}  v}{5 \sqrt{\hat{c}_{00} ^2+v^5}}-1
\end{equation}
while for the $\hat{f}_0^{20}$, $\hat{f}_3^{11}$ and $\hat{f}_0^{21}$, we have either 
non-illuminating analytic or numerical expressions. 
For that reason, we choose to plot them in Fig. \ref{CSapproxPlots}, instead of providing the explicit expressions.

\begin{figure}[h] 
   \centering
   \includegraphics[width=4.7cm]{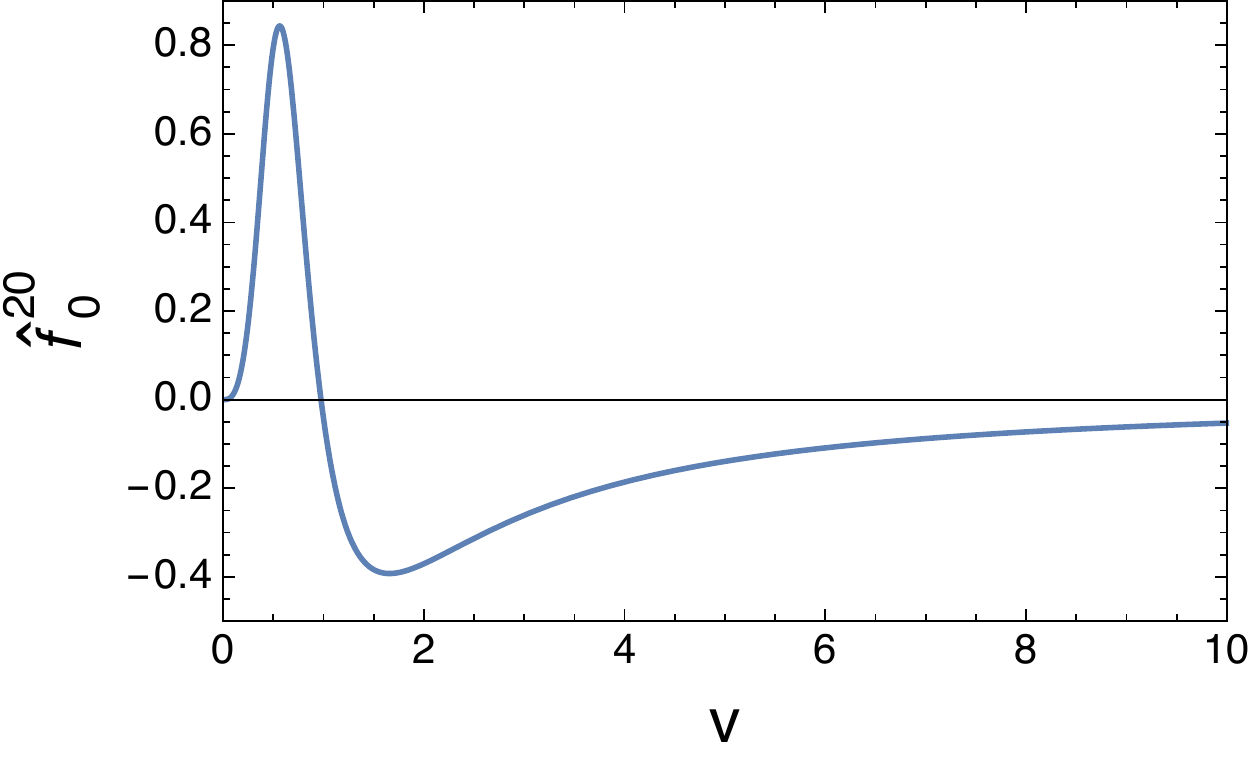}
   \hspace{0.2cm}
    \includegraphics[width=4.7cm]{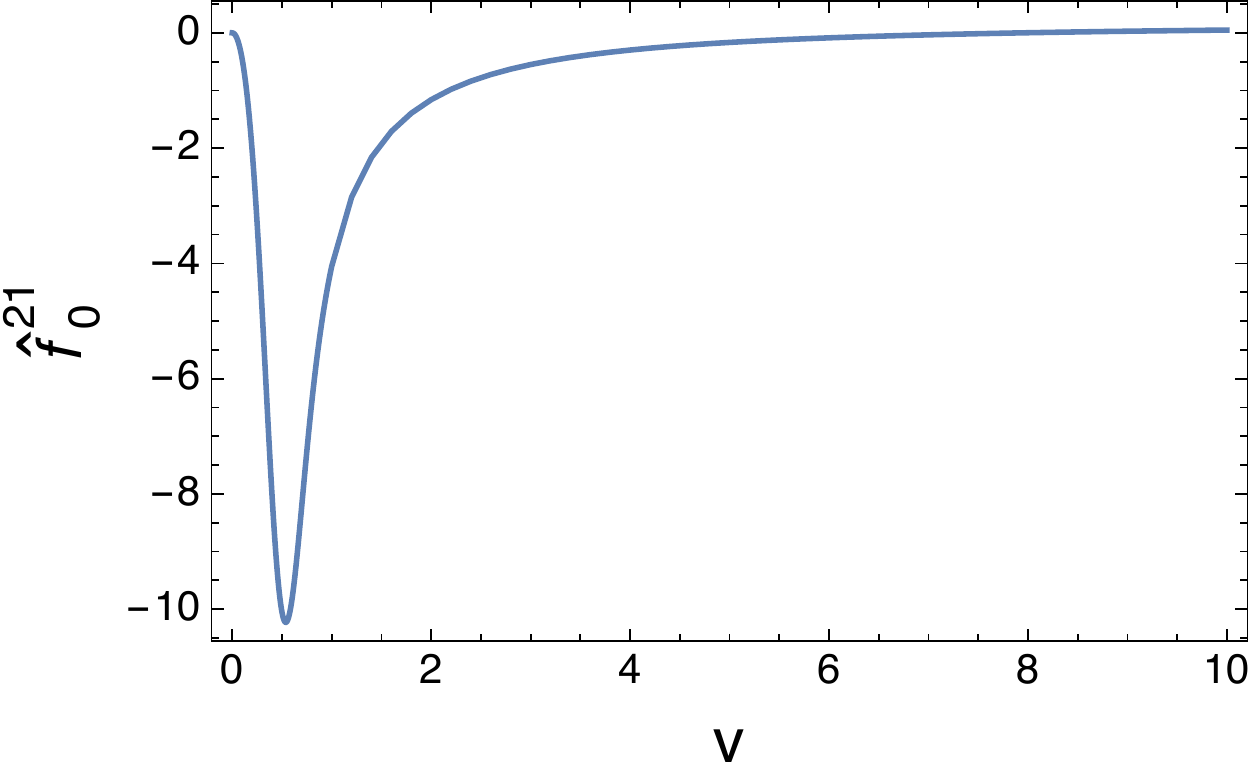}
     \hspace{0.2cm}
    \includegraphics[width=4.7cm]{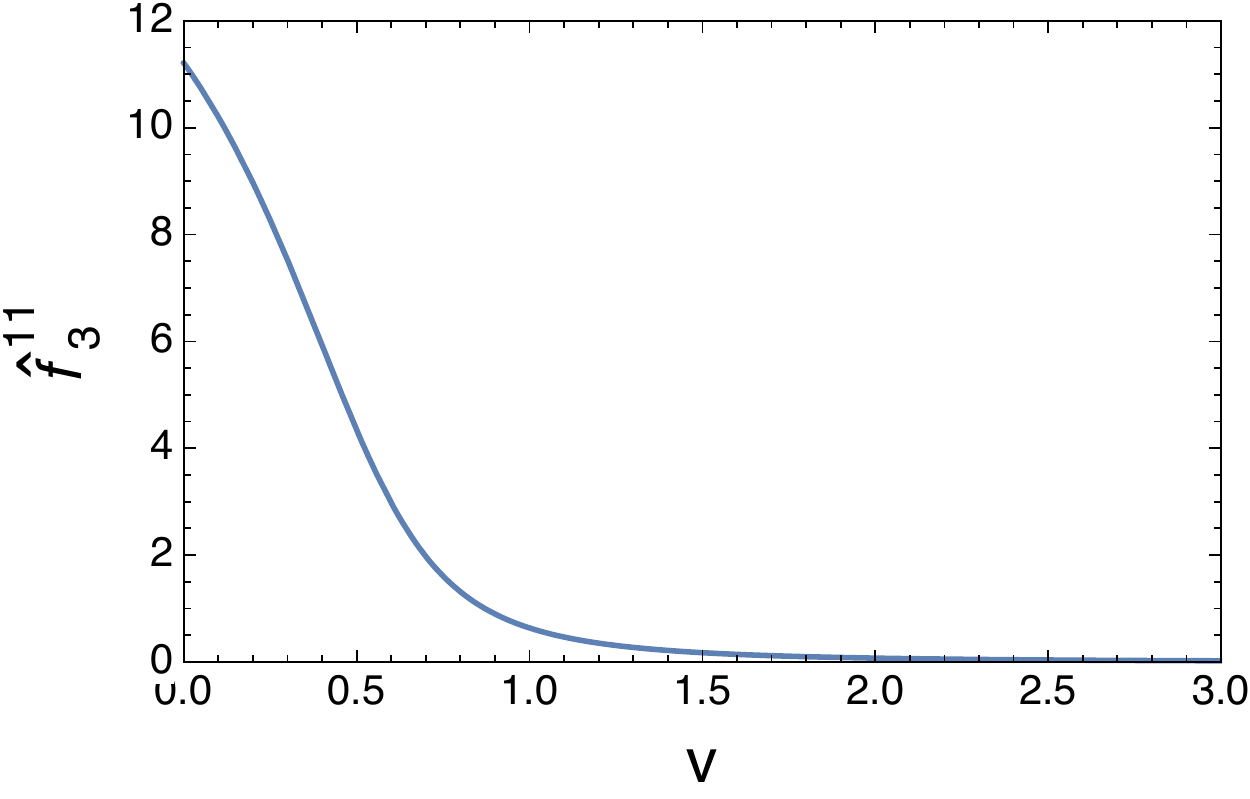}
       \caption{Plots for the functions $\hat{f}_0^{20}$, $\hat{f}_3^{11}$ and $\hat{f}_0^{21}$ that appear in the expansion
       \eqref{XSexpansions}.}
   \label{CSapproxPlots}
\end{figure}

Putting all these ingredients together in the equation for the Hamiltonian, it is again possible to obtain the expression for the free energy\footnote{In \eqref{FEIIapprox} we have reinserted $\mu$ and we have expressed the free energy as a function of $b$, $v_T$ and $\mu$.}
\begin{equation} \label{FEIIapprox}
\frac{\Omega_{||}}{\cal N} \simeq -\frac{2}{7} \, \frac{\mu^{7/2}}{Q_1^{5/2}}\,
\left(1 + \frac{v_T}{\mu}\right) \, - \, 
Q_3\sqrt{\mu}\, \left(1 + 0.78 \, \frac{v_T}{\mu}\right)\,b^2\, ,  
\end{equation}
where 
\begin{equation} 
Q_3 =  \frac{3}{2}Q_1^{5/2} + \frac{\Gamma\left(\frac{9}{10}\right)\Gamma\left(\frac{3}{5}\right)}{Q_1^{1/2}\sqrt{\pi}} 
\quad \& \quad 
Q_1 = \frac{\Gamma\left(\frac{3}{10}\right)\Gamma\left(\frac{6}{5}\right)}{\sqrt{\pi}} \, .
\end{equation}
Using the same notation as in \cite{Preis:2010cq}, the comparison with the zero temperature 
result (see the eq. (D.3) of that paper) is immediate. 



\end{document}